\documentclass{jfm}
\usepackage{graphicx}
\usepackage{caption}
\usepackage{subcaption}
\usepackage{tikz}
\usepackage{placeins}
\usepackage{float}
\usepackage{comment}
\usepackage{newtxtext}
\usepackage{newtxmath}
\usepackage{amsmath}
\usepackage{natbib}
\usepackage{hyperref}
\usepackage{booktabs}
\usepackage{siunitx} 
\hypersetup{
    colorlinks = true,
    urlcolor   = blue,
    citecolor  = black,
}

\newcommand{\RomanNumeralCaps}[1]

\title{Dynamics of singular-jet formation during the collapse of drop-impact craters.}

\author{
  Nivesh Mishra\aff{1}
  \and
  Anikesh Pal\aff{1}
  \corresp{\email{pala@iitk.ac.in}}
}

\affiliation{
\aff{1}Department of Mechanical Engineering, Indian Institute of Technology Kanpur, Kanpur 208016, U.P., India
}

\begin{document}
\maketitle

\begin{abstract}
The collapse of a drop-impact crater that bypasses the pinch-off singularity produces exceptionally thin, high-velocity jets, often referred to as singular-jets. We perform two-dimensional axisymmetric direct numerical simulations using Basilisk, employing adaptive mesh refinement to capture the intricate flow features. The numerical results for a drop diameter of $3.64$ mm at ambient pressure match the experiments well under different impact conditions. During the crater collapse, qualitatively different dimple shapes form, including an entrapped bubble beneath the dimple. Quantitatively, an entrapped bubble with a diameter of $\sim 210\,\mu\text{m}$ and a jet velocity of $61.
7~\mathrm{m\,s^{-1}}$ for the singular-jet case during cylindrical dimple collapse is obtained from simulations. Subtle morphological changes in crater evolution distinguish the jetting regime, with a critical angle of $109.2^\circ$ between the crater side wall and its base in the singular-jet case. The simulation provides insight into the internal flow fields, enabling us to delineate the observed phenomenon. The pressure and velocity fields reveal that the high jetting velocity is associated with the formation and rapid collapse of a nearly cylindrical dimple. The resulting pressure distribution drives a strongly upward flow near the dimple base, while the nearly parallel dimple side walls redirect radial liquid motion towards the axis. We extend the simulations into the post-entrainment region and identify another singular-jet near the upper boundary of the entrainment region. The jet reaches a velocity of $63.93~\mathrm{m\,s^{-1}}$ and exhibits morphological characteristics similar to the first singular-jet case. A regime map is constructed from simulations spanning Weber numbers ranging from $100$ to $200$ and Froude numbers from $45$ to $92$; the updated lower and upper boundaries for the entrainment region, based on singular-jet formation, are $We = 45.5\, Fr^{0.237}$ and $We = 53.5\, Fr^{0.251}$, respectively.

\end{abstract}


{\bf MSC Codes }  {\it(Optional)} Please enter your MSC Codes here

\section{Introduction}
\label{sec:Introduction}

\begin{figure}
\centering
\includegraphics[height=5cm,width=1.0\linewidth]{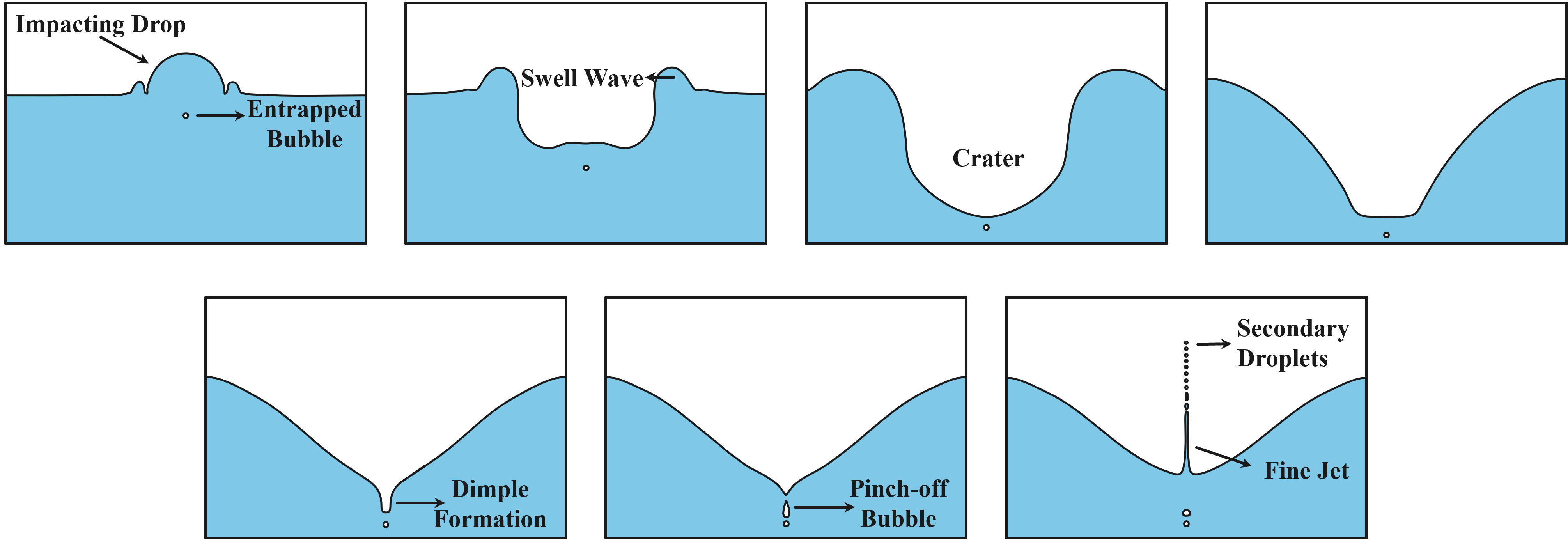}
\captionsetup{justification=centering}
\caption{Schematic diagram showing the different phenomena after a drop impacts a liquid pool.}
\label{fig:Process}
\end{figure}

Drop impact on a liquid pool has piqued researchers' interest because of its phenomenology, as shown in the initial work of \cite{worthington1908study}. It comprises a sequence of events, as illustrated in figure \ref{fig:Process}, that begins with the entrapment of a thin air sheet beneath the pool surface during drop impact \citep{franz1959splashes,oguz1990bubble,pumphrey1990entrainment}. The impact ejects a splash above the pool surface, known as a swell-wave, and forms a hemispherical cavity in the pool, referred to as a crater, which retracts after reaching a maximum depth and can form a dimple. Under suitable impact conditions, this may lead to a bubble pinch-off. After the pinch-off, it produces two jets that propagate in opposite directions. The downward jet drives the pinch-off bubble to the bottom of the pool until gravity dominates. At the same time, the upward jet rises above the undisturbed pool surface and splits into fine droplets once it reaches a certain height. These droplets are known as secondary droplets. The phenomenon can indeed be observed when a raindrop hits the sea surface \citep{tsimplis1989wave}, binder drops in the powder-bed technique \citep{sun2023droplet}, high-pressure atomized droplets in the galvanization process \citep{Bierbrauer2004}, or in an engine’s combustion chamber \citep{williams1973combustion}. The sequence of events accompanying a drop impact on a liquid pool is pertinent to various natural phenomena and industrial applications, such as the entrapped air during impact, which facilitates the exchange of gases and biomaterials between the atmosphere and water bodies, transporting contaminants into the bulk liquid in a chemical process \citep{blanchard1957bubble,esmailizadeh1986bubble}. It also finds relevance in inkjet printing \citep{yarin2006drop,lohse2022fundamental}. Raindrops that fall into puddles contribute to soil erosion and degrade soil health \citep{ellison1950soil}. The bubble formed owing to pinch-off acts as a driving force in volcanic eruptions \citep{gonnermann2007fluid}, and its bursting at the free surface mainly contributes to cell damage in bioreactors \citep{tran2016hydrodynamic,walls2017quantifying}. Bubble bursting boosts aroma dispersion and transports savory compounds to the nose, and has significant implications for the wine and beverage industries \citep{seon2017effervescence}. The upward jet, after reaching a particular height, starts to split into fine drops, which serve as nucleation sites for cloud formation (\citep{kientzler1954photographic,lewis2004sea,veron2015ocean}, transmitting germs, viruses, and other dangerous chemicals, such as DDT and pathogens \citep{Blanchard1970,walls2014moving}. The widespread occurrence of drop-impact crater collapse and its significance in both natural phenomena and industrial processes have attracted researchers for centuries.\\

Following impact, the initial dynamics leading to crater formation are broadly similar. However, the subsequent crater collapse are primarily governed by three non-dimensional numbers that serve as control parameters: the  Weber number, is the ratio of inertial to surface-tension force given by $We = \rho V^2 D / \sigma$; the Froude number, characterizes the ratio of inertial to gravitational forces can be expressed as, $Fr = V^2 / g D$; and the Reynolds number, quantifies the relative importance of inertial to viscous forces and is defined as $Re = \rho V D / \mu$, where $\rho$ is the density, ${V}$ is the velocity of the drop, $D$ is the drop diameter, $\sigma$ is the surface tension coefficient, $\mu$ is the dynamic viscosity, and $g$ is the acceleration due to gravity, resulting in distinct crater collapse behaviors. In several cases, the air bubble is entrapped, whereas in some cases, pinch-off is missed after crater collapse. This leads to the formation of distinct jets; among them, singular-jets are captivating. A singular-jet, owing to its thin cylindrical structure and extremely high jetting velocity, is of particular interest. Their jetting velocity is approximately $35-45$ times the drop-impact velocity. The jets emerge during the final stage of crater collapse; therefore, understanding crater dynamics is an important step for the underlying mechanism responsible for singular-jet formation. \\


When a drop impinges on the pool surface, it either coalesces \citep{marrucci1969theory,rodriguez1988penetration,cai1989phenomena} or splashes \citep{worthington1908study,rein1993phenomena,morton2000investigation,leng2001splash,fedorchenko2004some}. In coalescence, the drop is absorbed at the pool surface, forming a small crater that collapses vertically. Between coalescence and splashing, a transitional regime exists in which the coalescence phenomenon shifts toward partial coalescence \citep{rein1996transitional,chen2006partial}. The present study focuses on singular-jet formation; however, in coalescence or partial coalescence, jet formation after crater collapse is absent. \\


The impact impulse generates vorticity that grows with the crater, forming vortex rings that entrain the drop material into the pool. These vortex rings detach from the base as the crater begins to retract and penetrate deep into the pool \citep{thomson1886v,chapman1967formation}, and are closely associated with low impact drop velocities. They also differentiate between coalescence and splashing. As we move toward splashing, the vorticity generated by impact dissipates well in the liquid pool. A force balance between surface tension and dynamic pressure induced by impact yields a critical Weber number $(We_{sp})$, beyond which vortex rings are not formed \citep{cresswell1995drop}. \\


In splashing, the drop-impact creates a deep crater that collapses laterally, giving rise to a jet that breaks into numerous microdrops once it reaches a certain height. The impact pushes a liquid film, mainly the pool liquid, that is ejected vertically upward from the pool surface at the rim of the crater. At a low $We$, the film becomes stable and forms a swell-wave. The impact energy increases with $We$, which makes the ejected liquid film thin, leading to droplet pinch-off at the periphery of the splash, forming a crown-like profile above the pool surface \citep{pumphrey1989underwater}. Thereafter, as the $We$ exceeds a certain threshold, the impact energy becomes high enough to raise and bend the crown wall inward, forming a bubble canopy above the crater. On its closure, two jets are initiated, one in an upward direction and the other in a downward direction, which pierce the bubble canopy and interact with the central jet formed after crater collapse \citep{engel1966crater,engel1967initial,medwin1992anatomy,wang2023analysis}. \\

The splashing regime further divides into distinct sub-regimes, depending on the impact conditions of the drop.  These sub-regimes are bubble entrainment following crater collapse, the formation of a thick jet with or without a secondary drop, and thin-jet formation accompanied by myriad droplets. \cite{pumphrey1990entrainment} described different ways of bubble entrainment by drop-impact. The process begins when the crest of a capillary wave, generated on a drop surface by impact, contacts the pool surface and entrains a small packet of air, which surface tension splits into multiple bubbles. The entrainment was known as Mesler entrainment \citep{esmailizadeh1986bubble}. Subsequently, a single bubble was entrapped, followed by crater collapse for a narrow range of impact velocity and drop diameter, known as regular entrainment. \cite{pumphrey1990entrainment} define this region in the $We- Fr$ plane as being dominated by surface tension and gravity. They observed that a capillary wave from the crater-wall, upon reaching the bottom, forms a dimple, further causing bubble entrainment. \cite{oguz1990bubble} were the first to address regular bubble entrainment and elucidate it through simulation using the boundary integral method. They reported that for the upper bound of bubble entrainment, the drop material must spread over the crater surface. The retracting surface-tension force and the pressure balance on the drop surface yield $We \sim Fr^{1/4}$ as the upper bound for regular entrainment. \cite{oguz1990bubble} observe that the bubble does not entrap if the capillary wave reaches the bottom of the crater before the crater attains its maximum depth. In this case, the crater-base begins to lift, and the space between the side walls is filled with pool liquid. Similarly, they formulate $We \sim Fr^{1/5}$ as a lower bound for regular entrainment and conclude that the crater wall’s outward motion reverses at different positions, which is essential for bubble entrainment. The relations obtained by \cite{oguz1990bubble} agree with the experimental data of \cite{prosperetti1989underwater}.  Afterward, regular entrainment is inhibited, and the regime transitions to large-bubble entrainment, which primarily occurs for large drops that produce a splash above the crater. The crater spreads and attains a dish shape at its base, trapping most of its volume within a bubble. The process eventually exhibits irregular bubble entrainment \citep{pumphrey1990entrainment}. This irregularity in bubble entrainment is attributed to secondary drop ejection: from the jet tip, from the crown at the free surface, or from fully grown jet collapse at the pool surface. \\

After examining different reasons for bubble entrainment and the boundaries of regular entrainment, \cite{rein1993phenomena} experimentally demonstrated distinct transition regimes in jets following crater collapse. They found that after a $We_{sp}$, which defines the splashing regime, a thick central jet forms without a secondary droplet. As the regime shifts toward the lower limit of regular entrainment, the jet height increases, and the jet begins to shred secondary droplets. In a regular entrainment regime, they observed a high-speed, narrow, thin jet with multiple secondary droplets. Subsequently, they observed a jet similar to the jet before the regular entrainment regime; in this case, the jet was larger and formed a single secondary drop. Both the narrow, thin jet and bubble entrainment originate at the base of a crater, suggesting a connection between these phenomena. To address this ambiguity, \cite{rein1996transitional} conducted experiments within the entrainment region and qualitatively demonstrated that the bubble was not always entrained under the impact conditions of regular entrainment proposed by \cite{oguz1990bubble}. However, a thin, narrow jet consistently forms and becomes characteristic of a regular entrainment region. The impact initially causes bubble entrapment; however, changes in surface tension due to impurities or microparticles on the pool surface can lead to this phenomenon being overlooked.  The argument presented by \cite{rein1996transitional} indicates no direct link between regular entrainment and the formation of a narrow, thin jet. \cite{leng2001splash} experimentally found this and countered \cite{rein1996transitional} by showing that high-speed narrow jets form both before and after the entrainment region. Eventually, the entrainment region was a subset of the high-speed jet region. The lower and upper limits of the high-speed jet region were $We = 34.7 Fr^{0.145}$ and $We = 54 Fr^{0.25}$, respectively, obtained from best-fit curves to their experimental data. \cite{leng2001splash} redefined the lower limit of the bubble entrainment region as given by \cite{oguz1990bubble} from $We = 41.3 Fr^{0.179}$ to $We = 36.2 Fr^{0.186}$. The upper limit of bubble entrainment agrees well with the experiments. \\

The researcher categorized the entrainment and high-speed jet region. However, the reason why both phenomena occur across a range of impact conditions remained unclear until \cite{morton2000investigation} presented experimental and numerical studies on a $2.9$ mm water drop at impact velocities ranging from $ 0.8$ to $ 2.5\,\mathrm{m\,s^{-1}}$. \cite{morton2000investigation} cover the entire phenomenon across the pre-entrainment, entrainment, and post-entrainment regions. In the pre-entrainment region, they consider coalescence, whereas the entrainment and post-entrainment regions fall within the splashing region. In both the entrainment and post-entrainment regions, capillary waves are generated at the crater opening and propagate downward through the crater-wall. Compared with the entrainment region, the post-entrainment region forms a relatively larger and wider crater.  The wider crater delays capillary waves from reaching the crater-base, allowing the base to rise before the crater side wall collapses and leading to the formation of a thick jet.  As a characteristic feature of splashing, vortex rings are suppressed within the entrainment region. In contrast, the upward motion of the base generates radial pressure gradients that produce vorticity, resulting in the formation of a single vortex ring during crater collapse in the post-entrainment region. \cite{ray2015regimes} performed a similar numerical study, demonstrating crater evolution and jet formation for different cases using pressure iso-contour and vorticity plots. They plot a regime map of $We$ from $50$ to $300$ and $Fr$ from $25$ to $600$ to show different phenomena and find the upper limit of small-bubble entrapment with a long, thick jet to be $We = 63.1 Fr^{0.257}$. The existing literature suggested that for a thin and narrow jet, dimple formation at the base of the crater was essential, regardless of whether pinch-off occurs or not. \cite{eggers1997nonlinear} characterized bubble pinch-off as a free-surface singularity, whereas \cite{day1998self} demonstrated that bubble pinch-off dynamics followed self-similar inertial focusing. \cite{zeff2000singularity} proposed that the fastest jet originated from the inertial collapse of the crater, bypassed the pinch-off singularity, and lifted the dimple base before pinch-off. They included surface-tension effects to predict the jet's free-surface profile. Later, \cite{thoroddsen2007experiments} showed that bubble pinch-off was purely driven by inertial focusing and that surface tension became irrelevant. In subsequent work, \cite{thoroddsen2007microjetting} discussed microjet formation. They emphasized the essential role of capillary wave motion at the jet surface that produces drops from its tip. \cite{walls2015jet} investigated post-jet dynamics during bubble bursting and identified a regime in which viscous and gravitational forces govern the behavior of droplets that emanate from the jet tip. Beyond a critical Ohnesorge number $(Oh^{*})$, viscous effects dominated over surface tension, whereas the Bond number accounted for gravitational force, which governed the bubble size before rupture at the free surface. \cite{ganan2017revision} postulated a universal scaling for jet velocity and drop size during bubble bursting without considering gravitational effects and resolved the problem of curvature reversal. The flow configuration and energy conservation lead to the following scaling. Using the same framework, \cite{deike2018dynamics} proposed an alternative scaling that accounts for the gravitational effect to obtain an optimal Laplace number $(La)$, which compares the capillary inertial force with the viscous forces. Gravity influences the initial cavity shape and alters the capillary wavelength, thereby affecting the jet velocity. \cite{michon2017jet} performed a detailed quantitative investigation on jet dynamics across a wide range of impact conditions. They divided the phenomenon into two regimes: the singular-jet regime and the cavity-jet regime. In the singular-jet regime, the self-similar collapse of capillary waves is coupled with cavity retraction, producing a thin, narrow, high-speed jet. They were among the first to report variations in jet velocity with fluid viscosity. The jet velocity increased with viscosity to some extent, then decreased within this regime. The second regime was the cavity jet regime, characterized by a slow, thick jet with one or two secondary droplets. The jet velocity in this regime is proportional to the capillary velocity $V_c = \sqrt{\frac{\sigma}{\rho_l D_d}}$, independent of viscosity and impact velocity. The uncorrelated effect of capillary wave collapse at the base of the retracting cavity leads to splashing and, in some cases, entraps a bubble in the pool but never triggers a thin, fast jet.\\

Building on the work of \cite{zeff2000singularity}, \cite{thoroddsen2018singular} experimentally investigated the evolution of the fastest jet for drop diameters of $3.33~\mathrm{mm}$ and $4.2~\mathrm{mm}$ with impact velocities ranging from $U \sim 1.1\text{--}1.6~\mathrm{m/s}$. They observed distinct pinch-off morphologies across the impact-velocity range and measured a maximum jet velocity of $50~\mathrm{m/s}$ in a case where the dimple retracts without pinch-off. Considering the impact velocity associated with the fastest jet and its characteristics to bypass the pinch-off singularity, it corresponds to the lower boundary of the entrapment region proposed by \cite{pumphrey1990entrainment}. They plotted the dimple collapse at different time instants and reported a power-law coefficient, $\beta = 0.55$, slightly higher than the inertial coefficient $0.5$. This indicates a transition from capillary-to-inertial driven flow. For the fastest jet, they observed microbubbles at the base of the dimple, attributed to local cavitation from inertial focusing, which produced a high-pressure region at the base of the jet. The large change in curvature generates vorticity, which may also play a crucial role here \citep{thoraval2012karman}. The experimental study by \cite{thoroddsen2018singular} left several open questions, some of which were addressed by \cite{tian2023conical}, who used Gerris to simulate the phenomenon and reported a jet velocity of $320~\mathrm{m/s}$ for a viscosity of $7.3~\mathrm{cP}$ at ambient pressure at a point of emergence. \cite{tian2023conical} sought an upper bound on the jetting velocity by reducing ambient pressure, thereby reducing the air drag on the pinch-off drop and consequently increasing the jet velocity. Under reduced ambient pressure, the maximum jetting velocity was $137~\mathrm{m/s}$. They also observed a similar variation of the jetting velocity with viscosity at ambient pressure, consistent with  \cite{michon2017jet}. These jets were an order of magnitude faster and two orders of magnitude thinner than the jets considered in \cite{ganan2017revision}; therefore, their proposed scaling on singular-jets was not applicable. \\

Most studies on singular-jets, whether generated by bubble bursting at the free surface or by drop-impact crater collapse, have focused primarily on jet velocity under varying ambient pressure and fluid viscosity. \cite{tian2023conical} showed that the fastest jet followed a conical focusing mechanism in which an air sheet is entrapped between the dimple wall and the outer jet surface, thereby providing effective slip. Still, it remains unclear what causes the air entrapment, why it is restricted only to a narrow range of $We$, and what determines the different pinch-off shapes. We perform a comprehensive study of the flow fields, unlike the previous literature on singular-jet dynamics, during dimple collapse and their interplay with the exceptionally high-velocity of a singular-jet to seek answers to these questions. 
\cite{morton2000investigation} and \cite{ray2015regimes} investigated crater collapse across different entrainment regimes, focusing primarily on bubble entrainment. Their working regime includes the lower boundary of the entrainment regime, where singular-jet formation has been reported by \cite{thoroddsen2018singular,tian2023conical}. Despite this, they do not observe a singular-jet, and their analysis does not address the underlying flow mechanism of jet formation. \cite{ray2012bubble} report jetting velocities but do not attain the exceptionally high values as observed in the singular-jet regime. In this work, we conducted a detailed study of singular-jet formation during drop-impacted crater collapse across a range of  $We$ and $Fr$ numbers, with particular emphasis on crater evolution and jetting velocity. In particular, we examine the transition in jet morphology from a short, thick jet to a long, thin jet, then to a high-speed singular-jet, and finally to a lower-speed long, thick jet. We use the pressure and velocity fields during dimple collapse to determine how the evolving crater dynamics modify the liquid motion and produce a distinctly high jetting velocity. We also examine the temporal evolution of vorticity, given its role in bubble entrapment, as reported in \cite{morton2000investigation,ray2015regimes}, and its generation associated with rapid changes in interface curvature as reported by \cite{thoraval2012karman}. Furthermore, we extend the simulations into the post-entrainment regime to investigate if singular-jets exist beyond the entrainment region. If we do not find a singular-jet beyond the entrainment regime, we further examine the underlying mechanisms responsible for the suppression of jet velocity through flow-field analysis. \\

Section \S~\ref{sec:methodology} involves the governing equation and the applied scheme in detail, with the initial flow configuration and required refinement level used in this problem; validation of numerical schemes is reported in section \S~\ref{sec:validation} by comparing our results with the experimental data available in \cite{tian2023conical}. Section \S~\ref{sec:results} presents a detailed discussion of various aspects of the problem based on the results obtained from the simulation, and we conclude in section \S~\ref{sec:conclusion}.

\section{Numerical Methodology and Flow Configuration}
\label{sec:methodology}

\subsection{Governing Equations and Numerical Methods}
\label{sec:Governing_equation}

The governing equations for drop-impact on a liquid pool are the incompressible mass and momentum conservation equations, along with the volume-of-fluid advection equation for tracking the evolution of the gas-liquid interface.\\

\begin{equation}
\nabla \cdot \mathbf{U} = 0,
\label{eq:continuity}
\end{equation}
\begin{equation}
\rho \left( \frac{\partial \mathbf{U}}{\partial t}
+ (\mathbf{U} \cdot \nabla)\mathbf{U} \right)
= -\nabla P + \nabla \cdot (2\mu \mathbf{D})
+ \rho \mathbf{g} + \sigma \kappa \delta_s \hat{\mathbf{n}},
\label{eq:momentum}
\end{equation}
\begin{equation}
\frac{\partial f}{\partial t}
+ \nabla \cdot (f\mathbf{U})
= f \, \nabla \cdot \mathbf{U}.
\label{eq:vof}
\end{equation}
\\

Here, $\mathbf{U}$ is the velocity field, $P$ is the pressure, $\rho$ and $\mu$ are the fluid's varying density and viscosity, $\sigma$ is the surface tension coefficient, $\kappa$ is the mean curvature, $\delta_s$ is the delta function, and $\hat{\mathbf{n}}$ is the unit normal vector at the interface. The deformation tensor is $\mathbf{D} = [\nabla \cdot \mathbf{U} + \nabla \cdot \mathbf{U}^T]/2$, and $\mathbf{g}$ is the acceleration due to gravity. The volume of fluid $f$ represents the fraction of the cell volume occupied by liquid. The domain contains single-phase cells, which take the value $1$ for a cell filled with liquid and $0$ for an empty cell. For multiphase cells containing an interface, the value of $f$ lies between $0$ and $1$. \\
We non-dimensionalize the governing equation using the drop diameter $D$ and impact velocity $V$ as the respective length and velocity scales. The variables in the governing equations are scaled accordingly: $\mathbf{U^*} = \mathbf{U}/V$, $z^* = z/D$, $r^* = r/D$, $t^* = t/(D/V)$,  $P^* = P/(\rho_w V^2)$, $\kappa^* = \kappa D$, $\rho^* = \rho/\rho_w$, $\mu^* = \mu/\mu_w$, $\mathbf{D^*} = [\nabla \cdot \mathbf{U^*} + \nabla \cdot \mathbf{U^*}^T]/2$. 

\begin{equation}   
\nabla \cdot \mathbf{U^*} = 0, \label{eq:Non_continuity} 
\end{equation}
\begin{equation}
\rho^* \left( \frac{\partial \mathbf{U^*}}{\partial t^*} + (\mathbf{U^*} \cdot \nabla)\mathbf{U^*} \right) 
= -\nabla P^* + \frac{\nabla \cdot (2\mu \mathbf{D^*})}{Re} + \frac{\rho^*}{Fr} + \frac{ \kappa^* \delta_s \hat{\mathbf{n}}}{We}. 
\label{eq:Non_momentum} 
\end{equation}

Three key dimensionless parameters emerge owing to non-dimensionalization. The Weber number $We = \rho_w V^2 D / \sigma$; the Reynolds number, $Re = \rho_w V D / \mu_w$; and the Froude number, $Fr = V^2 / g D$, as described in section \S~\ref{sec:Introduction}. \\

We use Basilisk \citep{popinet2015quadtree}, which solves the time-dependent Navier-Stokes equation using a finite-volume formulation that is second-order accurate in both space and time. The finite-volume space discretization in Basilisk is based on a quadtree-adaptive cartesian mesh \citep{popinet2003gerris,popinet2015quadtree}, in which all the primitive variables, such as velocity fields and pressure, are defined at the cell centers. During refinement, the parent cell is recursively divided into four equal square child cells until we reach the maximum level of refinement. By default, the root cell (the base of the cell tree) is assigned a refinement level of $0$, and its level increments by $1$ in each successive generation. The framework employs adaptive mesh refinement to coarsen or refine the computational grid by computing the wavelet numerical discretization error \citep{van2018towards}. The error is evaluated at every time step for specially discretized fields, such as volume fraction, curvature, velocity, and vorticity. The mesh adapts multilevel refinement from the minimum level $L_{\min}$ to the maximum level $L_{\max}$ when the calculated error exceeds the prescribed threshold within the computational domain. \\

The advection terms in equation \ref{eq:Non_momentum} are discretized using a second-order upwind scheme \citep{bell1989second}, stable for $\mathrm{CFL} < 1$. The Crank–Nicolson method, which is unconditionally stable and second-order accurate in time, is employed to compute the viscous terms. The volume-of-fluid (VOF) method \citep{hirt1981volume} conserves mass but does not accurately capture the complex geometric features of the interface. Equation \ref{eq:vof} is advected using the advection scheme given in \cite{weymouth2010conservative}. To obtain an accurate interface representation, the interface is reconstructed using a piecewise-linear interface-capturing technique \citep{rudman1998volume}. Computing the surface tension terms using the continuum surface force (CSF) model proposed by \cite{brackbill1992continuum} is the most challenging aspect of the solver. The CSF formulation requires a continuous representation of the interface, which is readily available in interface-tracking methods such as the level set, front-tracking, and ghost-fluid method. However, such an explicit representation is absent in the VOF method. To address this limitation, \cite{francois2006balanced} and \cite{renardy2002prost} give an approximation in the CSF model by replacing the product of $\delta_s$ and the normal vector of the interface with the gradient of the volume fraction $(\delta_s \mathbf{n} = \nabla F)$. The interface curvature in the surface-tension term is approximated using a second-order-accurate height function proposed by \cite{popinet2009accurate} and \cite{popinet2018numerical}. For pressure, a multigrid solver is used to solve the pressure Poisson equation and project the divergence-free velocity fields. The implementation details are provided in  \cite{fuster2018all, Popinet2013}. \\

\subsection{Initial Flow Configuration}\label{sec:Flow_configuration}

\begin{figure}
\centering
\includegraphics[width=0.96\linewidth]{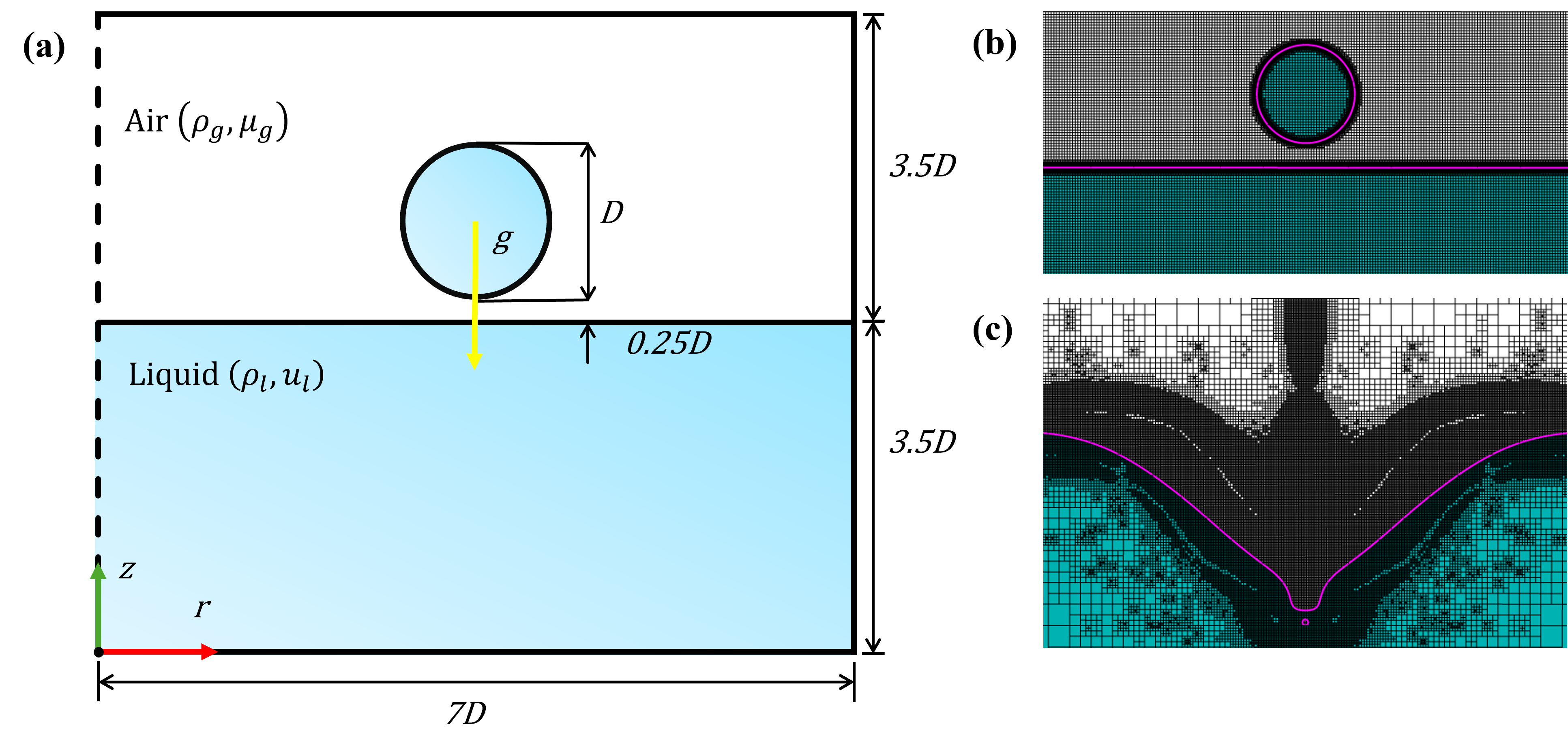}
\caption{\footnotesize
Schematic of the computational domain and mesh refinements for the axisymmetric simulations: (a) computational domain; (b) initial mesh refinement, with \(L_{\max}=12\) across the drop–pool interface and \(L_{\min}=10\) in the remaining domain; and (c) refined mesh with \(L_{\max}=14\) in the entire domain at the onset of dimple formation, when the estimated errors exceed the prescribed thresholds, \(\omega_{\mathrm{err}}\) and \(\kappa_{\mathrm{err}}\). The air–water interface is marked by the magenta line.}
\label{fig:domain_setup}
\end{figure}

\cite{thoroddsen2018singular} and \cite{tian2023conical} investigated the formation of singular-jets experimentally and numerically. A water-glycerin mixture for both the drop and pool liquids, with density $\rho_w = 1140\,\mathrm{kg\,m^{-3}}$, dynamic viscosity $\mu_w = 7.3\,\mathrm{cP}$, and surface tension coefficient $\sigma = 0.068\,\mathrm{N\,m^{-1}}$. The diameter of the drop is $D = 3.64\,\mathrm{mm}$, and the impact velocity ranges from $V \simeq 1.28 - 1.53\,\mathrm{m\,s^{-1}}$, is used in the present axisymmetric simulations. A square computational domain with a side length of $\mathit{L} = 7\mathit{D}$. The free surface of the pool is located at a distance $\mathit{H} = 3.5\mathit{D}$ from the bottom, providing sufficient space for crater evolution and for the drop that emanates from the tip of the jet, without any interaction with the boundaries. The drop is placed at $\delta = 0.25\mathit{D}$, as shown in figure \ref{fig:domain_setup}(a), just above the free surface of the liquid pool. This setup entraps a thin air sheet when the drop hits the pool surface. 
\\

We impose a no-slip boundary condition at the bottom $(z = 0)$ and the right $(r = 7D)$ boundaries \citep{tian2023conical}(supplementary material): 
\begin{gather}
\mathbf{U}\cdot\mathbf{n}=0. \label{eq:No_Slip} 
\end{gather}
A free-surface boundary condition is imposed at the top boundary$(z = 7D)$: 
\begin{gather}
\mathbf{n}\cdot\nabla \mathbf{U}=0. \label{eq:Neumann_BC} 
\end{gather}
and a symmetric boundary condition is imposed at the left boundary $(r = 0)$:
\begin{gather}
\mathbf{U}\cdot\mathbf{n}=0,
(\mathbf{n}\cdot\nabla\mathbf{U})\times\mathbf{n}=0.
\end{gather}
For pressure, a homogeneous Neumann boundary condition is applied at all boundaries except at the top $(z = 7D)$, where the pressure equals the atmospheric pressure $P_{atm}$. 
\begin{gather}
\mathbf{n}\cdot\nabla p = 0. \label{eq:Pressure_BC} 
\end{gather}
\\ 
In adaptive mesh refinement, we use vorticity and curvature to compute the error of a specially discretized field, with threshold values of $\omega_{err} = 10^{-3}$ and $\kappa_{err} = 10^{-4}$, respectively. Initially, the domain is discretized at the minimum refinement level $L_{\min}$, except for the region around the pool and the drop interface, which are discretized at the maximum refinement level $L_{\max}$, shown in figure~\ref{fig:domain_setup}(b). Thereafter, adaptive refinement was gradually employed to ensure numerical stability and keep the residuals within the prescribed tolerance limits, indicating that the pressure and velocity fields are converging properly.
\\

\subsection{Minimum and maximum levels of refinement}\label{sec:Refinemet_level}

Direct numerical simulation (DNS) of drop-impact on a liquid pool requires resolving the smallest length and time scales to capture secondary droplets originating at the jet tip and the entrapment of an air sheet as the drop hits the pool surface. We conduct numerous tests to determine the optimal minimum and maximum refinement levels for our study. The minimum and maximum refinement levels best suited for this problem are $L_{\min} = 9$ and $L_{\max} = 14$. A refinement level below $9$ introduces spurious splashing during impact and distorts the crater morphology. An adequate maximum refinement level is also required to resolve dimple formation and prevent it from unwanted pinch-offs. It must also resolve to capture every drop emanating from the jet tip, which plays a significant role in the formation of a singular-jet during crater collapse. We perform a grid-independence study (see Appendix ~\ref{appA} for details), which suggests that a maximum refinement level $L_{\max} = 12$ is required to capture subtle features of impact dynamics. These include initial air-bubble entrapment when the drop hits the pool and the radially inward movement of the crater side walls to form a dimple at the crater-base. However, once dimple retraction begins, $L_{\max} = 12$ is insufficient to resolve the subsequent jet-formation dynamics. Therefore, we employ $L_{\max} = 14$, as shown in figure~\ref{fig:domain_setup}(c), to capture the jet formation and drop emanation from the jet tip during dimple collapse. Resolving the entire process at $L_{\max} = 14$ is very challenging and time-consuming. Therefore, based on physical morphology and the dynamics of the problem, the simulation is performed in two different stages. Initially, $L_{\max}$ starts at $12$ up to dimple formation, after which the refinement level increases to $14$ and remains there until the first secondary drop leaves the crater. This mesh-refinement strategy is computationally efficient and ensures that intricate features and physics are properly captured. In an axisymmetric simulation, the base grid is refined at $L_{\min} = 10$, corresponding to $(2^{10})^2$ cells, while $L_{\max} = 14$ is equivalent to $(2^{14})^2$ cells, yielding a finite grid spacing of $\sim 1.5\,\mu\mathrm{m}$.


\section{Validation}\label{sec:validation}
Basilisk has been extensively used to study the dynamics of liquid drops impacting a liquid pool and has been validated for various multiphase problems. To validate our numerical approach in the present study, we compared our results qualitatively and quantitatively with the experimental data of \cite{tian2023conical}. Table~\ref{tab:Impact_condition} shows the impact conditions for the cases used in this study. In the previous section, we discussed the feasible value of $L_{\min}$ and $L_{\max}$ for this problem, the time step $(\Delta t)$ calculated using the Courant-Friedrichs-Lewy (CFL) condition with a value of 0.5, and computed the capillary time step $\Delta t_\sigma \le 0.5 \sqrt{\frac{\rho_m \, \min(\Delta r,\Delta z)^3}{\pi \, \sigma}}$,  given by \cite{brackbill1992continuum}, where $\rho_m$ is the mean density, $\Delta r$ and $\Delta z$ are the grid spacings in the $r$ and $z$ directions, respectively, and $\sigma$ is the surface tension coefficient. There are two other time steps based on viscosity and gravity.  We chose the minimum value of $\Delta t$ among all of them because it must be sufficiently small to resolve the phenomenon accurately. The time step is on the order of $10^{-6}$ when the dimple at the crater-base begins to retract. \\

\begin{table}
\centering
\small
\setlength{\tabcolsep}{10pt}
\renewcommand{\arraystretch}{1.2}

\begin{tabular}{c c c c c}
Cases & $V$ (m\,s$^{-1}$) & $\mathrm{We}$ & $\mathrm{Fr}$ & $\mathrm{Re}$ \\
\\[-6pt]        

We100Fr45.8       &   1.280    &   100     &   45.88   &   727.60
\\
We113Fr51.8       &   1.360    &   113     &   51.80   &   773.07
\\
We115Fr52.7       &   1.370    &   115     &   52.70   &   778.76
\\
We117Fr53.3       &   1.380    &   117     &   53.33   &   784.44 
\\
We119Fr54.5       &   1.396    &   119     &   54.57   &   793.54 
\\
We120Fr55.0       &   1.402    &   120     &   55.04   &   796.95 
\\
We122Fr56.0       &   1.414    &   122     &   56.00   &   803.77 
\\
We123Fr56.4       &   1.420    &   123     &   56.46   &   807.18
\\
We127Fr58.2       &   1.442    &   127     &   58.23   &   819.68 
\\
We130Fr59.7       &   1.460    &   130     &   59.70   &   829.92 
\\
We134Fr61.5       &   1.482    &   134     &   61.50   &   842.42
\\
We140Fr64.2       &   1.514    &   140     &   64.20   &   860.61
\\
We143Fr65.5       &   1.530    &   143     &   65.55   &   869.71 
\\
We150Fr68.1       &   1.560    &   150     &   68.15   &   886.92
\\
We155Fr70.8       &   1.593    &   155     &   70.80   &   905.94
\\
We157.3Fr72.4     &   1.605    &   157.3   &   72.41   &   912.63
\\
We159Fr72.9       &   1.614    &   159     &   72.95   &   917.55
\\
We160Fr73.5       &   1.619    &   160     &   73.50   &   920.43
\\
We170Fr78.1       &   1.670    &   170     &   78.10   &   948.76
\\
We190Fr86.7       &   1.760    &   190     &   86.74   &   1003.02
\\
We200Fr91.7       &   1.810    &   200     &   91.74   &   1029.08
\\

\end{tabular}

\caption{Impact conditions for numerical simulation used to study the mechanics of singular-jet formation for a drop diameter of $3.64\,\mathrm{mm}$.}
\label{tab:Impact_condition}
\end{table}

\subsection{Qualitative Validation}\label{sec:Qualitative_validation}

\begin{figure}
\centering
\hspace{-0.07\textwidth}
\begin{subfigure}[t]{0.10\textwidth}
    \centering
    \begin{tikzpicture}
    \node[anchor=north west] at (0,0)
    {\includegraphics[height=2.0cm]{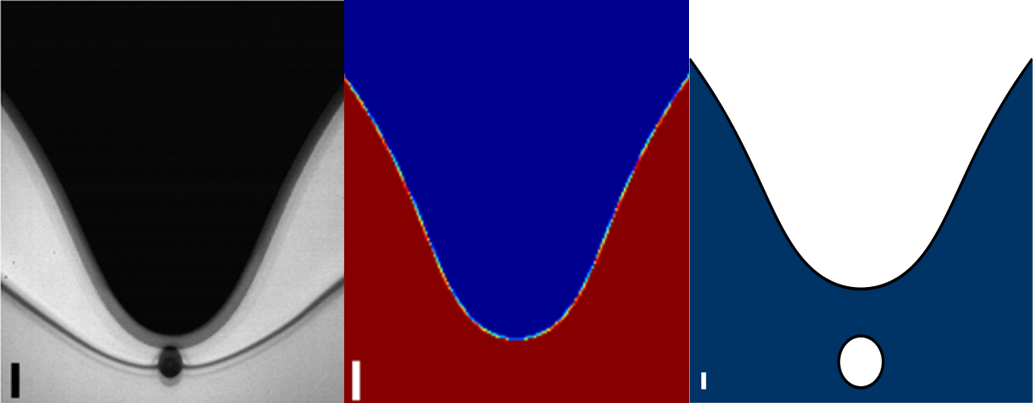}};
    \node[anchor=north west,font=\bfseries,fill=white,inner sep=0.0pt] at (-0.35,-0.1) {(a)};
    \end{tikzpicture}
\end{subfigure}\hfill
\begin{subfigure}[t]{0.10\textwidth}
    \centering
    \begin{tikzpicture}
    \node[anchor=north west] at (0,0)
    {\includegraphics[height=2.0cm]{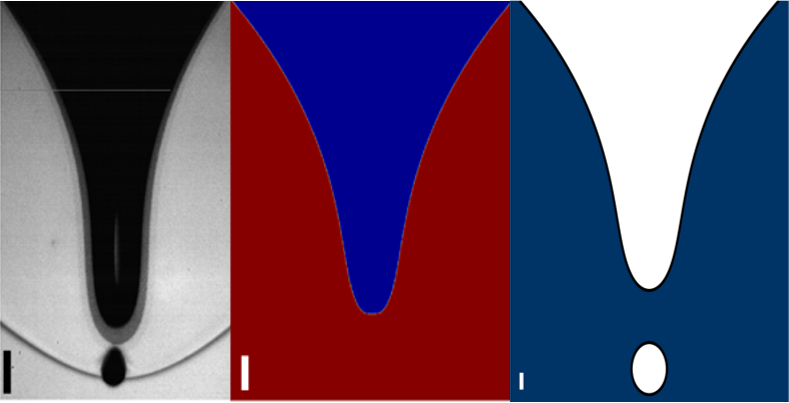}};
    \node[anchor=north west,font=\bfseries,fill=white,inner sep=0.0pt] at (-0.35,-0.1) {(b)};
    \end{tikzpicture}
\end{subfigure}\hfill
\hspace{-0.1\textwidth}
\begin{subfigure}[t]{0.10\textwidth}
    \begin{tikzpicture}
    \node[anchor=north west] at (0,0)
    {\includegraphics[height=2.0cm]{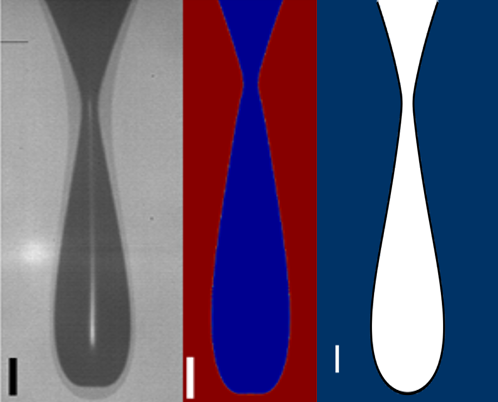}};
    \node[anchor=north west,font=\bfseries,fill=white,inner sep=0.0pt] at (-0.35,-0.1) {(c)};
    \end{tikzpicture}
\end{subfigure}
\hspace{0.15\textwidth}
\par\medskip

\hspace{-0.1\textwidth}
\begin{minipage}[t]{0.65\textwidth}
    \vspace{0pt} 

    \begin{subfigure}[t]{0.48\textwidth}
        \centering
        \begin{tikzpicture}
        \node[anchor=north west] at (0,0)
        {\includegraphics[height=2.2cm]{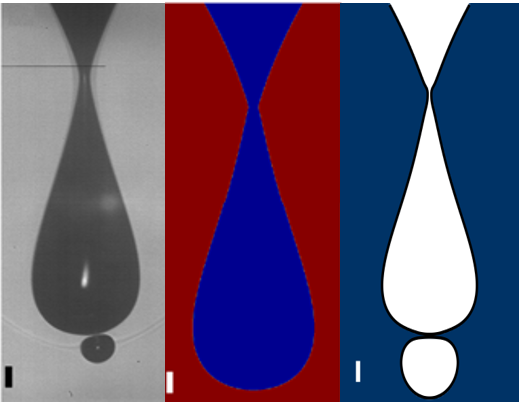}};
        \node[anchor=north west,font=\bfseries,fill=white,inner sep=0.0pt] at (-0.35,-0.1) {(d)};
        \end{tikzpicture}
    \end{subfigure}
    \hfill
    \hspace{0.08\textwidth}
    \begin{subfigure}[t]{0.48\textwidth}
        \centering
        \begin{tikzpicture}
        \node[anchor=north west] at (0,0)
        {\includegraphics[height=2.2cm]{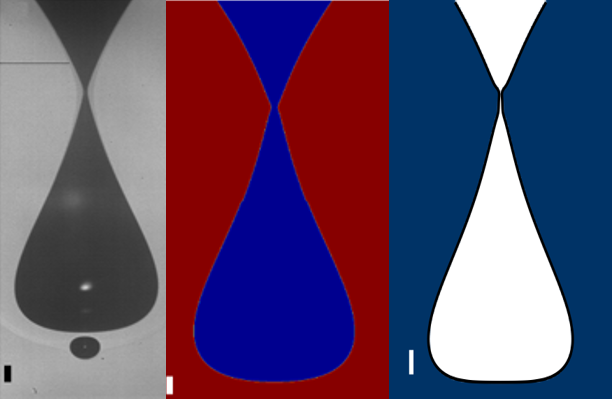}};
        \node[anchor=north west,font=\bfseries,fill=white,inner sep=0.0pt] at (-0.35,-0.1) {(e)};
        \end{tikzpicture}
    \end{subfigure}

    \par\medskip

    \hspace{0.04\textwidth}
\begin{subfigure}[t]{0.7\textwidth}
    \centering
    \begin{tikzpicture}
        \node[anchor=north west] at (0,0)
        {\hspace{-0.6cm}\includegraphics[height=9.0cm]{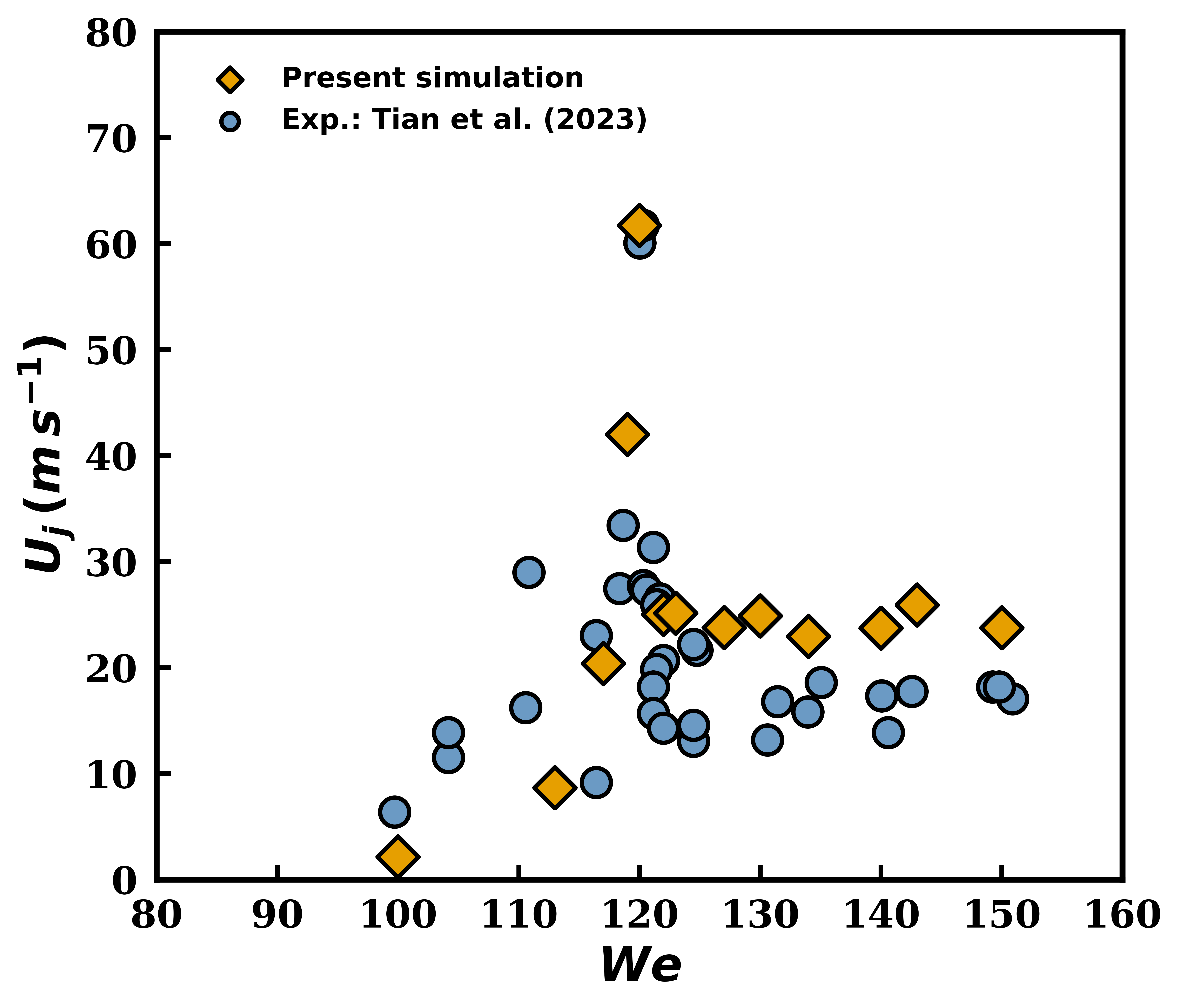}};

        \node[anchor=north west,font=\bfseries,fill=white,inner sep=0.0pt]
        at (-0.2,-0.2) {(g)};
    \end{tikzpicture}
\end{subfigure}
\end{minipage}
\hfill
\begin{minipage}[t]{0.32\textwidth}
    \vspace{0pt} 

    \begin{subfigure}[t]{\textwidth}
        \centering
        \begin{tikzpicture}
        \node[anchor=north west] at (0,0)
        {\includegraphics[width=0.90\textwidth,height=11.0cm]{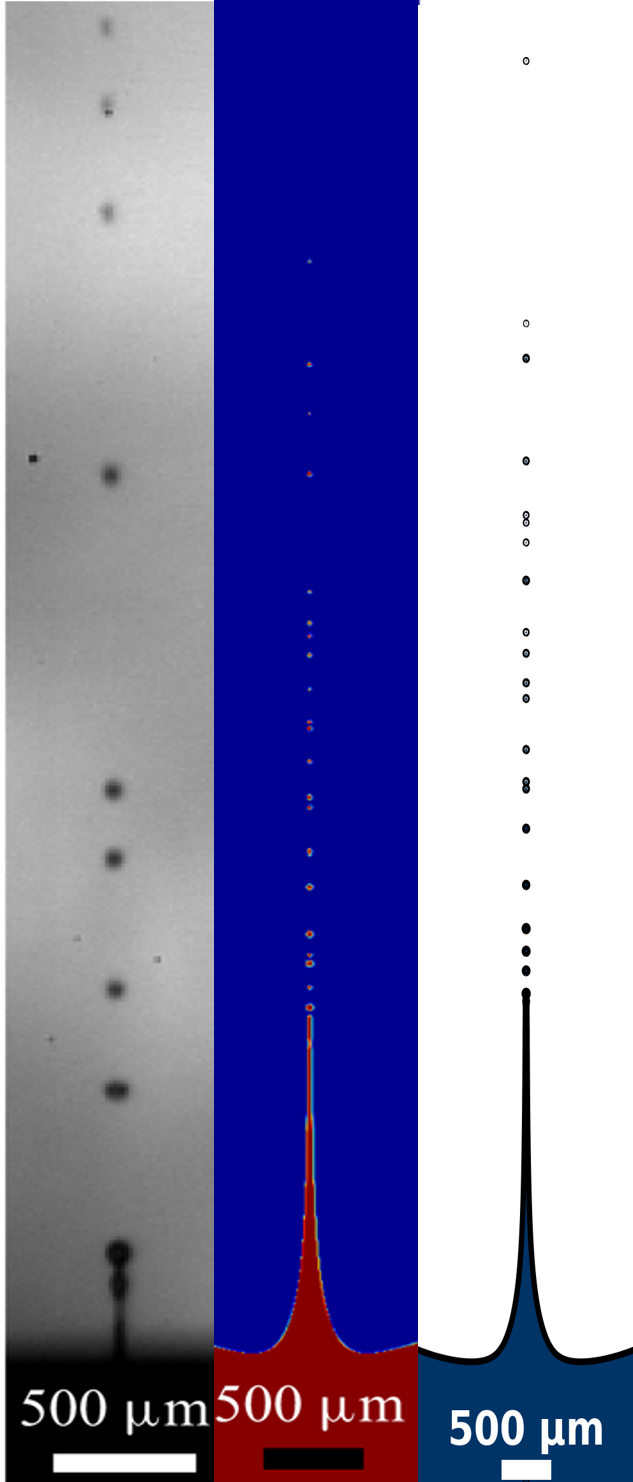}};
        \node[anchor=north west,font=\bfseries,fill=white,inner sep=0.0pt] at (-0.35,-0.1) {(f)};
        \end{tikzpicture}
    \end{subfigure}
\end{minipage}

\caption{\footnotesize 
Comparison of the present numerical results with the experiments and simulations of \cite{tian2023conical}. Panels (a)–(e) show a qualitative comparison of different dimple and pinch-off shapes. Panel (f) shows the complete evolution of the singular-jet. Panel (g) quantitatively compares the corresponding jetting velocity $(U_j)$ with the experimental measurements. The experimental measurements have an uncertainty of approximately \(\pm 4~\mathrm{m\,s^{-1}}\), arising from image smearing over approximately three pixels in determining the drop displacement from the post-processed images.}
\label{fig:Validation}
\end{figure}

Figure \ref{fig:Validation} shows the physical dimple profiles formed at the base during crater collapse under different impact conditions reported in the experiments of \cite{tian2023conical}. A conical-shaped dimple is observed in figure \ref{fig:Validation}(a) and transforms to a more elongated conical shape as shown in figure \ref{fig:Validation}(b), with an increase in impact velocity. In both cases, the crater retracts without pinch-off and continues until we get a singular-jet. Thereafter, we observe a pinch-off, with distinct shapes for the different cases. In figure \ref{fig:Validation}(c), we observe a teardrop pinch-off. Similarly, we observe a heavier and wider base pinch-off in figure \ref{fig:Validation}(d). Following this, the pinch-off becomes triangular, as shown in figure \ref{fig:Validation}(e). Our qualitative results capture all the distinctive features, including the entrapped bubble beneath the dimple, which was absent in the numerical results of \cite{tian2023conical}. The entrapped bubble size is slightly larger than that observed experimentally \citep{thoroddsen2018singular}, but the pinched-off volume resembles the experiments very well. \\

\subsection{Quantitative Validation}\label{sec:Quantitative_validation}
In addition to different dimple profiles, we compute the diameter of the entrapped bubble and the jetting velocity. 
We quantitatively validate our numerical approach with the experiments of \cite{tian2023conical}. \cite{thoroddsen2018singular} evaluate the jet velocity by computing the distance traveled by the first drop to emerge from the crater, which originates at the tip of the narrow jet due to Rayleigh instability, and dividing it by the time it takes to travel that distance. We follow the same method to compute the corresponding jet velocities $(U_j)$ for the cases listed in Table~\ref{tab:Impact_condition}, with selected results in figure \ref{fig:Validation}(g) within the experimental range given by \cite{tian2023conical}. The experimental results are marked with solid circles, whereas our numerical results are represented by solid diamonds. The maximum jetting velocity obtained in the present study for the singular-jet case in which the dimple retracts without pinch-off is $61.7\mathrm{ms^{-1}}$. Our simulated jet velocity depicts the same variation with $We$. We observe the same isolated peak velocity for a narrow range of $We$ as those reported by \cite{tian2023conical} for a liquid viscosity of $7.3\,\mathrm{cP}$ at ambient pressure. Furthermore, $20$ tiny microdroplets originating from the jet tip, for the singular-jet case as shown in figure \ref{fig:Validation}(f), are consistent with numerical and experimental observations. The jetting velocity after the singular-jet case is slightly higher than the experimental observations. In the experimental measurements, the drop displacement was determined by the post-processed images, which involves an uncertainty of approximately three pixels for the largest droplet, leading to image smearing and resulting in a variation of $ \pm 4$ m/s in the measurement of jetting velocity (\cite{tian2023conical} supplementary material).

Air is entrapped as a bubble when the drop interacts with the pool. This feature was absent in the simulations of \cite{tian2023conical}. In the present work, we observe the entrapped bubble with a diameter of approximately $210\,\mu\text{m}$, as shown in figure \ref{fig:Validation}. \cite{thoroddsen2018singular} also reported an entrapped bubble diameter beneath the crater dimple of approximately $80\,\mu\text{m}$, although their impact conditions differed from those considered here. Furthermore, \cite{thoroddsen2003air} report that the maximum diameter of the entrapped bubble can be $300\,\mu\text{m}$. A large volume of air is entrapped in the pool when we use $L_{\min} < 9$  and $L_{\max} < 12$ during impact. This is attributed to the formation of a large bubble beneath the crater interface, which collapses before the crater reaches its maximum depth.  As the level of refinement increases, the volume of entrained air and the size of the entrapped bubble decrease. We find that $L_{\min} = 10$ and $L_{\max} = 12$ capture the initial physics and switch to $L_{\max} = 14$ to reproduce jet speeds similar to the experiments of \cite{tian2023conical}.

\section{Results and discussion}\label{sec:results}

\subsection{Morphological transitions of the jet cross-flow regimes}\label{sec:results-1}

\begin{figure}
\centering

\begin{minipage}{\textwidth}
\centering
\begin{tikzpicture}
\node[anchor=north west,inner sep=0] at (0,0)
{\includegraphics[width=\linewidth,height=0.10\textheight]{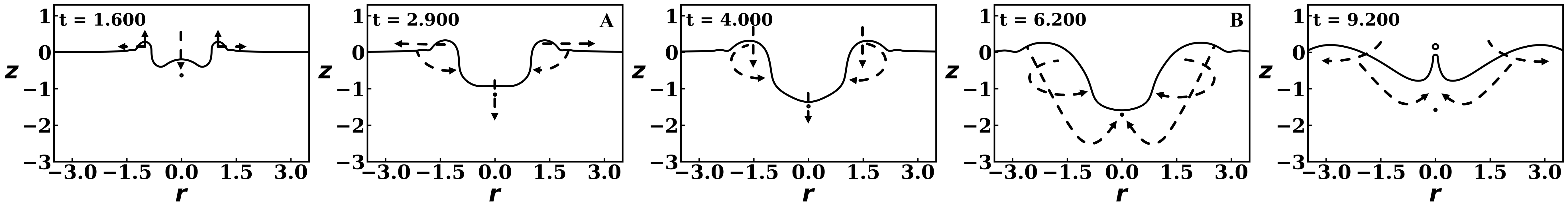}};
\node[anchor=north west,font=\bfseries,fill=white,inner sep=0.0pt] at (-0.2,-0.1) {(a)};
\end{tikzpicture}
\end{minipage}

\vspace{1mm}

\begin{minipage}{\textwidth}
\centering
\begin{tikzpicture}
\node[anchor=north west,inner sep=0] at (0,0)
{\includegraphics[width=\linewidth,height=0.10\textheight]{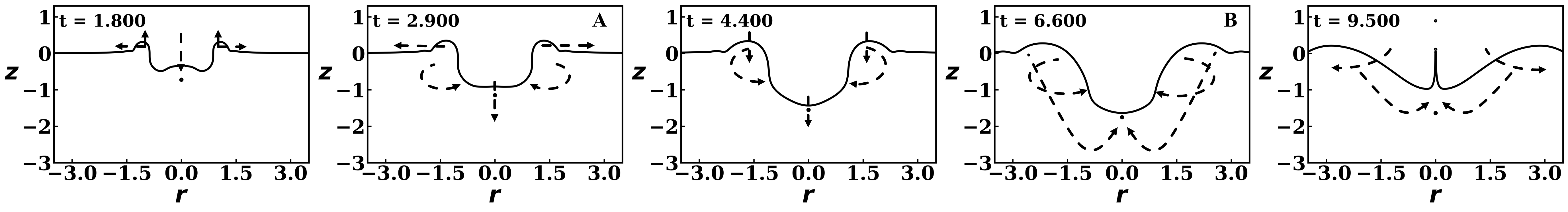}};
\node[anchor=north west,font=\bfseries,fill=white,inner sep=0.0pt] at (-0.2,-0.1) {(b)};
\end{tikzpicture}
\end{minipage}

\vspace{1mm}

\begin{minipage}{\textwidth}
\centering
\begin{tikzpicture}
\node[anchor=north west,inner sep=0] at (0,0)
{\includegraphics[width=\linewidth,height=0.10\textheight]{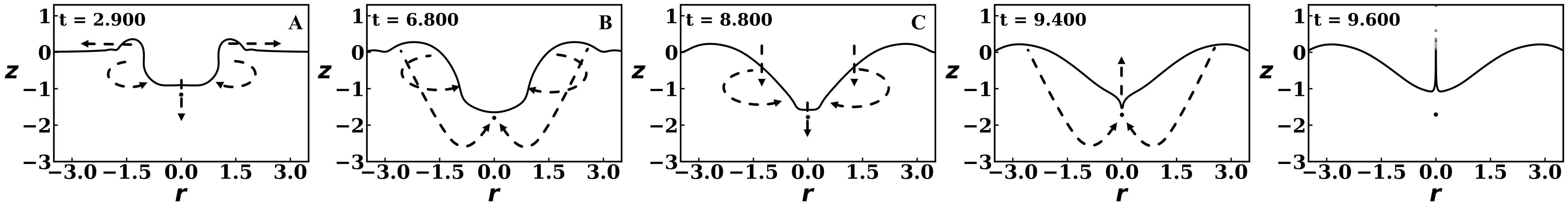}};
\node[anchor=north west,font=\bfseries,fill=white,inner sep=0.0pt] at (-0.2,-0.1) {(c)};
\end{tikzpicture}
\end{minipage}

\vspace{1mm}

\begin{minipage}{\textwidth}
\centering
\begin{tikzpicture}
\node[anchor=north west,inner sep=0] at (0,0)
{\includegraphics[width=\linewidth,height=0.10\textheight]{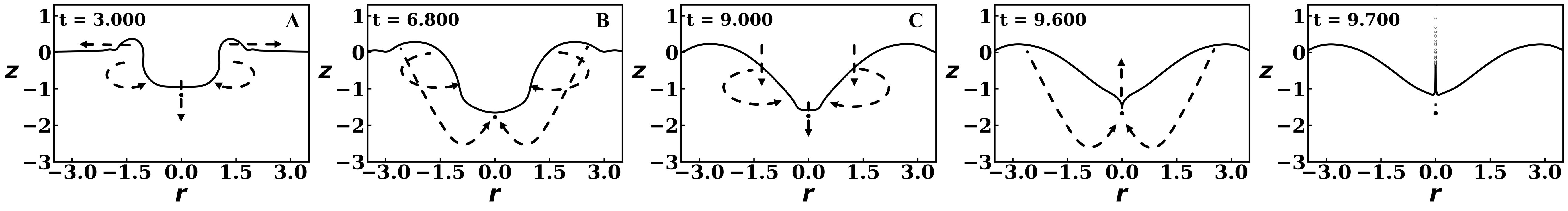}};
\node[anchor=north west,font=\bfseries,fill=white,inner sep=0.0pt] at (-0.2,-0.1) {(d)};
\end{tikzpicture}
\end{minipage}

\vspace{1mm}

\begin{minipage}{\textwidth}
\centering
\begin{tikzpicture}
\node[anchor=north west,inner sep=0] at (0,0)
{\includegraphics[width=\linewidth,height=0.10\textheight]{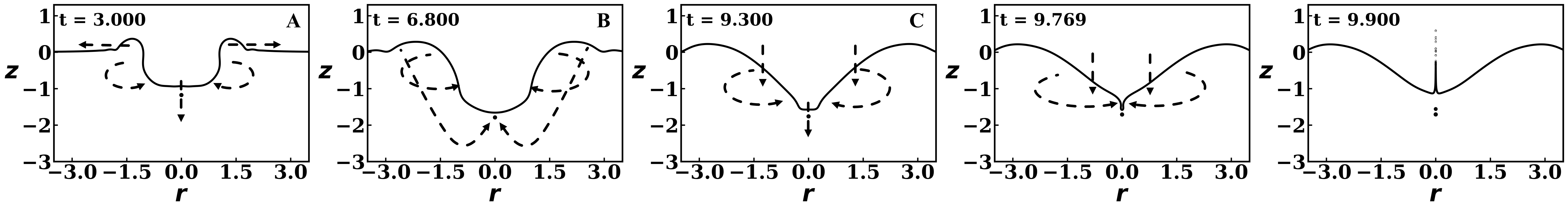}};
\node[anchor=north west,font=\bfseries,fill=white,inner sep=0.0pt] at (-0.2,-0.1) {(e)};
\end{tikzpicture}
\end{minipage}

\vspace{1mm}

\begin{minipage}{\textwidth}
\centering
\begin{tikzpicture}
\node[anchor=north west,inner sep=0] at (0,0)
{\includegraphics[width=\linewidth,height=0.10\textheight]{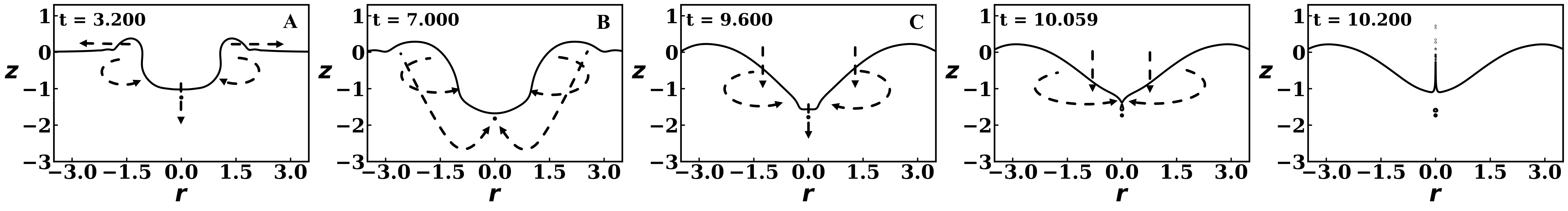}};
\node[anchor=north west,font=\bfseries,fill=white,inner sep=0.0pt] at (-0.2,-0.1) {(f)};
\end{tikzpicture}
\end{minipage}

\vspace{1mm}

\begin{minipage}{\textwidth}
\centering
\begin{tikzpicture}
\node[anchor=north west,inner sep=0] at (0,0)
{\includegraphics[width=\linewidth,height=0.10\textheight]{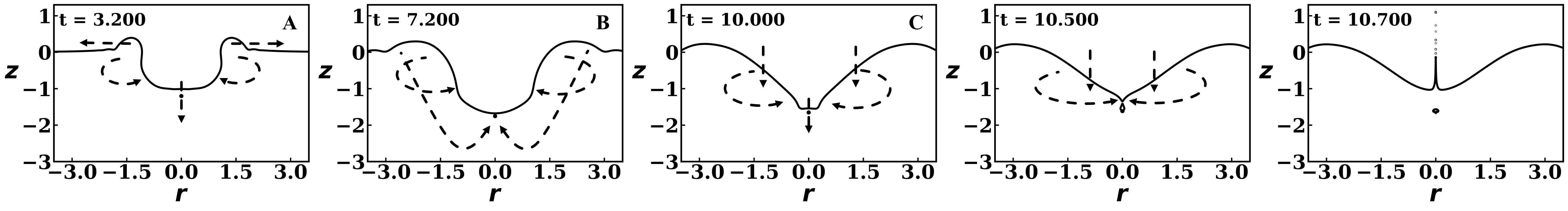}};
\node[anchor=north west,font=\bfseries,fill=white,inner sep=0.0pt] at (-0.2,-0.1) {(g)};
\end{tikzpicture}
\end{minipage}

\vspace{1mm}

\begin{minipage}{\textwidth}
\centering
\begin{tikzpicture}
\node[anchor=north west,inner sep=0] at (0,0)
{\includegraphics[width=\linewidth,height=0.10\textheight]{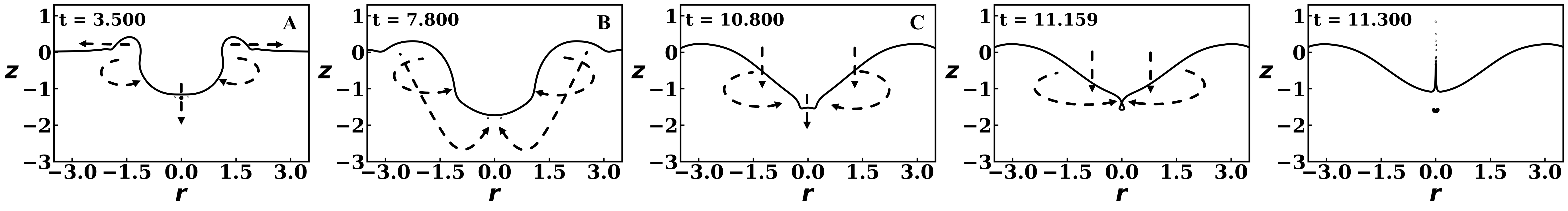}};
\node[anchor=north west,font=\bfseries,fill=white,inner sep=0.0pt] at (-0.2,-0.1) {(h)};
\end{tikzpicture}
\end{minipage}

\vspace{3mm}

\caption{\footnotesize
Temporal evolution of the swell-wave, crater retraction, and jet formation:
(a) a short, thick jet with a secondary droplet (P1), $We=100$, $Fr=45.8$;
(b) a short, thin jet with multiple secondary droplets (P2), $We=113$, $Fr=51.8$;
(c) a long, thin jet with multiple secondary droplets (P3), $We=117$, $Fr=53.3$;
(d) a high-speed singular-jet with microbubble entrapment (P4), $We=120$, $Fr=55.0$;
(e) a thin jet with a small bubble (P5), $We=123$, $Fr=56.4$;
(f) a thin jet with a teardrop-shaped bubble (P6), $We=127$, $Fr=58.2$;
(g) a thin jet with a heavier and wider base bubble (P7), $We=134$, $Fr=61.5$;
and (h) a thin jet with a triangular bubble (P8), $We=143$, $Fr=65.5$.
}

\label{fig:crater_evolution}

\end{figure}

To elucidate the effect of impact conditions on crater collapse and subsequent jet formation, we demonstrate the temporal evolution of the air-liquid interface following drop-impact in different cases within the high-speed jetting regime \citep{leng2001splash} in figure~\ref{fig:crater_evolution}.  The different jet phenomena with their flow patterns are marked by dashed lines at different time instants. The short thick jet with a secondary drop phenomenon (P1), the short thin jet with multiple secondary drop phenomena (P2), the long thin jet with multiple secondary droplet phenomena (P3), the high-speed singular-jet with microbubbles entrapment phenomenon (P4), the thin jet with small bubble entrapment (P5), the thin jet with teardrop bubble entrapment (P6), the thin jet with a heavier and wider base bubble entrapment (P7), and the thin jet with a triangular bubble entrapment (P8). See the supplementary movies S1-S8 corresponding to P1-P8 respectively. We further divide these jet phenomena into three different regimes based on their characteristics.  \\

\FloatBarrier

\begin{figure}
\centering

\begin{minipage}{0.49\textwidth}
\centering
\begin{tikzpicture}
\node[anchor=north west,inner sep=0] at (0,0)
{\includegraphics[width=\linewidth,height=0.22\textheight]{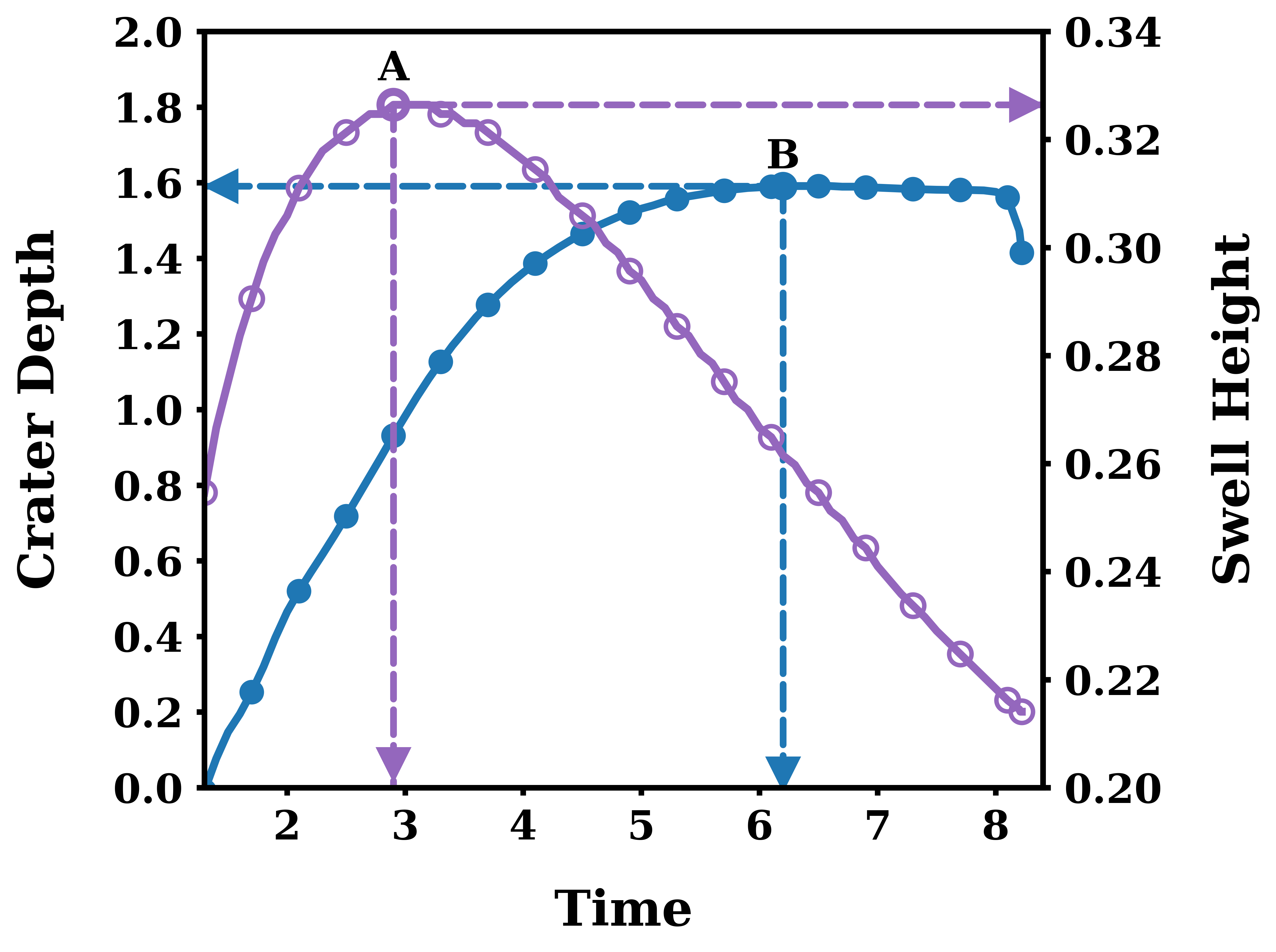}};
\node[anchor=north west,font=\bfseries,fill=white,inner sep=1pt] at (-0.1,-0.05) {(a)};
\end{tikzpicture}
\end{minipage}
\hfill
\begin{minipage}{0.49\textwidth}
\centering
\begin{tikzpicture}
\node[anchor=north west,inner sep=0] at (0,0)
{\includegraphics[width=\linewidth,height=0.22\textheight]{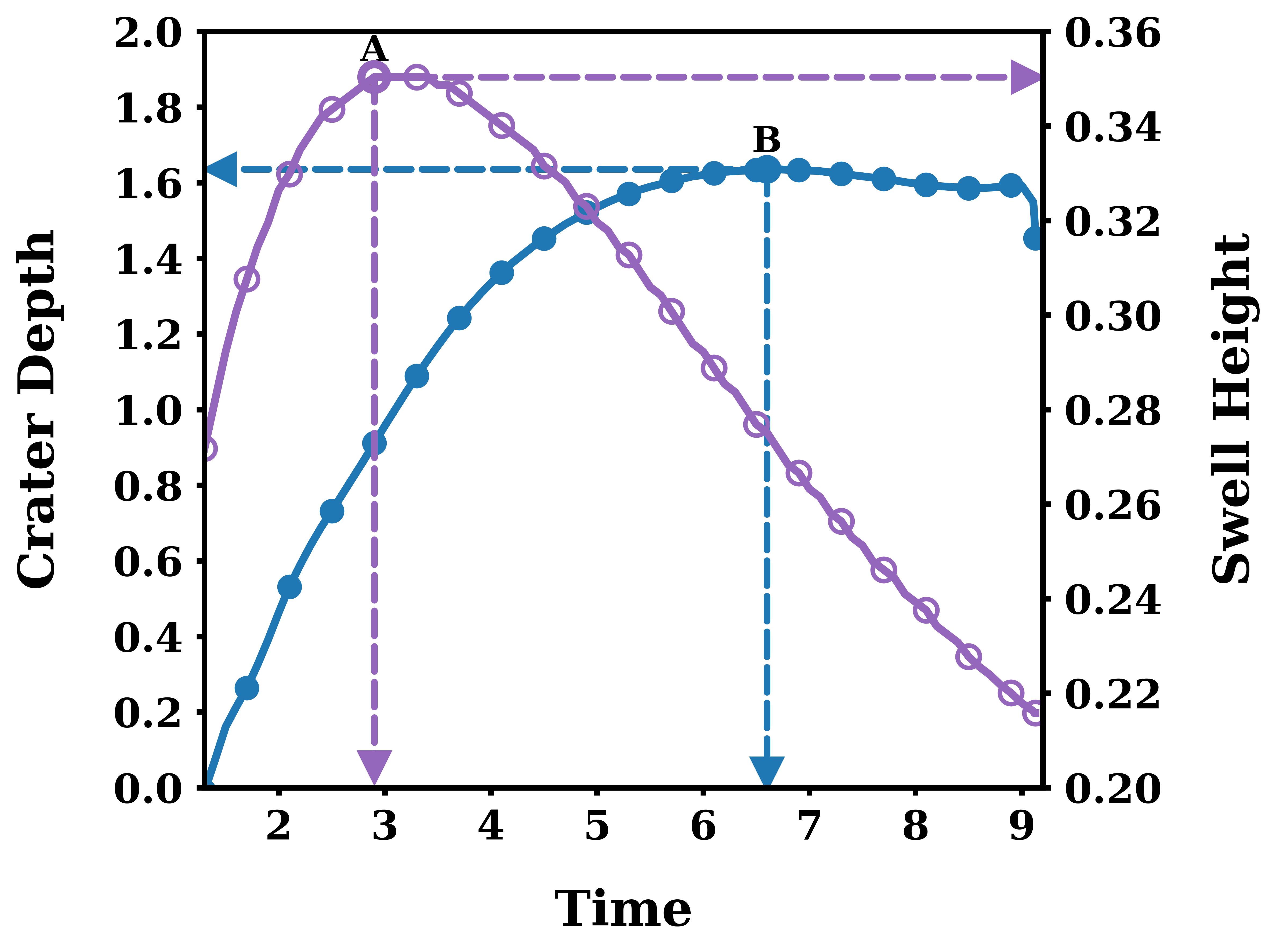}};
\node[anchor=north west,font=\bfseries,fill=white,inner sep=1pt] at (-0.1,-0.05) {(b)};
\end{tikzpicture}
\end{minipage}

\vspace{1mm}

\begin{minipage}{0.49\textwidth}
\centering
\begin{tikzpicture}
\node[anchor=north west,inner sep=0] at (0,0)
{\includegraphics[width=\linewidth,height=0.22\textheight]{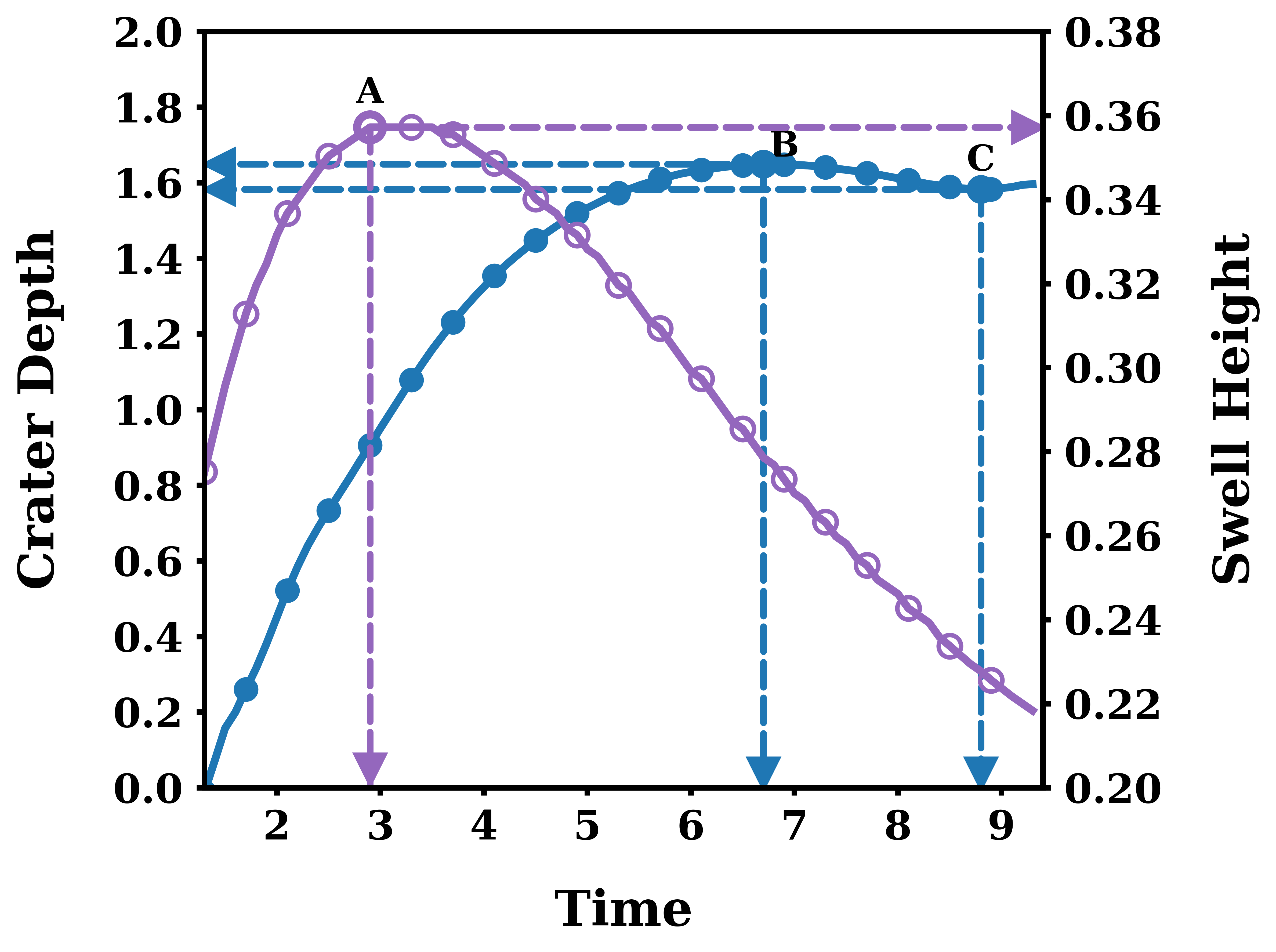}};
\node[anchor=north west,font=\bfseries,fill=white,inner sep=1pt] at (-0.1,-0.05) {(c)};
\end{tikzpicture}
\end{minipage}
\hfill
\begin{minipage}{0.49\textwidth}
\centering
\begin{tikzpicture}
\node[anchor=north west,inner sep=0] at (0,0)
{\includegraphics[width=\linewidth,height=0.22\textheight]{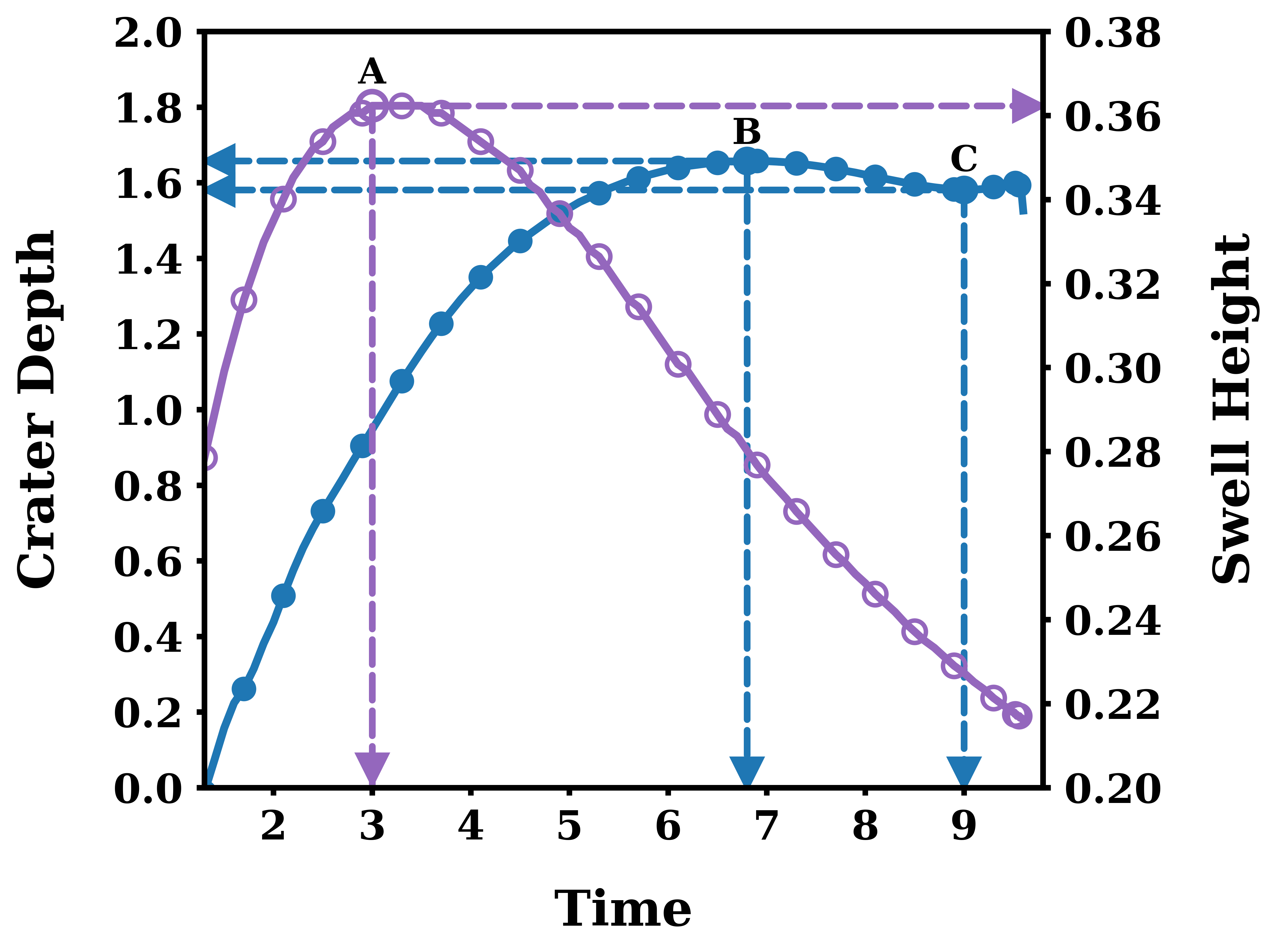}};
\node[anchor=north west,font=\bfseries,fill=white,inner sep=1pt] at (-0.1,-0.05) {(d)};
\end{tikzpicture}
\end{minipage}

\vspace{1mm}

\begin{minipage}{0.49\textwidth}
\centering
\begin{tikzpicture}
\node[anchor=north west,inner sep=0] at (0,0)
{\includegraphics[width=\linewidth,height=0.22\textheight]{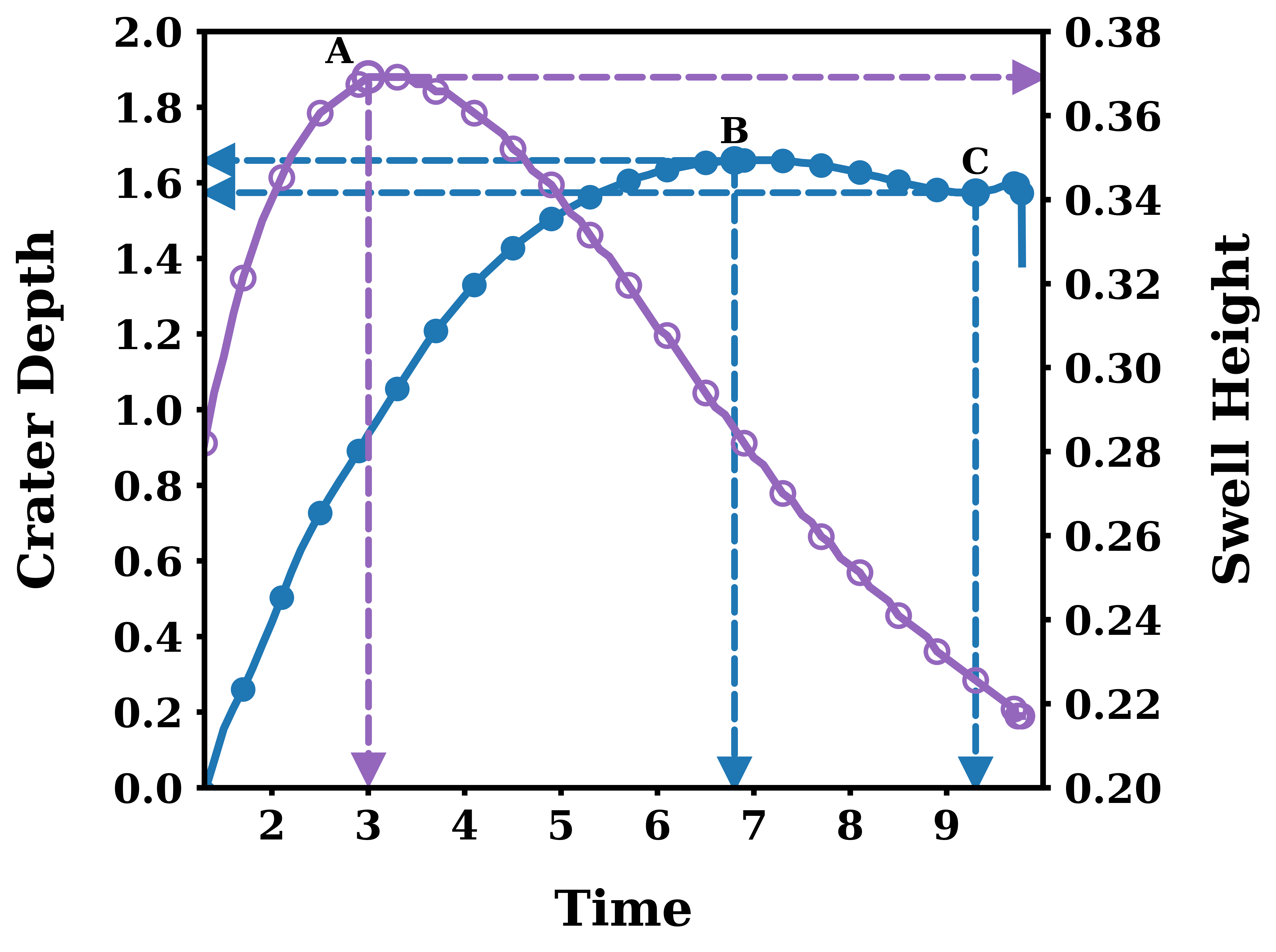}};
\node[anchor=north west,font=\bfseries,fill=white,inner sep=1pt] at (-0.1,-0.05) {(e)};
\end{tikzpicture}
\end{minipage}
\hfill
\begin{minipage}{0.49\textwidth}
\centering
\begin{tikzpicture}
\node[anchor=north west,inner sep=0] at (0,0)
{\includegraphics[width=\linewidth,height=0.20\textheight]{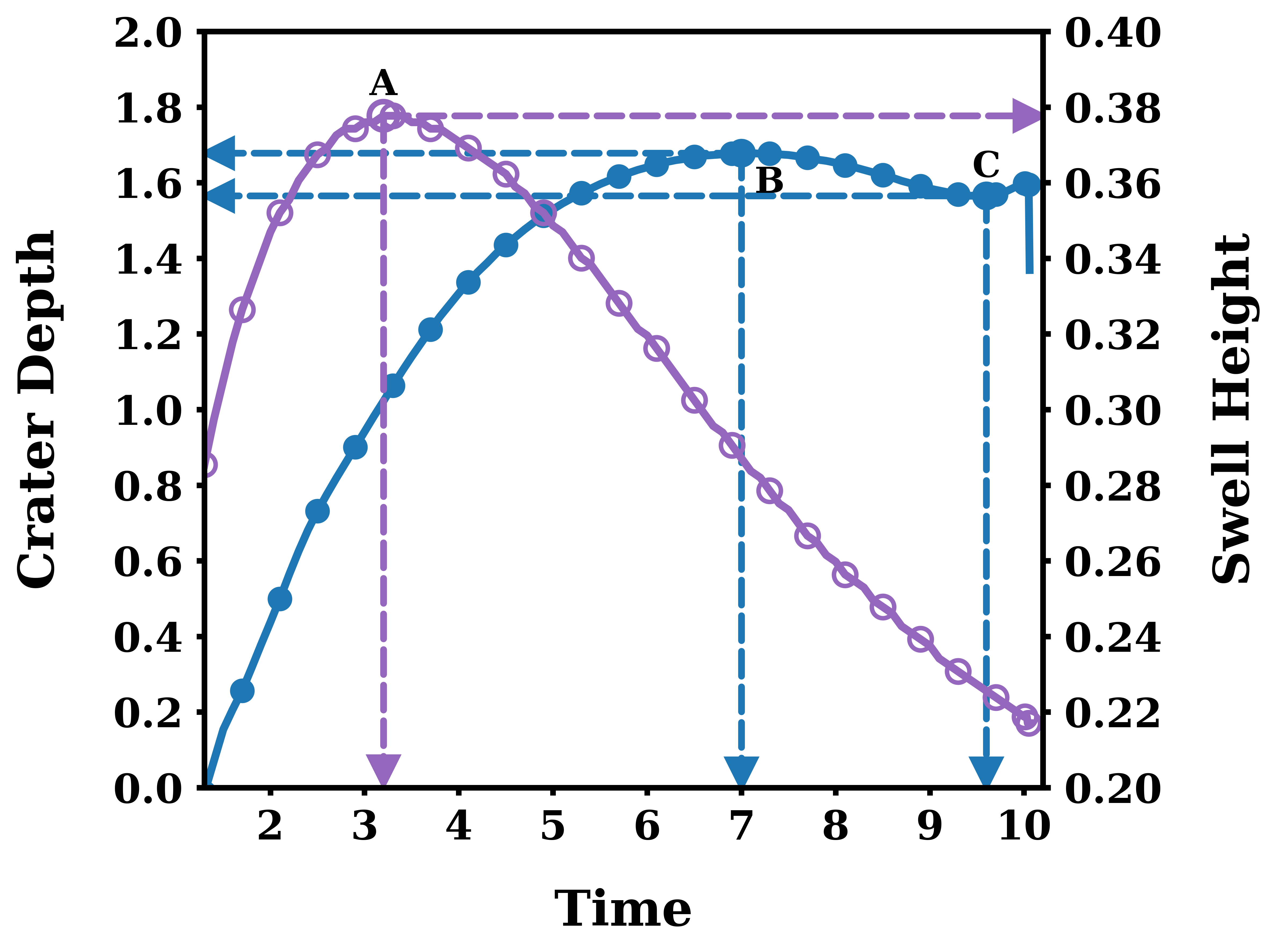}};
\node[anchor=north west,font=\bfseries,fill=white,inner sep=1pt] at (-0.1,-0.05) {(f)};
\end{tikzpicture}
\end{minipage}

\vspace{1mm}

\begin{minipage}{0.49\textwidth}
\centering
\begin{tikzpicture}
\node[anchor=north west,inner sep=0] at (0,0)
{\includegraphics[width=\linewidth,height=0.22\textheight]{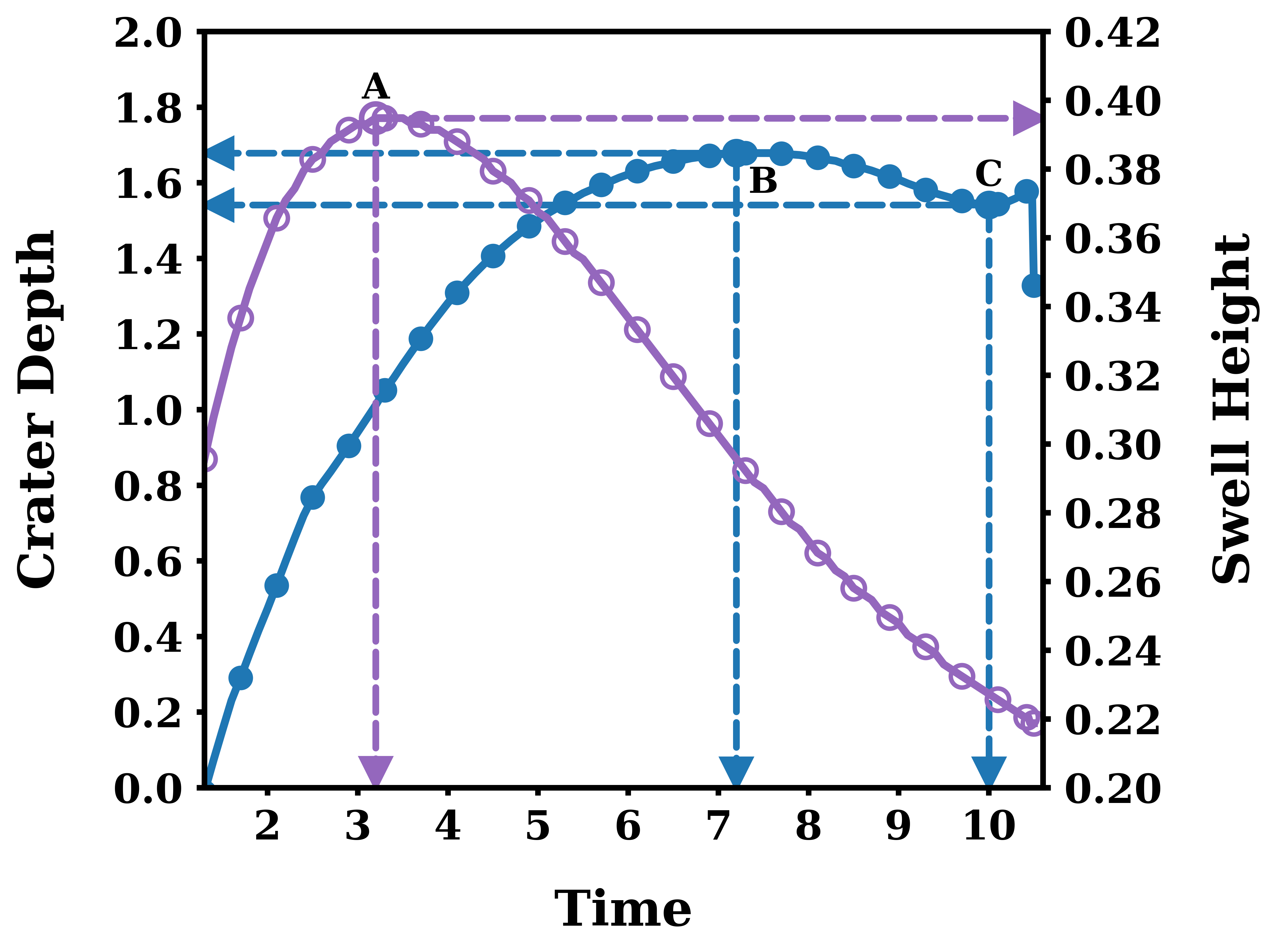}};
\node[anchor=north west,font=\bfseries,fill=white,inner sep=1pt] at (-0.1,-0.05) {(g)};
\end{tikzpicture}
\end{minipage}
\hfill
\begin{minipage}{0.49\textwidth}
\centering
\begin{tikzpicture}
\node[anchor=north west,inner sep=0] at (0,0)
{\includegraphics[width=\linewidth,height=0.22\textheight]{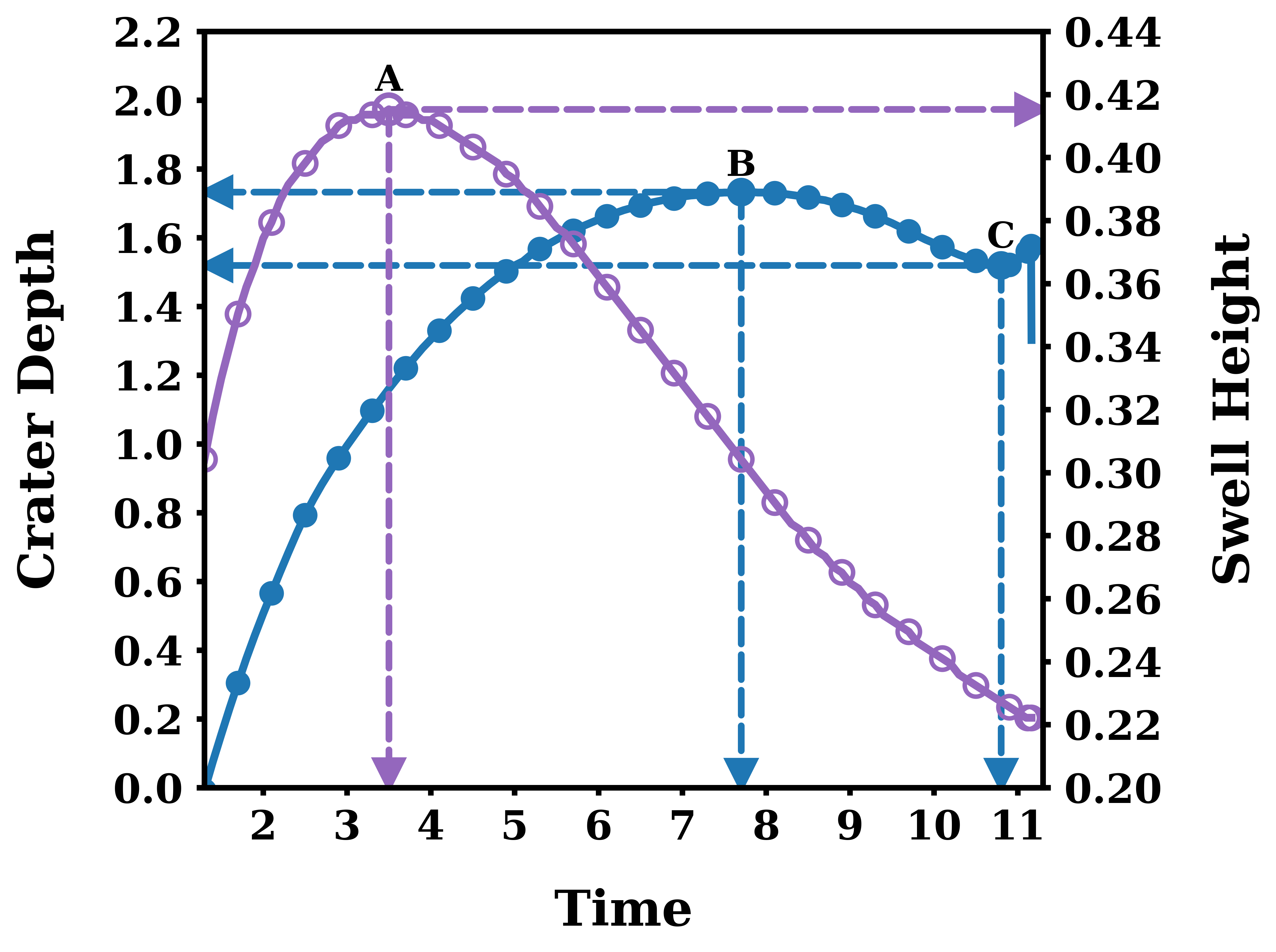}};
\node[anchor=north west,font=\bfseries,fill=white,inner sep=1pt] at (-0.1,-0.05) {(h)};
\end{tikzpicture}
\end{minipage}

\vspace{3mm}

\caption{\footnotesize
Swell-wave and crater-depth variation during crater retraction and jet formation during:
(a) a short, thick jet with a secondary droplet (P1);
(b) a short, thin jet with multiple secondary droplets (P2);
(c) a long, thin jet with multiple secondary droplets (P3);
(d) a high-speed singular-jet with microbubble entrapment (P4);
(e) a thin jet with a small bubble entrapment (P5);
(f) a thin jet with a teardrop-shaped bubble entrapment (P6);
(g) a thin jet with a larger bubble entrapment having a wider base (P7);
and (h) a thin jet with a triangular bubble entrapment (P8).
}

\label{fig:crater_depth}
\end{figure}

In all of the aforementioned cases, the crater evolves through three transitional stages: (a) crater formation accompanied by the generation of swell-waves on the pool surface, (b) development of swell-waves and crater side walls, and (c) retraction of crater-bases, which leads to jet formation. The statistical representation of the expansion and retraction of swell-waves and crater-depth is shown in figure~\ref{fig:crater_depth}. The markers in figures~\ref{fig:crater_evolution} and \ref{fig:crater_depth} represent the maximum swell-wave height at point A and the maximum crater-depth at point B. In some cases, the crater exhibits a transient re-expansion marked by point C, leading to oscillations of the cavity base before the onset of jet formation. When a drop impinges on a liquid pool at a velocity that exceeds the threshold and enters the splashing regime, it forms a deep crater and pushes the swell-waves radially outward. The height of the swell-waves increases to a maximum value (A), and at this instant, craters form a U-shaped profile. Afterward, the swell-waves above the undisturbed pool surface retract, pushing the pool liquid towards the crater. This drives the crater side wall radially inward and initiates crater retraction. The crater side wall retracts, but its base continues to expand until it reaches its maximum depth (B). Following this, the crater-base becomes flat. Thus far, the process is common in all the aforementioned cases. Beyond this point, the flow undergoes distinct morphological changes that form the basis for the classification proposed in this study as further discussed in section \S~\ref{sec:results-1}. 
\\

The pre-singular-jet regime comprises the short, thick jet with a secondary drop (P1), the short, thin jet with multiple secondary droplets (P2), and the long, thin jet with multiple secondary droplets (P3). These cases are collectively classified as regime $\mathrm{I}$, and this notation is adopted throughout the rest of the study. For $We = 100$, and $ Fr = 45.8 $, a short, thick jet with a secondary droplet is shown in figure~\ref {fig:crater_evolution}(a). In this case, before the crater side wall collapses, the crater-base begins to move upward. The crater-base forms a wide V-shaped profile, with a large separation between the side walls before jet formation. This leads to a thick jet that emits a large secondary drop once the jet reaches nearly half of the crater-depth.  Figure~\ref {fig:crater_depth}(a) shows that the maximum amplitude of the swell-wave and crater-depth are $0.326D$ and $1.59D$, respectively, at time units $2.9$ and $6.2$. Similarly, a short, thin jet with multiple secondary droplets forms for $ We = 113 $ and $ Fr = 51.8 $ as shown in figure~\ref {fig:crater_evolution}(b). The corresponding swell-waves attain their maximum amplitude of $0.350D$ and crater-depth of $1.63D$ at time units $2.9$ and $6.7$, respectively, as shown in figure~\ref {fig:crater_depth}(b). Owing to the higher impact velocity, the increased impact energy generates larger swell-waves and a deeper crater, delaying their subsequent retraction compared to P1. The delay in retraction allows the crater side walls to move inward, thereby reducing their separation. The crater-base begins to rebound before the side walls collapse. It forms a thin jet that emanates the first secondary drop at the same jet height as P1. The jet undergoes capillary breakup through the Rayleigh–Plateau instability, producing multiple secondary droplets. These secondary droplets are considerably smaller and, therefore, experience a lower drag force, thereby increasing the jet velocity. In a long thin jet with multiple secondary droplet phenomenon (P3) for $We = 117$ and $ Fr = 53.3 $, the crater-base and side walls retract synchronously. This results in the formation of a central dimple at the crater-base, as shown in figure \ref{fig:crater_evolution}(c). The swell-waves evolve similarly, with a maximum magnitude of $0.357D$ at $2.9$ time units, attaining the maximum crater-depth of $1.65D$ at $6.7$ time units. The central dimple at the crater-base undergoes re-expansion, increasing the crater-depth again after $8.8$ time units, as indicated by point C in figure \ref{fig:crater_depth}(c). Following this expansion, contraction, and re-expansion of the crater-base, the central dimple recedes and forms a very fine jet. The jet subsequently breaks into multiple secondary drops shortly after reaching a much smaller height than in P1 and P2. \\

Upon further increasing the impact velocity, we identify the singular-jet regime designated as regime $\mathrm{II}$. It is characterized by a high-speed singular-jet with microbubble entrapment (P4) for $We = 120$ and $Fr = 55$. The dynamics in figure \ref{fig:crater_evolution}(d) closely resemble those in P3, with a similar magnitude of the maximum swell-wave height, crater-depth, and subsequent re-expansion, as shown in \ref {fig:crater_depth}(d). However, the central dimple becomes cylindrical, and the side wall at the base of the crater is nearly straight. The crater-base undergoes re-expansion while the dimple side walls retract inward. The central dimple base begins to move upwards at the point where the crater side walls are on the verge of collision. Immediately after the central dimple collapses, a secondary drop emanates from the tip of a highly slender, steep singular-jet. The phenomenon consists of several secondary droplets. The first secondary drop travels very fast, covering the crater-depth shortly after pinch-off. The singular-jet continues to grow, while the inward retraction of the central dimple side walls converges and misses the pinch-off singularity. Instead, the center dimple collapse results in the formation of microbubble at its base. \\


After the singular-jet regime, the crater dynamics undergo a qualitative transition to a bubble-entrainment regime $\mathrm{III}$. The central dimple forms at the base of the crater, detaches from the surrounding cavity to form a bubble, and initiates a downward jet that pierces the bubble and propagates further into the liquid pool beneath the cavity. The regime consists of the thin jet with small bubble entrapment (P5) for $We = 123$ and $Fr = 56.4$, the thin jet with teardrop bubble entrapment (P6) for $We = 127$ and $Fr = 58.2$, the thin jet with a heavier and wider base bubble entrapment (P7) for $We = 134$ and $Fr = 61.5$, and the thin jet with a triangular bubble entrapment (P8) phenomenon for $We = 143$ and $Fr = 65.5$. Across this regime, both the swell-wave height and crater-depth are greater than those observed in  regimes $\mathrm{I}$ and $\mathrm{II}$, shown in figures \ref{fig:crater_depth}(e-h), consistent with the increasing impact energy with the $We$. The increased crater dimensions are accompanied by the delay in crater-base retraction and the subsequent re-expansion. The delay in crater collapse increases as we enter regime $\mathrm{III}$ and approach the thick-jet regime \citep{oguz1990bubble}. Despite the transition to regime $\mathrm{III}$, the jet velocity is higher than in regime $\mathrm{I}$ but decreases to one-third of the value attained in regime $\mathrm{II}$. In the thin-jet phenomenon with small bubble entrapment (P5), the maximum swell-wave amplitude and crater-depth are, $0.37D$ and $1.66D$, respectively, at time units $3.0$ and $6.8$. The cylindrical center dimple becomes elongated, and its depth increases beyond point C relative to P4, which is $0.09D$ as shown in figure \ref{fig:crater_depth}(e). The elongation delays the dimple collapse and promotes air-bubble entrapment. In this case, during re-expansion, the base starts moving downward again, and the side wall of the dimple continues to move inward. The motion of the dimple side wall leads to a collision at a point above the dimple base, as shown in figure \ref{fig:crater_evolution}(e), entraining an air bubble and ejecting a high-speed, thin jet in both upward and downward directions, which breaks into numerous secondary drops. Beyond regime $\mathrm{II}$, both the maximum crater-width and depth, as well as the dimple depth during re-expansion, increase significantly. In the thin-jet phenomenon with teardrop bubble entrapment (P6) and in the thin-jet phenomenon with a heavier, wider-base bubble entrapment (P7), the maximum amplitude of the swell-waves are $0.377D$ and $0.394D$, at time units $3.2$ in both cases, whereas the maximum crater-depths are $1.68D$ and $1.67D$, respectively, at time units $7.0$ and $7.2$, respectively. The center dimple is elongated and forms a wider hanging-drop-like shape before pinch-off, as illustrated in figures \ref{fig:crater_evolution}(f) and \ref{fig:crater_evolution}(g), respectively. Again, the pinch-off generates two jets; the upward jet fragments into multiple secondary droplets, and from this point onward, we see the downward jet pierce the entrapped bubble in the pool. The thickness of a jet increases as it moves upward. In both P6 and P7, the number of secondary drops formed at the jet tip decreased compared to P5. \\

In all cases considered above, the time instant for the onset of swell-wave is nearly the same, while the crater-base retraction also begins at the same instants, except for phenomenon P8 in figure \ref{fig:crater_evolution}(h). Because of its high impact velocity, the drop-impact splashes the pool liquid high, driving it radially outward and creating a deep crater. The maximum swell-wave amplitude and crater-depth are $0.415D$ and $1.73D$, respectively, at time units $3.5$ and $7.8$. The largest difference between the maximum crater-depth and the crater-depth before re-expansion is $0.21D$, observed in phenomenon P8, as shown in figure \ref{fig:crater_depth}(h). We have not seen any morphological changes in crater evolution in phenomena P3, P4, and P5. Their crater depths before and after the re-expansion phase are nearly the same, but changes in the dynamics of the central dimple at the bottom of the crater significantly alter the jet speed in each case. A peak was observed for regime II, as shown in figure \ref{fig:Validation}(g). The crater dynamics in the present study are qualitatively similar to those reported by \cite{ray2015regimes}; the corresponding impact conditions occupy distinct regions. The impact conditions considered by \cite{ray2015regimes} span $We = 50-300$ and $Fr = 25-600$, whereas the present simulations are confined to $We = 100-200$ and $Fr = 45-92$. In their study, they do not report thin, cylindrical, high-velocity jets comparable to those obtained in the present study. The jetting velocities reported in \cite{ray2012bubble} are not close to the ones we are getting in the singular-jet case. 
For pre-entrainment cases, the maximum amplitude of the swell-wave and crater-depth reported by \cite{ray2015regimes} are $0.4D$ and $2.27D$, respectively, and for the entrainment cases, they become $0.53D$ and $2.2D$, respectively. In the present study, among cases without bubble entrainment, the singular-jet case shows the maximum swell-wave amplitude and crater-depth of $0.357D$ and $1.65D$, respectively. The bubble entrainment cases show a maximum swell-wave of $0.415D$ and a crater-depth of $1.73D$ for $We = 143$. 
\\ 


\subsection{Geometrical characterization of crater and jet}\label{sec:results-2}

\begin{figure}
    \centering
    \includegraphics[width=\linewidth,height=0.45\textheight]{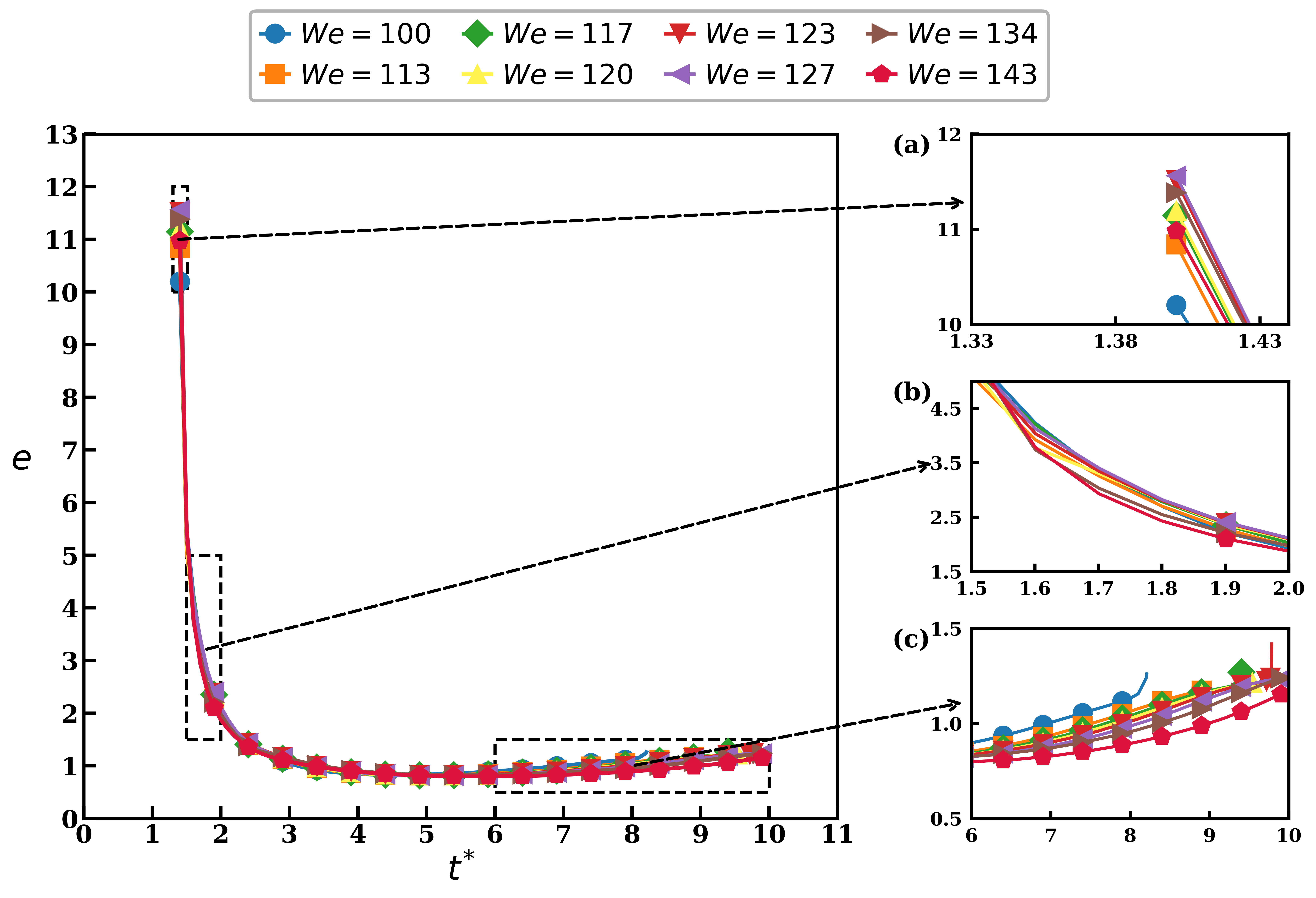}    
    \caption{The e-ratio for the different cases characterizes the crater aspect ratio throughout its evolution up to the crater collapse. An e-ratio of unity corresponds to a hemispherical crater, whereas $e < 1$ and $e > 1$ indicate prolate and oblate crater geometries, respectively.}
    \label{fig:e_ratio}
\end{figure}

We quantitatively examine the physical changes observed in section \S~\ref{sec:results-1} to elucidate the underlying dynamics of crater collapse and jet formation. Figure \ref{fig:e_ratio} shows the temporal variation of the e-ratio ($\textit{e}$) \citep{leng2001splash}, defined as the ratio of the crater-width at the free surface to twice the crater-depth, for different cases across regimes. The e-ratio, $e=1$, corresponds to a hemispherical crater with the same radial and axial displacements, whereas $e < 1$ and $e > 1$ indicate a prolate and an oblate cavity, respectively. In a prolate shape, the cavity is elongated toward the poles, whereas in an oblate cavity, it spreads more in the radial direction. \\


\begin{figure}
    \centering
    \includegraphics[width=\linewidth,height=0.25\textheight]{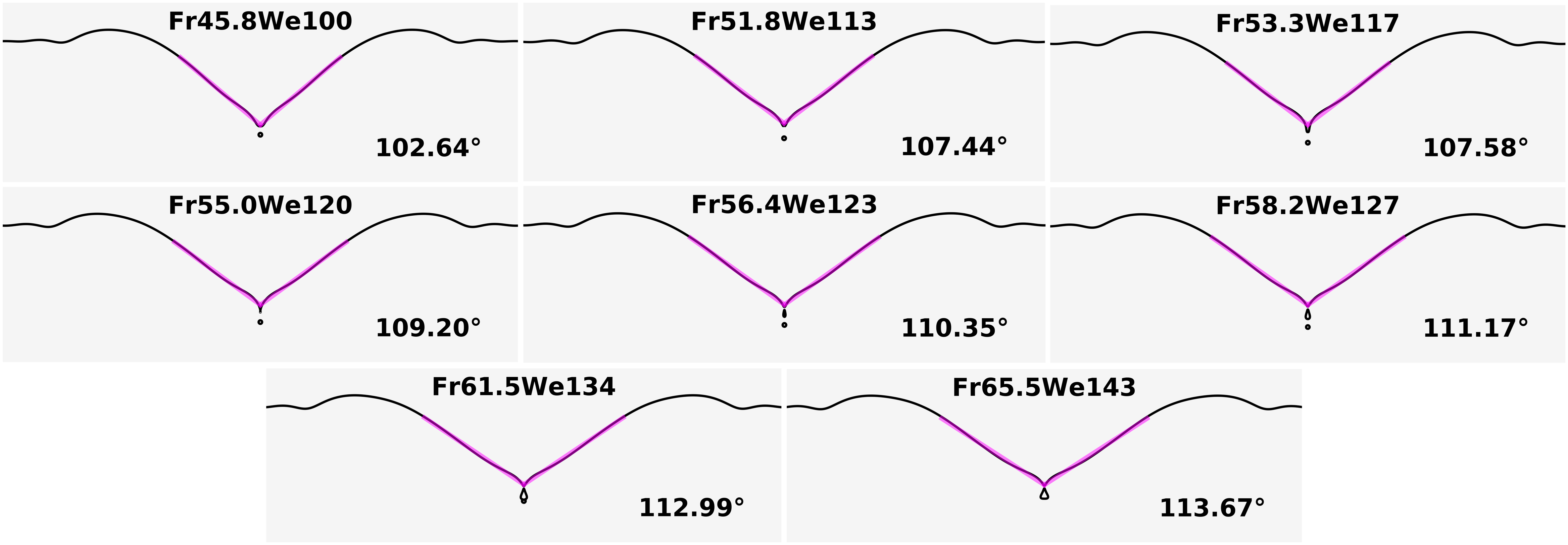}
    \caption{\footnotesize
Variation of the swell angle immediately before crater collapse for the different impact conditions, characterizing the inclination of the crater side wall relative to the crater-base.
}
    \label{fig:swell_angle}
\end{figure}

Following a drop-impact, the impact energy deforms the free surface radially, producing swell-waves that propagate outward across the pool surface and axially, forming a deep crater in the pool. Initially, the e-ratio attains relatively large values because the crater opening is wider than its depth, making the crater shape oblate. In figure \ref{fig:e_ratio}(a), the e-ratio increases with increasing $We$, then decreases as the $We$ further increases. To explore possible reasons for this behavior, we need to revisit the crater-depth and swell-wave dynamics shown in figure \ref{fig:crater_depth}. The crater-depth characterizes the axial extent of the cavity, whereas the outward displacement associated with the swell-wave contributes to its radial displacement. The greater impact inertia associated with a higher $We$ allows the drop to penetrate deeper into the pool and splash the pool liquid upward, forming higher swell-waves rather than displacing it radially. The combined effect of an increased crater-depth and a relatively reduced crater-width is responsible for a dip in the e-ratio at higher $We$, as shown in figure \ref{fig:e_ratio}(a). Thereafter, the cavity continues to expand axially, leading to a further reduction in the e-ratio. The expansion of the crater is nearly identical across all cases up to $1.5$ time units; their corresponding e-ratios overlap, as shown in figure \ref{fig:e_ratio}. Subsequently, the crater-width and depth increase at the same rate, which results in the e-ratio approaching unity and remaining constant. An enlarged view is shown in figure \ref{fig:e_ratio}(b), in which the e-ratio follows a similar trend to that observed in figure~\ref{fig:e_ratio}(a): the e-ratio at higher $We$ is lower than in the other cases. The crater acquires a hemispherical shape; thereafter, the cavity continues to expand axially towards its maximum depth, the e-ratio falls below unity, prompting the crater to transition from a hemispherical to a prolate shape. Figure \ref{fig:e_ratio}(c) illustrates that as the crater begins to retract, the e-ratio increases and eventually exceeds unity as the crater-base rises and forms an upward jet. At higher $We$, craters retract slowly owing to their larger dimensions and delay the increase in e-ratio compared to the other cases. \\

The e-ratio quantifies the temporal evolution of the crater aspect ratio and helps to predict the crater's shape before collapse. The emergence of a jet occurs during the final stage of crater collapse, and its velocity strongly depends on how the crater collapses. In figure~\ref{fig:Validation}(f), we observe that for the fastest singular-jets, the crater dimple retracts without pinch-off. In regime $\mathrm{I}$, the separation between the crater side walls limits the jet velocity, whereas the side walls collapse, resulting in a pinch-off that is responsible for the jet velocity reduction in regime III. To quantify the transition in dimple shapes associated with different jet velocities, we evaluate the swell-wave angle for cases across the three regimes, as shown in figure~\ref{fig:swell_angle}. The swell-wave angle is evaluated just before the crater collapse and is defined as the inclination of the crater side walls relative to the crater-base. Following the impact, capillary waves are generated and propagate in different directions. Several waves travel toward the crater-base along the side walls; some of them move radially outward across the pool surface, while a few travel along the drop surface. The most pronounced capillary wave emerges when swell-waves interact with the crater side wall, forming a sharp corner that moves along the crater-depth.  The motion of these capillary waves governs the swell-wave angle, thereby influencing crater dynamics. We will discuss the role of capillary waves in section \S~\ref{sec:results-3}. Following their maximum elevation, the swell-waves propagate radially outward and subsequently descend along the crater side wall as indicated by the dashed lines in figure~\ref{fig:crater_evolution}. Swell-waves exert forces that drive the crater side wall radially inward toward the axisymmetric axis, increasing the swell-wave angle as the $We$ increases. At $We = 100$ and $Fr = 45.8$, the swell-waves become weak after $8$ time units, as shown in figure~\ref{fig:crater_depth}(a). Thereafter, the force exerted by swell-waves is insufficient to drive the crater side walls radially inward, resulting in a minimum swell angle among all the cases. Consequently, the crater side walls remain well separated, while the crater-base retracts and lifts independently, resulting in a low jet velocity. Afterward, the swell-waves become weak with a noticeable delay, allowing the crater side walls to continue moving radially inward and reducing the gap between the opposing side walls before the crater-base begins to lift. The reduced gap between the crater side walls enables synchronous collapse, enhancing the upward acceleration of the crater-base and thereby promoting faster jet formation. In the singular-jet case at $We = 120$ and $Fr = 55$, an optimum swell angle of $109.2^\circ$ yields the maximum jet velocity, and the side walls become nearly vertical before collapse. In regime $\mathrm{III}$, we observe a higher swell angle, indicating dominance of radial inward motion that drives the crater-base downward, as depicted in figures~\ref{fig:crater_evolution} and~\ref{fig:crater_depth}. The crater side walls collide above the crater-base, resulting in pinch-off and leading to the formation of two counter-propagating jets. The critical swell angle for the bubble entrainment in the references is $109^\circ$. The optimum swell angle in the present singular-jet case to avoid the pinch-off singularity is $109.2^\circ$, which is in good agreement with the existing literature \citep{leng2001splash,ray2015regimes}.\\

\begin{figure}
    \centering   \includegraphics[width=0.80\linewidth,height=0.45\textheight]{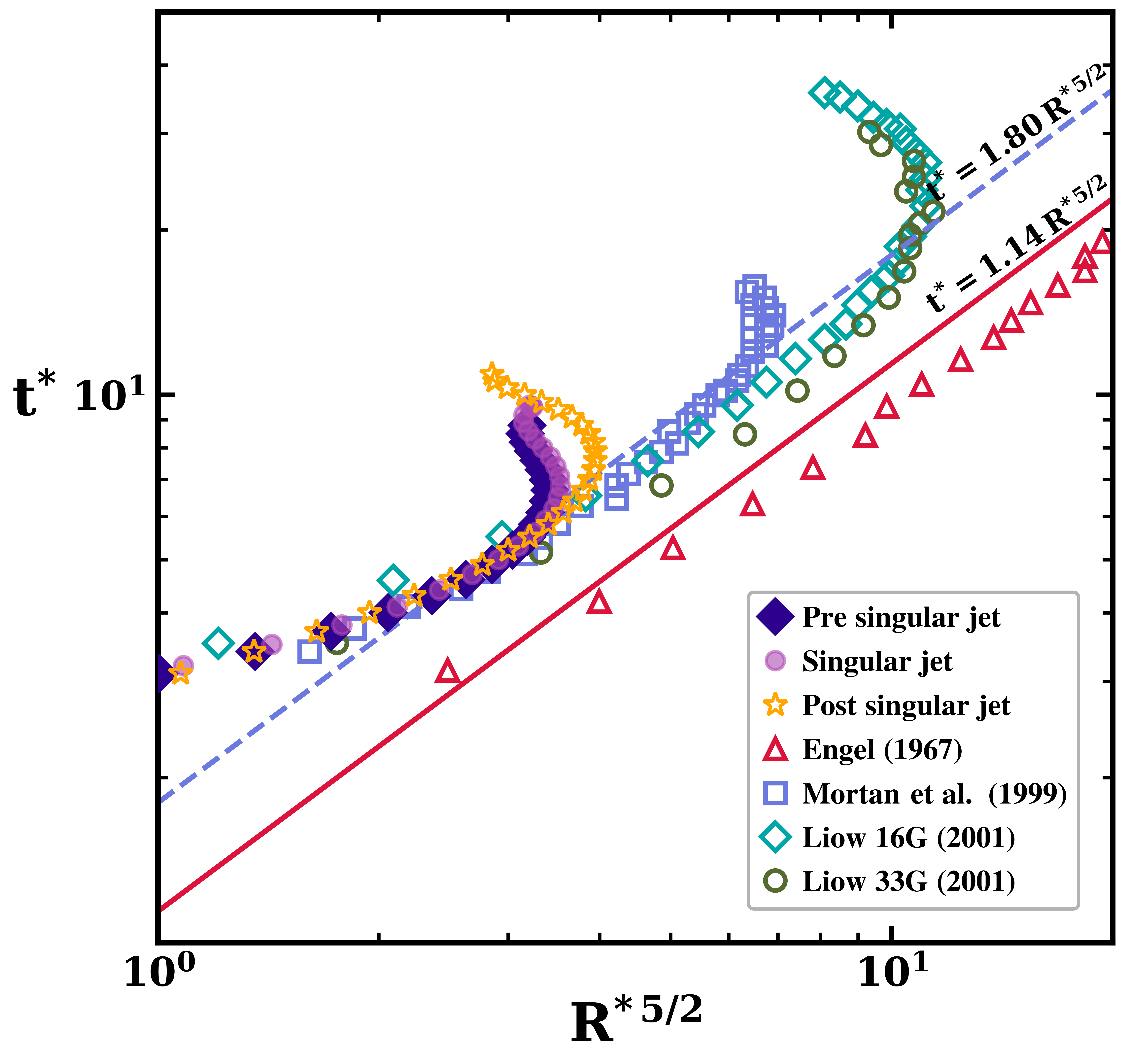}    
  \caption{\footnotesize
Variation of dimensionless crater-depth with time under different impact conditions. The solid and dashed lines represent the least-squares fits of \cite{engel1967initial}, $t^{*}=1.14R^{*5/2}$, and \cite{morton2000investigation}, $t^{*}=1.8R^{*5/2}$, respectively. Experimental results for the $16G$ and $33G$ needle-drop cases of \cite{leng2001splash} at a fall height of $800 mm$ are shown for $Fr=387$, $We=756$, $d=3.77mm$ and $Fr=600$, $We=368$, $ d=2.12 mm$, respectively. The present simulations correspond to $We=113$, $Fr=51.8$ in regime $\mathrm{I}$, $We=120$, $Fr=55.0$ in regime $\mathrm{II}$, and $We=143$, $Fr=65.5$ in regime $\mathrm{III}$.}
    \label{fig:liow_ratio}
\end{figure}


\cite{engel1967initial} analytically derived an expression for the cavity-depth rate using an energy balance. The potential energy comprises the crater’s gravitational energy, the swell-wave above the undisturbed pool level, together with the potential energy generated by the surface. The potential energy is balanced by the kinetic energy of the liquid surrounding the cavity, in the cylindrical wave, as well as the dissipated energy. This dissipated energy is half the energy of the impinging drop. The liquid drop retains the remaining half, which remains available for its subsequent motion. Using a similar energy balance approach, \cite{leng2001splash} derived an expression for the kinetic energy of the cavity at its maximum depth by incorporating the crater-depth proposed by \cite{engel1967initial}, as given by the following equation: 
\begin{equation}
E_k = \pi r R^3 \left(\frac{dR}{dt}\right)^2,
\label{eq:kinetic_energy_max_depth}
\end{equation}
where $R$ is the crater-depth. They observed that the kinetic energy of an expanding cavity remains nearly constant in the early stages of cavity formation. Based on this observation, they derived the scaling relation $t \propto R^{5/2}$, assuming that the cavity kinetic energy remains constant, where $t$ is the instantaneous time step, and $R$ is the corresponding cavity depth. In figure~\ref{fig:liow_ratio}, we present some of the cases in each regime, along with those from the previous studies by \cite{leng2001splash,engel1967initial,morton2000investigation}. The least-squares fit to the data of \cite{engel1967initial} yields the scaling relation $t^{*} = 1.14\,R^{* 5/2}$, represented by the solid line, whereas the corresponding fit to \cite{morton2000investigation} data gives $t^{*} = 1.8\,R^{* 5/2}$, shown by the dashed line. The data from the present study show excellent agreement with those of \cite{morton2000investigation} and \cite{leng2001splash} during the cavity-expansion stages. In the initial stage and after the crater expansion, when it begins to retract, the results do not follow the power law. \\   

\begin{figure}
    \centering
    \includegraphics[width=\linewidth,height=0.30\textheight]{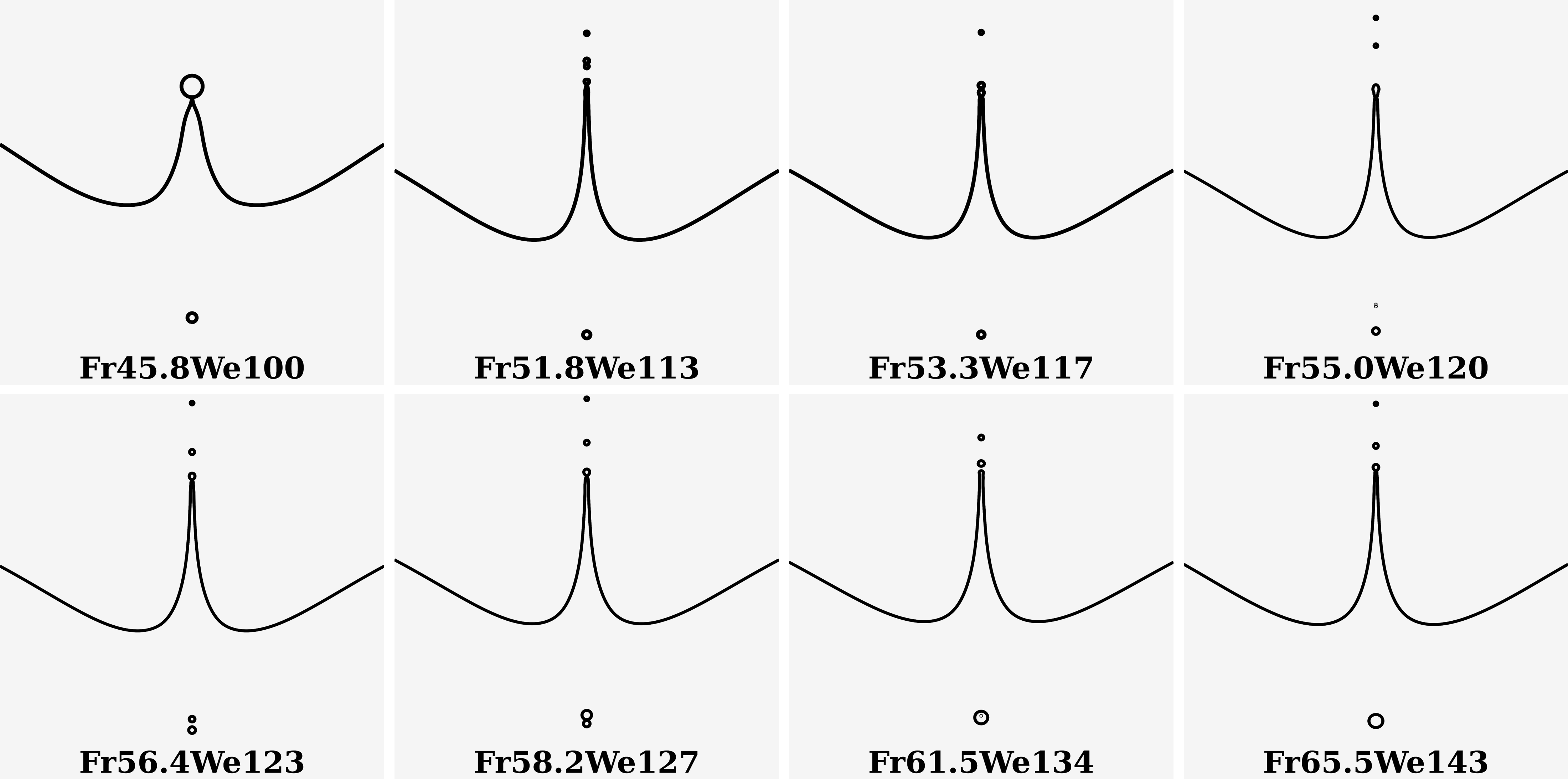}
    \caption{Jet profiles at the instant of emission of the last secondary droplet from the jet tip for different We and Fr across the different regimes.}
    \label{fig:jet_profile}
\end{figure}

\begin{figure}
    \centering
    \includegraphics[width=\linewidth,height=0.45\textheight]{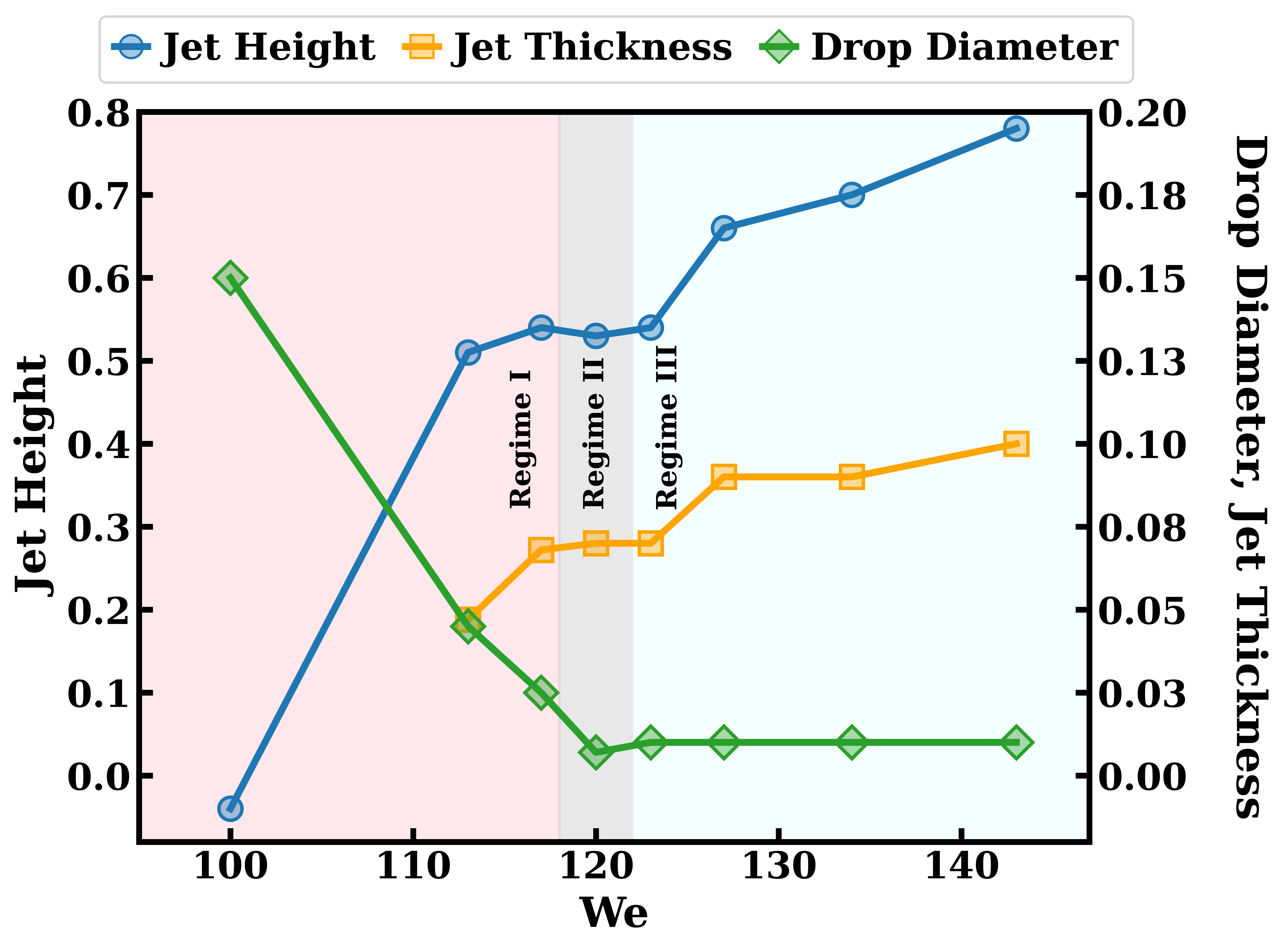}    
    \caption{\footnotesize
Statistical characteristics of the jets for different $We$ and $Fr$ across the different regimes, evaluated at the instant of emission of the last secondary droplet from the jet tip: (a) jet height measured above the undisturbed pool surface; (b) jet thickness defined as the width of the jet base at the undisturbed pool surface; and (c) diameter of the first secondary droplet emitted from the jet tip.
}
    \label{fig:jet_height_thickness}
\end{figure}

To quantify the observed changes in the jet morphology, figure~\ref{fig:jet_profile} shows the jet profile at the instant when the last secondary drop is emitted from the jet tip for different $We$ and $Fr$, while the corresponding statistical characterization is presented in figure~\ref{fig:jet_height_thickness}. The jet profiles are qualitatively similar across all cases except for $We = 100$, which produces a short, thick, fully developed jet. In contrast, a thin jet profile is observed that is still growing at the instant when the last secondary drop is emitted from the jet tip. Three geometric parameters are considered: the jet height, defined as the maximum height of the jet above the undisturbed pool surface at the instant when the last secondary droplet is emitted from the jet tip; the jet thickness, defined as the jet width measured at the undisturbed pool level at the same instant; and the diameter of the first secondary droplet emitted from the jet tip. In figure~\ref{fig:jet_height_thickness}, the jet height for $We = 100$ is negative, and its thickness is undefined because the jet has not yet risen above the undisturbed pool level at the instant when the last secondary droplet pinches off from the jet tip. The jet height increases gradually with the $We$ in regime $\mathrm{I}$. In regime $\mathrm{II}$, the jet height attains a plateau as $We$ increases. We observe a small dip in the jet height for the singular-jet case at $We=120$, where we obtain the fastest jet. The dip in jet height is attributed to the jet breaking into numerous secondary droplets, indicating that a substantial fraction of the jet liquid is redistributed into droplets rather than remaining in the continuous jet column. The formation of secondary droplets decreases in regime $\mathrm{III}$. Owing to the large cavity formed at higher $We$, as shown in figure~\ref{fig:crater_depth}, the jet rebounds to a substantial height, resulting in a progressive increase in the jet height with $We$. The jet thickness exhibits a similar trend: it gradually increases in regime $\mathrm{I}$, remains nearly constant in regime $\mathrm{II}$, and then increases again with the $We$ in regime $\mathrm{III}$.\\


The next parameter is the diameter of the first secondary drop emitted from the jet tip, which is crucial for characterizing the subsequent jet dynamics. In the present study, we compute the jet velocity using the first drop to emerge from the crater, emitted from the jet tip. The larger the secondary drop diameter, the more air drag it experiences as it travels after being pinched off from the jet tip. As shown in section \S~\ref{sec:results-1}, the number of secondary droplets and the instant at which they detach from the jet tip both evolve with the $We$. The number of secondary droplets increases as the jet becomes thinner. The maximum number of secondary droplets occurs in the singular-jet case within regime $\mathrm{II}$, as shown in figure~\ref{fig:Validation}(g). The secondary drop emerges owing to capillary waves that propagate along the jet surface, following dimple collapse. After reaching the jet apex, these waves reflect towards the jet base, where they interact with waves traveling from the base toward the jet apex. The interaction reverses the wave motion, triggering an oscillation, generating a surface instability that induces the neck formation, and causing the jet to break into secondary droplets. \\

The crater morphology and jet-formation analysis discussed in section \S~\ref{sec:results-1} cannot adequately explain the exceptionally high jet velocity in the singular-jet case. 
We observe that cases around the regime $\mathrm{II}$  have very similar overall morphology and statistical data sets. Despite these similarities, the jet velocity exhibits a pronounced local maximum in the singular-jet case. To address this ambiguity, we quantitatively investigate the pressure, velocity, and vorticity fields in different cases and seek reasons for the high-speed in the singular-jet case. \\

\subsection{Pressure, velocity, and vorticity evolution across the jetting regimes}\label{sec:results-3}
Despite similar qualitative and quantitative characteristics, the jets near regime $\mathrm{II}$ have different velocities. In regime $\mathrm{III}$, bubble pinch-off accounts for the reduction in the jet velocity reported by \cite{thoroddsen2018singular}.  However, the underlying mechanism reducing the jet velocity in regime $\mathrm{I}$ remains unclear. The high-velocity of the singular-jet has been attributed to a conical focusing mechanism, as discussed by \cite{tian2023conical}. In a conical focusing mechanism, the air sheet is entrapped between the radially converging dimple and the rising jet. The air layer trapped between the dimple boundaries prevents the emerging jet from coming into direct contact with the surrounding liquid. It provides an effective slip in the converging channel, thereby protecting it from viscous stress. During dimple collapse, different flow conditions arise that cause pronounced changes in jet velocity across cases.  Still, it remains unclear which factor triggers these mechanisms over a narrow range of impact conditions. In earlier studies \citep{pumphrey1990entrainment, morton2000investigation, berberovic2009drop,ray2015regimes} and in the above discussion, capillary waves govern important aspects of crater collapse and subsequent jet formation. \cite{pumphrey1990entrainment} were among the first to observe capillary wave propagation along the crater and their influence on the subsequent dynamics.  \cite{morton2000investigation} also emphasized that capillary waves strongly influence crater morphology. They show that for bubble-entrainment cases, the capillary wavefront reaches the maximum crater-depth at an instant when the crater side wall retracts. This forms a sharp corner along the crater surface, as indicated by the arrow in figure~12 of \citet{morton2000investigation}. Capillary wave propagation creates a substantial pressure gradient across the interface \citep{berberovic2009drop}. A high-pressure region above the wavefront and a low-pressure region below it, providing a method to track capillary-wave motion. \cite{ray2015regimes} used pressure iso-contours to capture capillary wave motion, using the same pressure signature provided by \cite{berberovic2009drop} for the bubble entrainment study during drop impact on a liquid pool. Here, we use pressure and velocity fields to elucidate how flow-field structures differ across the jetting regimes, considering $P1$ and $P3$ in regime I, $P4$ in regime II, and $P6$ and $P8$ in regime III. For the short, thick jet with a secondary drop phenomenon $(P1)$ at $We=100$ and $Fr=45.8$, the capillary waves are initiated during the formation of the swell-waves as depicted in figure~\ref{fig:pressure_evolution}(a). At the intersection of the swell-wave and the undisturbed pool surface, the first capillary wave emerges. It propagates radially outward along the free surface. The second wave is generated when the crater-base contacts the swell-wave side wall. In the pressure contours, the low-pressure region along the crater side wall indicates a capillary wave traveling toward the crater-base, which influences crater dynamics. The capillary wave within the crater moves both radially and axially. In $P1$, immediately after impact, capillary waves propagate to the center $r=0$. At time unit $2.9$, when the swell-waves reach their maximum height,  the crater side walls incline toward the central axis, and the capillary wave starts traveling axially at the bottom of the crater. \\
At maximum crater-depth, the axial propagation of the capillary wave stops. Subsequently, the wave turns toward the central axis $(r=0)$. However, the capillary waves attenuate before reaching the central axis and become too weak to form the pointed conical dimple required for thin high-velocity jets. A high-pressure zone is confined up to the crater-base. This elevates the crater-base, which results in thick jet formation. The simulation demonstrates that a high-pressure region persists within the entrapped air bubble. Owing to the high-pressure in the entrapped bubble, pressure gradients form at the crater-base, impeding the capillary-wave propagation. In figure~\ref{fig:pressure_evolution}(b), for the long, thin jet with multiple secondary droplet phenomena $(P3)$, corresponding to $We = 113$ and $Fr = 53.3$, the first capillary waves move radially outward immediately after impact. This increases the crater-width near the junction of the crater side wall and base. Following this, the crater stabilizes, with the side walls remaining nearly straight at the instant of maximum swell-wave height. Beyond the maximum crater-depth, the capillary waves persist and continue to propagate towards the central axis $(r=0)$. A corner is formed at the dimple base as capillary waves propagate along the crater side wall, but are not strong enough to bring the side walls close together. A high-pressure region beneath the dimple base forms and extends along the dimple side walls. However, the crater-base still begins to lift, and the crater side walls remain separated with a finite gap. \\

The singular-jet case $(P4)$ at $We=120$ and $Fr=55$ exhibits the most pronounced pressure variation around the dimple shown in figure~\ref{fig:pressure_evolution}(c). When the dimple base is on the verge of lifting, a high-pressure region develops beneath the base along the side walls. This pressure variation is associated with local curvature deformation at the crater-base. The pressure contours indicate that capillary waves are confined to forming a dimple and driving the dimple side wall toward $r=0$. 
The entrapped air sheet between the jet and the inner crater surface prevents direct liquid–liquid contact thereby limiting the influence of viscous stresses on the jet. Furthermore, \cite{thoroddsen2018singular} reported that the final dimple collapse is independent of surface tension. This suggests 
that singular-jet dynamics is governed by the pressure field that drives the liquid upward, leading to air entrapment in the dimple. In contrast, the literature shows that the interplay between surface tension and viscous stress forms cusps that entrap air, thereby generating a lubrication pressure as the air tries to escape. This pressure pushes the side wall, preventing the cusp from closing completely and causing air entrapment.  \\

\begin{figure}
\centering

\begin{minipage}{0.32\textwidth}
\centering
\begin{tikzpicture}
\node[anchor=north west,inner sep=0]
{\includegraphics[
    width=\linewidth,
    height=0.95\textheight,
    keepaspectratio
]{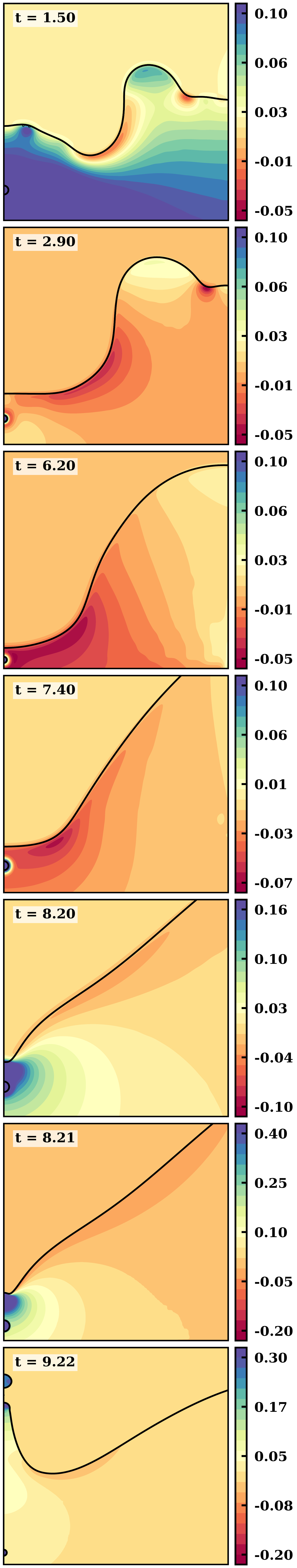}};
\node[anchor=north west,font=\bfseries,fill=white,inner sep=1pt]
at (-0.60,-0.05) {(a)};
\end{tikzpicture}

\end{minipage}
\hfill
\begin{minipage}{0.32\textwidth}
\centering
\begin{tikzpicture}
\node[anchor=north west,inner sep=0]
{\includegraphics[
    width=\linewidth,
    height=0.95\textheight,
    keepaspectratio
]{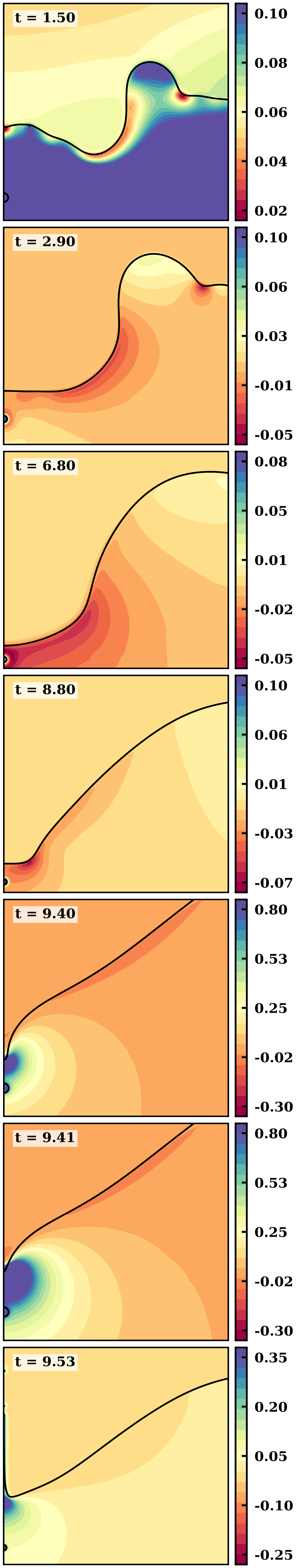}};
\node[anchor=north west,font=\bfseries,fill=white,inner sep=1pt]
at (-0.60,-0.05) {(b)};
\end{tikzpicture}

\end{minipage}
\hfill
\begin{minipage}{0.32\textwidth}
\centering
\begin{tikzpicture}
\node[anchor=north west,inner sep=0]
{\includegraphics[
    width=\linewidth,
    height=0.95\textheight,
    keepaspectratio
]{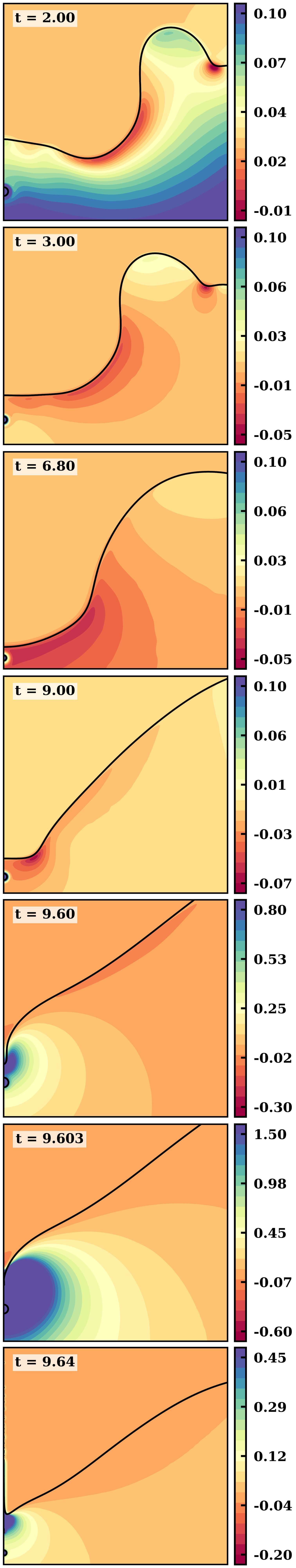}};
\node[anchor=north west,font=\bfseries,fill=white,inner sep=1pt]
at (-0.60,-0.05) {(c)};
\end{tikzpicture}

\end{minipage}

\label{fig:pressure_evolution}
\end{figure}

\clearpage
\begin{figure}

\centering

\hspace*{0.10\textwidth}

\begin{minipage}{0.32\textwidth}
\centering
\begin{tikzpicture}
\node[anchor=north west,inner sep=0]
{\includegraphics[
    width=\linewidth,
    height=0.95\textheight,
    keepaspectratio
]{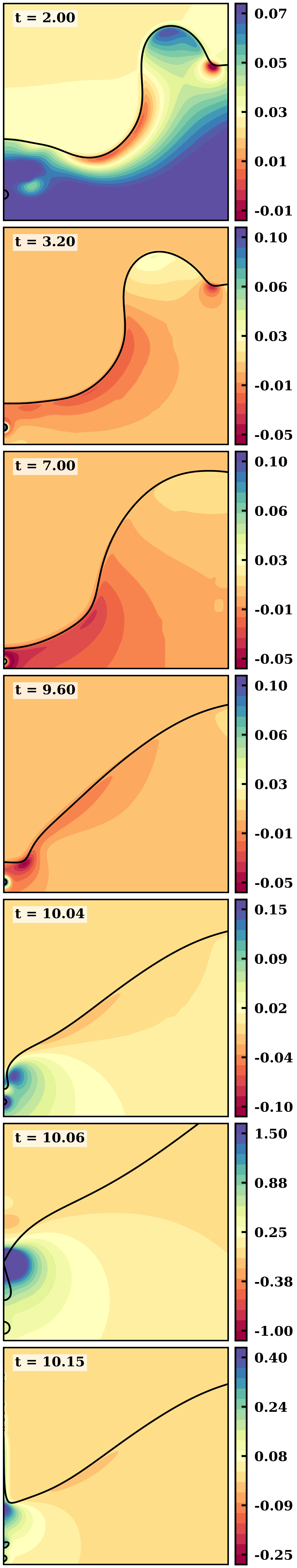}};
\node[anchor=north west,font=\bfseries,fill=white,inner sep=1pt]
at (-0.60,-0.05) {(d)};
\end{tikzpicture}
\end{minipage}
\hspace{0.04\textwidth}
\begin{minipage}{0.32\textwidth}
\centering
\begin{tikzpicture}
\node[anchor=north west,inner sep=0]
{\includegraphics[
    width=\linewidth,
    height=0.95\textheight,
    keepaspectratio
]{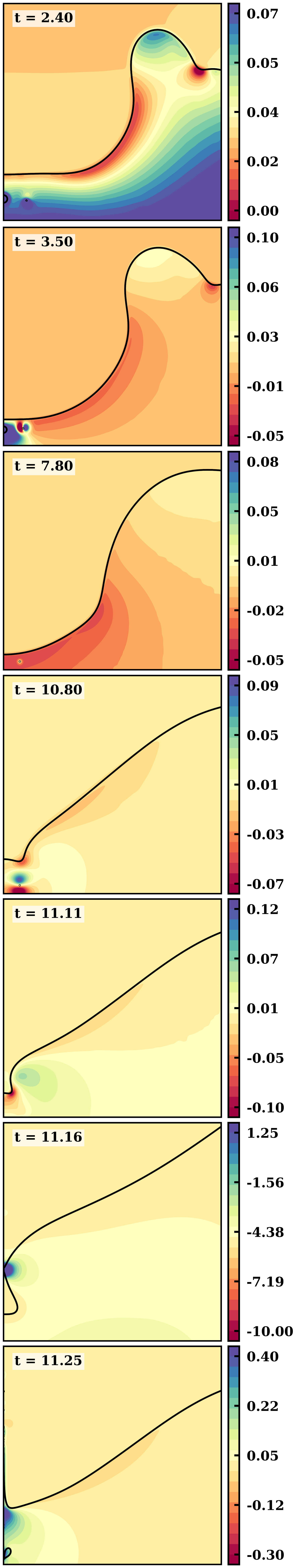}};
\node[anchor=north west,font=\bfseries,fill=white,inner sep=1pt]
at (-0.60,-0.05) {(e)};
\end{tikzpicture}
\end{minipage}

\caption{\footnotesize
Temporal evolution of the pressure field for different $We$ cases at characteristic stages of the impact dynamics: development of the swell-wave, maximum swell-wave height, maximum crater-depth, crater retraction and dimple formation, dimple collapse, onset of jet formation, and emission of the first secondary droplet. 
The cases shown are:
(a) $We=100$,
(b) $We=117$,
(c) $We=120$,
(d) $We=127$,
and (e) $We=143$.
}
\label{fig:pressure_evolution}
\end{figure}


After the singular-jet regime, pinch-off occurs in regime $\mathrm{III}$. To examine the pressure variation for this regime, we consider $P6$ at $We = 127$ and $Fr = 58.2$ and $P8$ at $We = 143$ and $Fr = 65.5$, shown in figures~\ref{fig:pressure_evolution}(d) and (e). The increased impact energy produces a wider crater, where capillary waves move radially away from the central axis, displacing the crater-wall towards the domain boundary. At the instant of maximum swell-height, for $P6$ at $3.2$ and for $P8$ at $3.5$,  the crater side wall attains an outward inclination away from the central axis. The degree of inclination increases with the $We$. In both cases, the point where the crater side wall meets the crater-base moves more rapidly outward than the swell-wave. This outward motion is more pronounced for P8 in figure \ref{fig:pressure_evolution}(e). Despite the larger cavity dimension, the capillary wave propagates faster. It reaches the crater-base when it attains maximum crater-depth at time units $7$ and $7.8$ for $P6$ and $P8$, respectively, followed by the onset of crater-base retraction and subsequent dimple formation. Consistent with \cite{morton2000investigation}, for bubble entrainment, the capillary waves cover the maximum crater-depth and reach close to the central axis $(r=0)$ during the crater retraction. The capillary wave reaches the crater-base at $9.6$ and $10.8$, followed by crater-base retraction at $10.05$ and $11.14$ for $P6$ and $P8$, respectively. Unlike the singular-jet case, where the high-pressure region develops beneath the dimple base, $P6$ and $P8$ have high-pressure regions along the crater side walls. This results in different pinch-offs. The pinch-off shapes for $We =  127$ and $143$ correspond to the teardrop- and triangular-shaped bubble entrapment. In both cases, the high-pressure region that develops between the converging dimple side walls deforms the crater-base shown in figures~\ref{fig:pressure_evolution}(d)-(e) at time units of $10.04$ and $11.11$, respectively. For $Wr=127$, the pressure field pushes the initially flat crater-base downwards. This results in a teardrop pinch-off. In contrast, for $We=143$, the W-shaped dimple profile is flattened as the pressure subdues its central peak, yielding a triangular pinch-off. Thus, the pinch-off dynamics reflect the pressure-driven deformation of the dimple.\\

Next, we plot the axial and radial velocity fields, along with streamlines, to resolve the momentum transport during crater collapse. These fields identify recirculation structures and the axial and radial momentum distribution in the emerging jet, as shown in figure~\ref {fig:vel_stream} for different $We$ and $Fr$. We show the velocity fields for maximum crater-depth, crater-base retraction, and dimple formation, followed by jet emergence from the crater. Figure~\ref{fig:vel_stream}(a) shows the axial velocity field, radial velocity field, and streamline contours of $P1$, respectively. At time $6.2$, the axial-velocity contour for maximum crater-depth exhibits localized upward motion along the cavity side walls. Radial velocity develops around swell-waves, pushing them outward. This indicates that the momentum generated during the drop-impact primarily governs cavity deformation rather than jet formation. It also indicates the onset of lateral cavity contraction. From the beginning of crater collapse at time unit $7.5$, the radial velocity that triggers inward motion in the side walls at the crater-base is weaker than the axial velocity. When the swell-wave gradually weakens, the radial velocity fields become localized around the cavity neck, while the axial velocity fields increase along the central axis at time $8.21$. This increase promotes an early upward motion of the crater-base while the side walls remain widely apart, leading to an incomplete conversion of radial momentum into axial momentum. The streamlines show two weak, symmetric recirculating vortices beneath the cavity. This indicates that circulation not only regulates the liquid motion but also concentrates momentum. This eventually leads to a relatively thicker jet with lower velocity at a time unit of $9.22$. \\

The pressure contour shown in figure~\ref{fig:pressure_evolution}(b) is owing to the wider crater at the interaction between the crater-base and side wall. The capillary wave propagates to a larger radial distance than we observed in $P1$. This indicates that, in $P1$, the momentum is primarily focused on increasing the crater-depth, whereas a shift in momentum promotes radial expansion in the subsequent cases. In $P3$, the swell-waves are weakened when dimple formation begins. A weak inward radial velocity develops around the dimple, driving its side walls toward the central axis. Meanwhile, the crater side walls continue to move radially outward, as shown in figure~\ref{fig:vel_stream}(b) at time unit $8.8$. The radial contraction remains confined to the crater-base, which eventually reduces the separation between the dimple side walls as compared to $P1$ for a time unit of $9.41$. But they are still far enough from the central axis to provide sufficient transfer of momentum toward the emerging jet. The corresponding streamlines exhibit a circulation that extends to the dimple base, and streamlines converge more sharply toward the dimple. In $P4$, at a time unit of $9.6$, the capillary waves persist for a longer duration and propagate closer to the central axis. This results in the dimple side wall undergoing a stronger radial contraction than those in $P3$, as shown in figure~\ref{fig:vel_stream}(c). The cylindrical dimple, formed prior to collapse, confines the liquid motion predominantly in the radial direction, thereby keeping the dimple side wall nearly parallel to the central axis. A synchronous retraction and uplift of the crater-base, leading to the complete conversion of radial momentum into axial momentum that emerges as the fastest jets. At the onset of jet formation, a substantial increase in axial velocity is observed, consistent with the velocity field owing to the conical-focusing mechanism reported by  \cite{tian2023conical}. Although the streamline patterns remain qualitatively similar to $P3$, the streamlines are more concentrated toward the central axis and become nearly normal to the crater surface during dimple collapse. When the jet emerges, these streamlines align with the jet axis, confirming that the flow transitions from cavity collapse to jet propagation. Figures~\ref{fig:vel_stream}(d) and (e) show the velocity variation for $P6$ and $P8$, where pinch-off takes place. In cases $P1$, $P3$, and $P4$, the dimple exhibits only radial motion.  In contrast, $P6$ and $P8$ develop a sharp axial fluid motion around the corner formed where the dimple side wall meets the crater side wall at time units $9.6$ and $10.8$, respectively. The axial motion promotes neck formation, eventually leading to pinch-off. Along the crater side walls, two large vortices produce intense streamline crowding that shifts toward the axis, causing streamlines to follow longer paths before approaching the centerlines. Later, when pinch-off occurs, the streamlines converge and become more diffuse within an entrapped bubble. This indicates that significant momentum is transferred in an entrapped bubble  rather than forming a jet.\\

\begin{figure}

\begin{tikzpicture}
\node[anchor=north west,inner sep=0] at (0,0)
{\includegraphics[width=\linewidth,height=0.42\textheight]{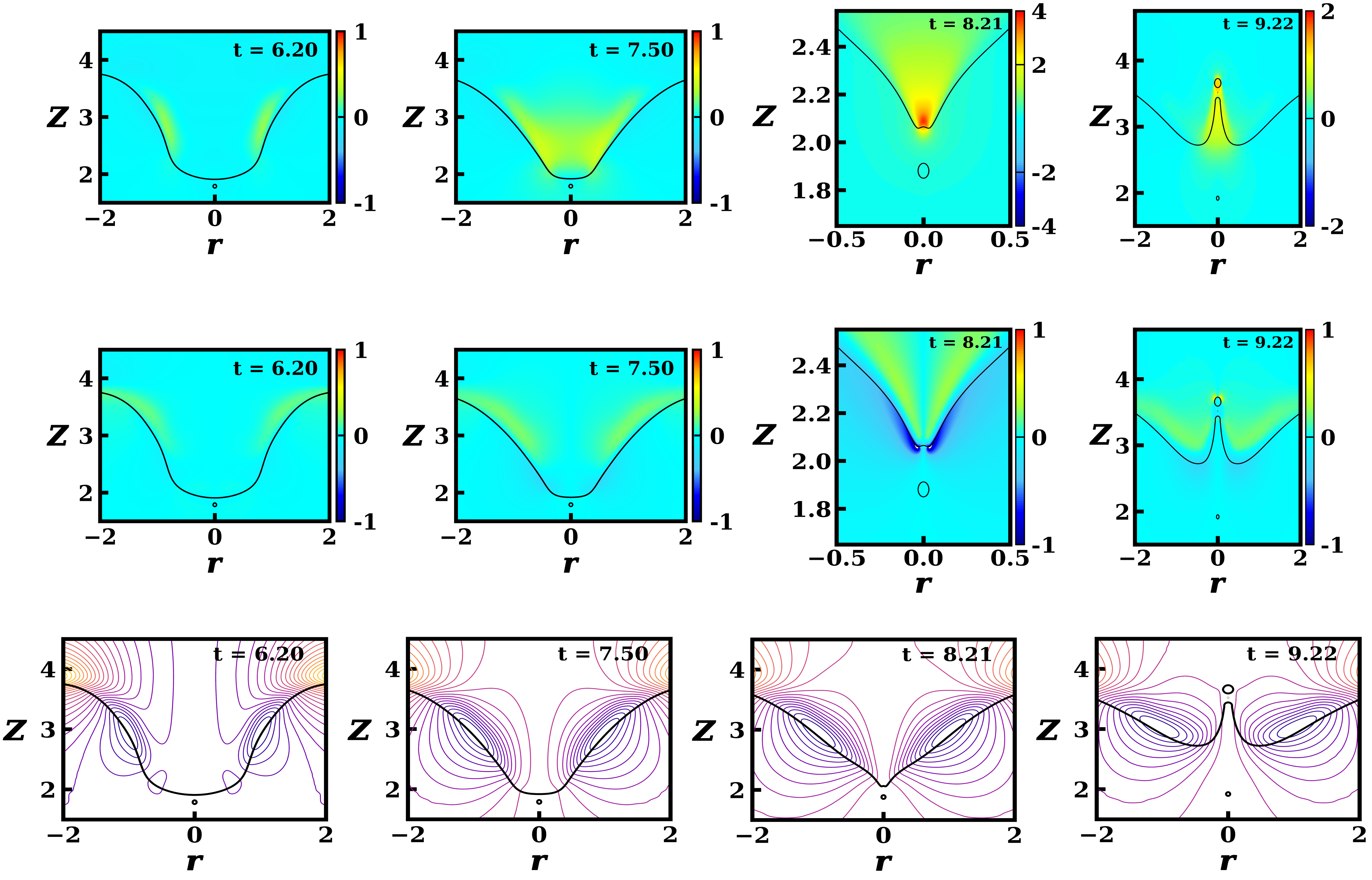}};
\node[anchor=north west,font=\bfseries,fill=white]
at (-0.2,-0.1) {(a)};
\end{tikzpicture}

\vspace{6mm}

\begin{tikzpicture}
\node[anchor=north west,inner sep=0] at (0,0)
{\includegraphics[width=\linewidth,height=0.42\textheight]{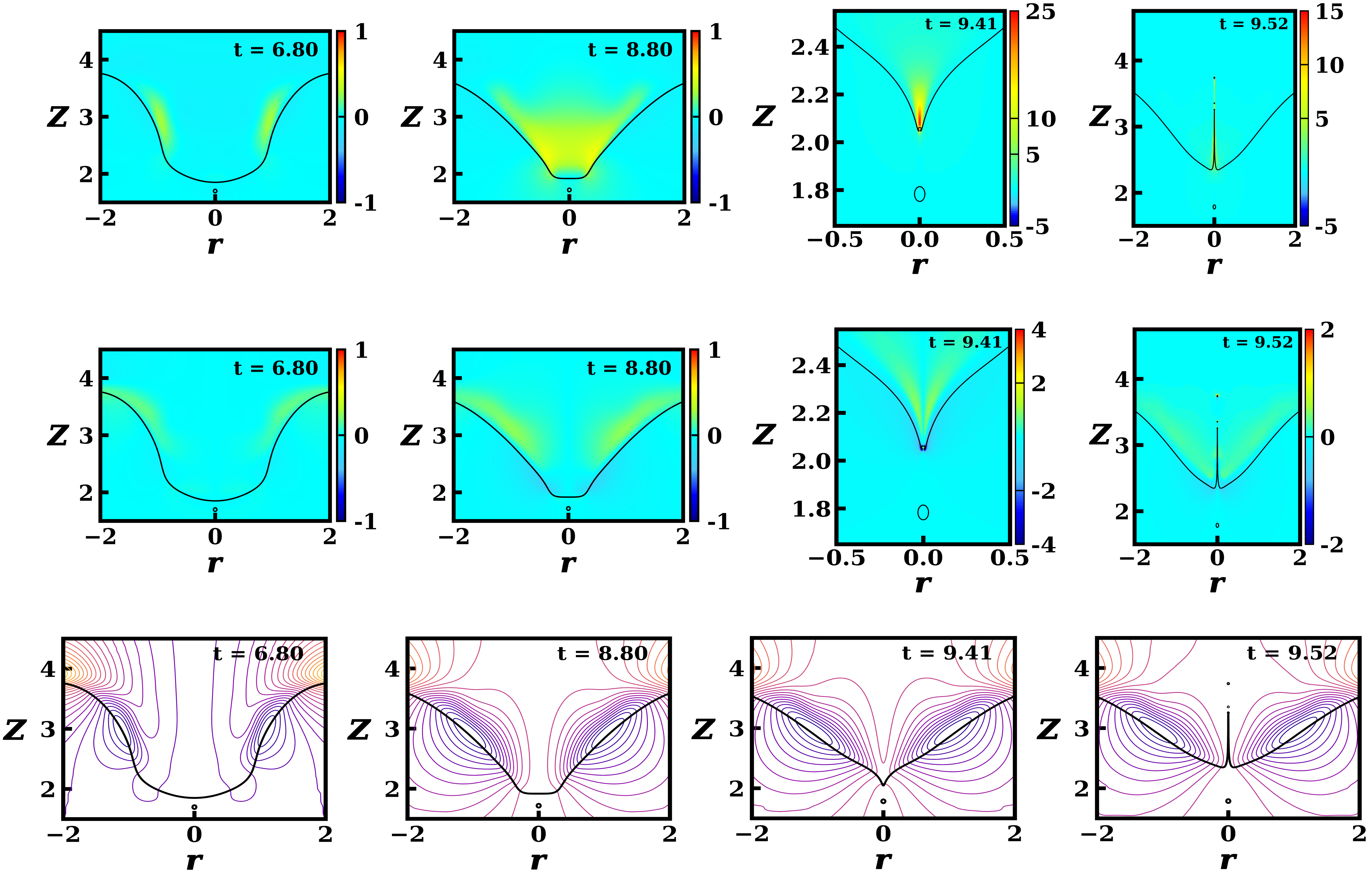}};
\node[anchor=north west,font=\bfseries,fill=white]
at (-0.2,-0.1) {(b)};
\end{tikzpicture}

\end{figure}


\begin{figure}

\begin{tikzpicture}
\node[anchor=north west,inner sep=0] at (0,0)
{\includegraphics[width=\linewidth,height=0.42\textheight]{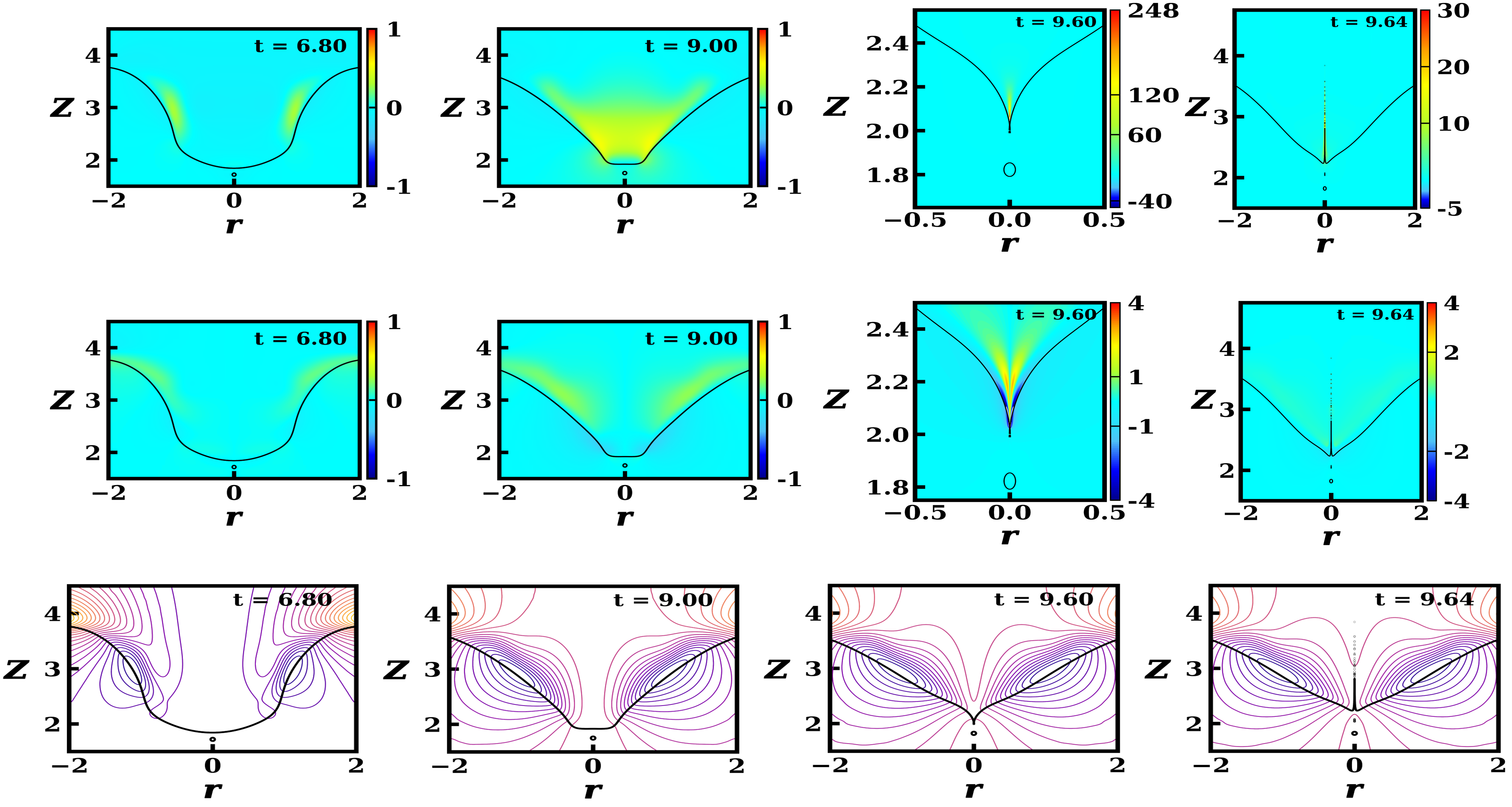}};
\node[anchor=north west,font=\bfseries,fill=white]
at (-0.2,-0.1) {(c)};
\end{tikzpicture}

\vspace{6mm}

\begin{tikzpicture}
\node[anchor=north west,inner sep=0] at (0,0)
{\includegraphics[width=\linewidth,height=0.42\textheight]{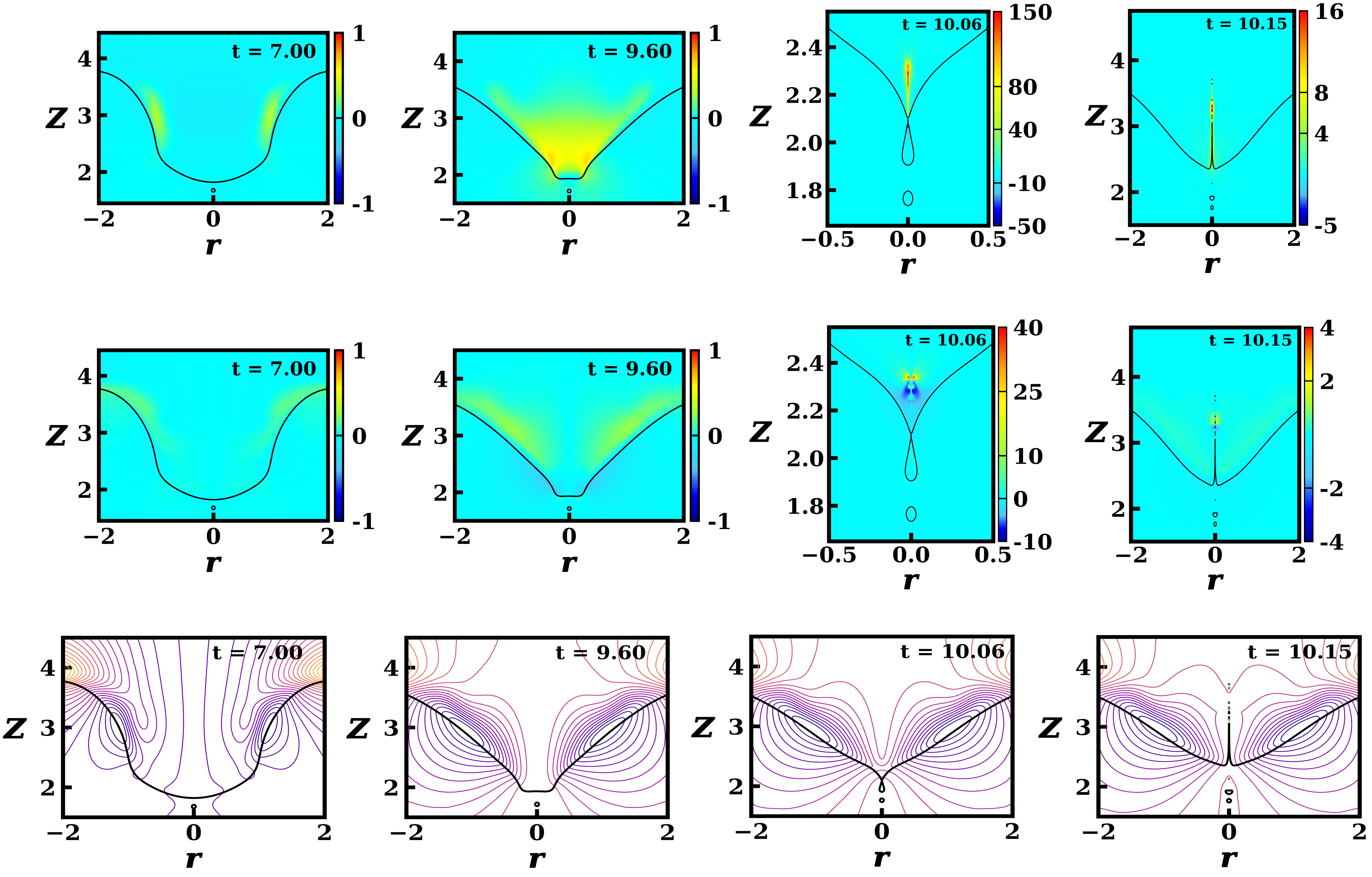}};
\node[anchor=north west,font=\bfseries,fill=white]
at (-0.2,-0.1) {(d)};
\end{tikzpicture}

\end{figure}


\begin{figure}

\begin{tikzpicture}
\node[anchor=north west,inner sep=0] at (0,0)
{\includegraphics[width=\linewidth,height=0.42\textheight]{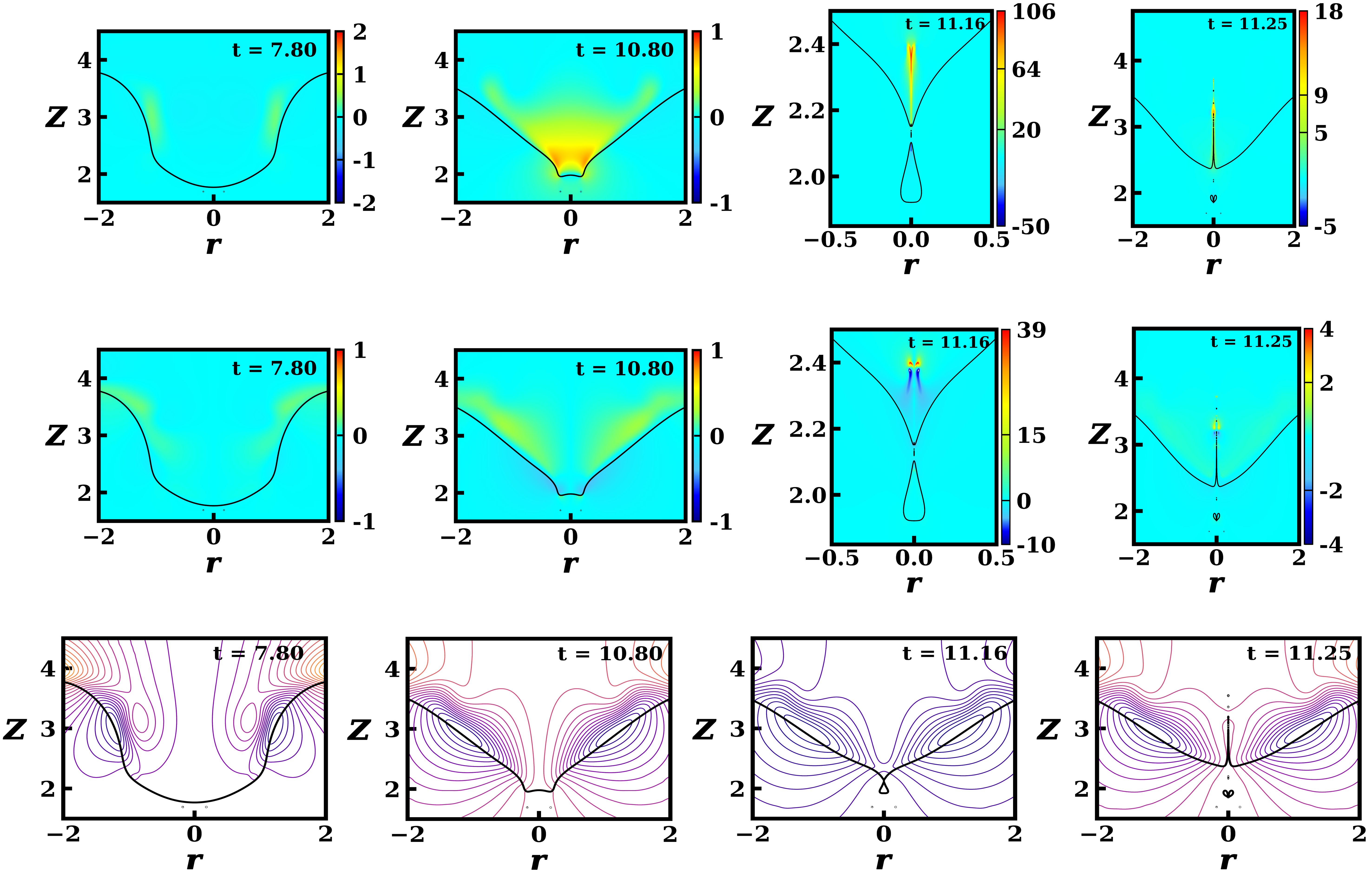}};
\node[anchor=north west,font=\bfseries,fill=white]
at (-0.2,-0.1) {(e)};
\end{tikzpicture}

\caption{\footnotesize
Axial and radial velocity fields and streamlines for representative cases across the different jetting regimes:
(a) the short, thick jet with a secondary droplet (P1);
(b) the long, thin jet with multiple secondary droplets (P3);
(c) the high-speed singular-jet with microbubble entrapment (P4);
(d) the thin jet with teardrop-shaped bubble entrapment (P6);
and (e) the thin jet with triangular bubble entrapment (P8).
}
\label{fig:vel_stream}

\end{figure}

\begin{figure}

\begin{tikzpicture}
\node[anchor=north west,inner sep=0] at (0,0)
{\includegraphics[width=\linewidth,height=0.25\textheight]{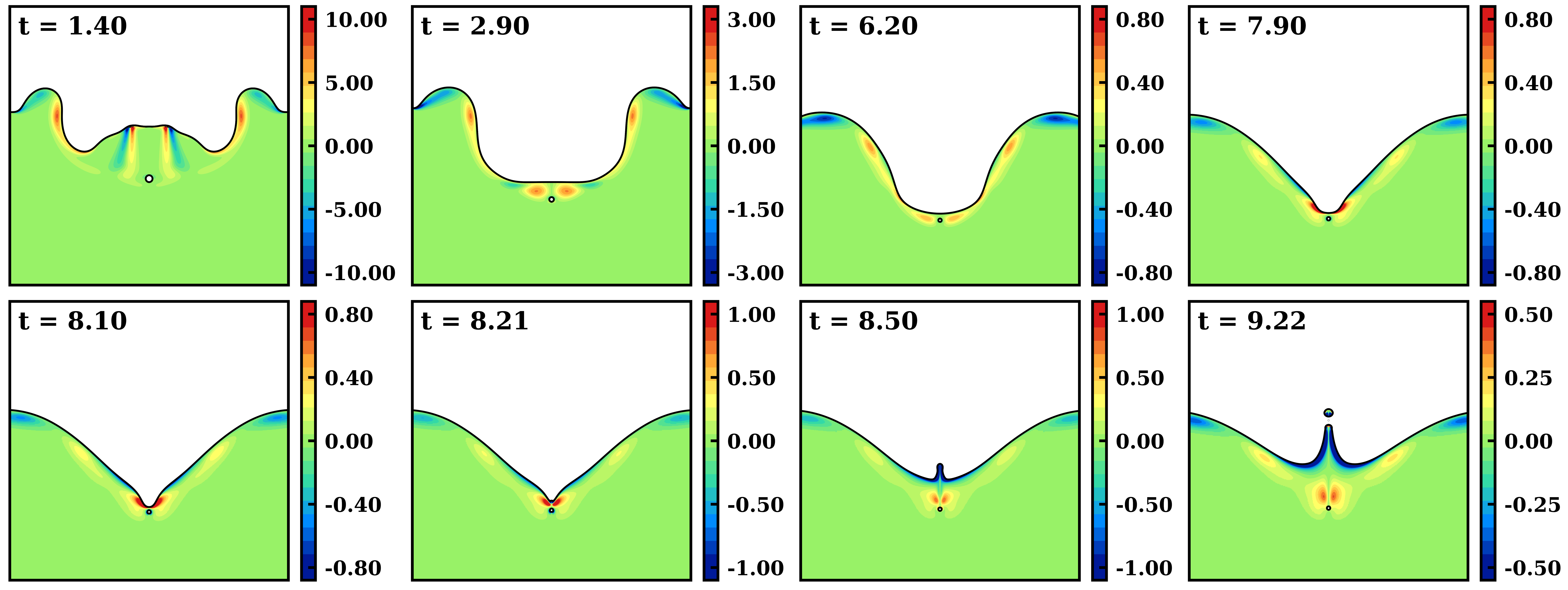}};
\node[anchor=north west,font=\bfseries,fill=white]
at (-0.6,-0.1) {(a)};
\end{tikzpicture}
\begin{tikzpicture}
\node[anchor=north west,inner sep=0] at (0,0)
{\includegraphics[width=\linewidth,height=0.25\textheight]{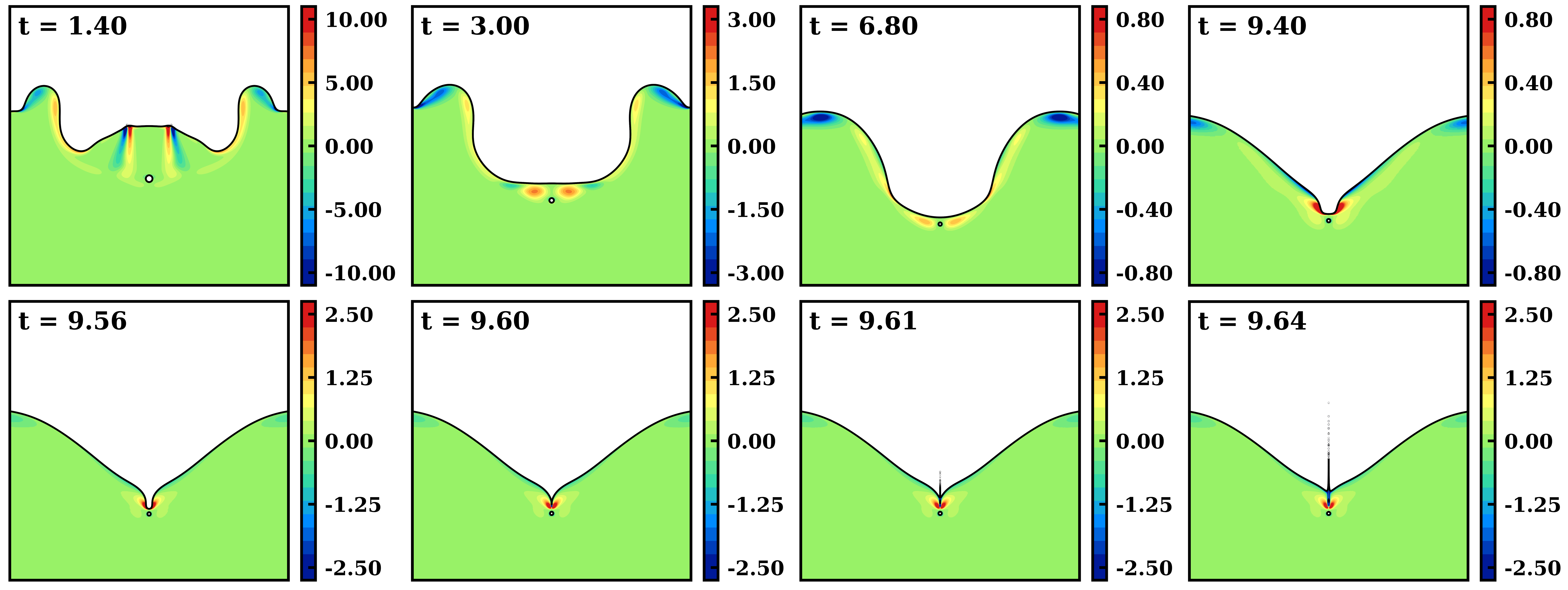}};
\node[anchor=north west,font=\bfseries,fill=white]
at (-0.6,-0.1) {(b)};
\end{tikzpicture}
\begin{tikzpicture}
\node[anchor=north west,inner sep=0] at (0,0)
{\includegraphics[width=\linewidth,height=0.25\textheight]{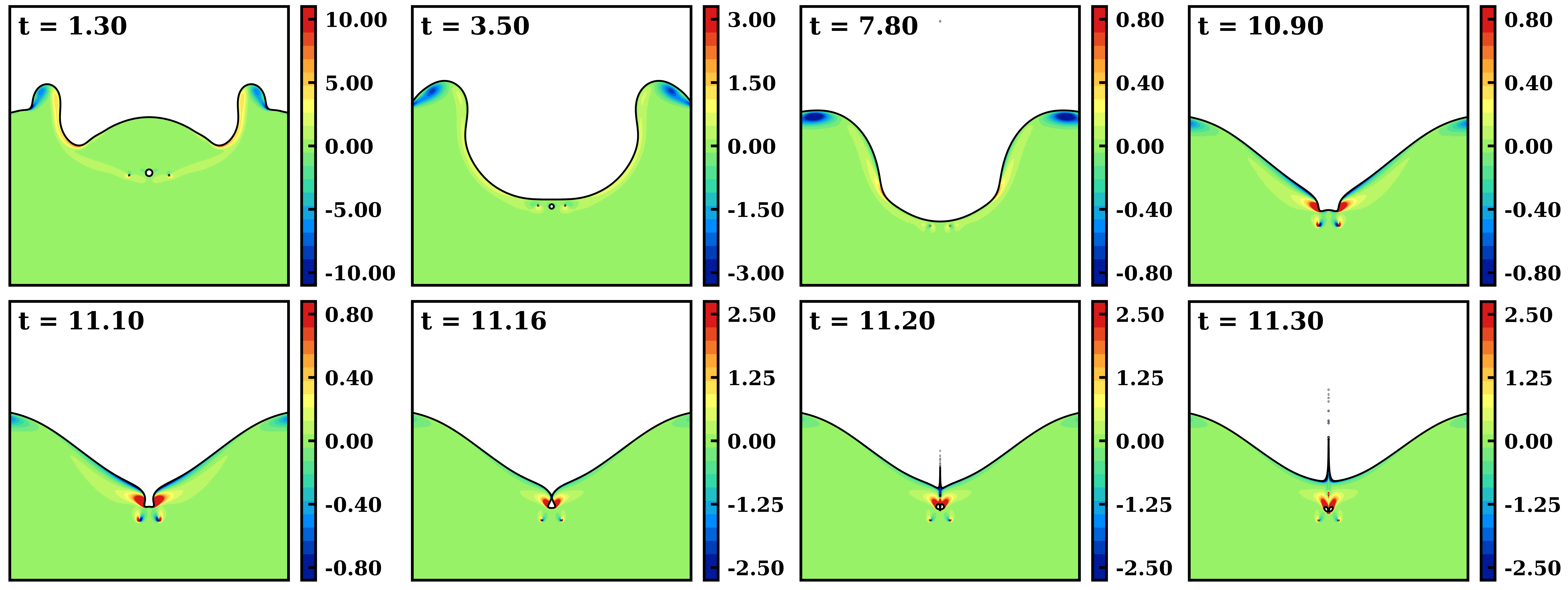}};
\node[anchor=north west,font=\bfseries,fill=white]
at (-0.6,-0.1) {(c)};
\end{tikzpicture}

\caption{\footnotesize
Vorticity contours in the liquid region for:
(a) the short, thick jet with a secondary droplet (P1);
(b) the high-speed singular-jet with microbubble entrapment (P4);
and (c) the thin jet with triangular bubble entrapment (P8).
}
\label{fig:vortticity_contor}
\end{figure}

\cite{thoroddsen2018singular} suggested that curvature change generates vorticity, which plays a crucial role in singular-jet formation. In figure~\ref{fig:vortticity_contor}, we examine vorticity distribution within the liquid region across the jetting regimes to assess its contribution to singular-jet formation. Vorticity generated upon impact propagates with the expanding crater. The positive vorticity is associated with vortex ring formation, and the negative vorticity is concentrated near the swell-waves. When the crater expands and reaches maximum depth, the vorticity becomes negligible. Subsequently, the dimple formation again develops positive vorticity associated with a local pressure drop caused by the rapid change in curvature. Following this, the dimple collapses and a jet forms. A negative vorticity develops at the jet base, whereas the positive vorticity generated during dimple formation is advected downward, as shown in figure~\ref{fig:vortticity_contor}(a) and (b). For bubble entrainment, positive vorticity is associated with the upper surface of the entrapped bubble, as shown in figure~\ref{fig:vortticity_contor}(c). For $We=100$ in figure~\ref{fig:vortticity_contor}(a), the vorticity remains broadly distributed with no pronounced concentration near the base of the emerging jet. Contrastingly, for $We=120$ in figure~\ref{fig:vortticity_contor}(b), the positive vorticity remains significantly localized around the rapidly collapsing cavity and jet-forming region. With a further increase in $We$, the vorticity is more distributed toward the entrainment bubble and is not associated with the jet base, as shown in figure~\ref{fig:vortticity_contor}(c) for $We=143$. Beyond this $We$, the vorticity fields remain qualitatively similar in different regimes. This suggests that singular-jet formation does not depend on vorticity. The present simulations span a subset of the parameter range considered by \cite{morton2000investigation,ray2015regimes}. They classified the pre-entrainment, entrainment, and post-entrainment regimes based on vortex-ring formation. The vorticity distributions in the present study are consistent with their observation, particularly in the bubble entrainment regime. Thus, the singular-jet forms because of pressure variations induced by curvature changes, rather than by a monotonic increase in vorticity as $We$ increases. \\

\subsection{Post-entrainment singular-jet formation}\label{sec:results-4}

\begin{figure}
    \centering
    \includegraphics[width=0.90\linewidth,height=0.40\textheight]{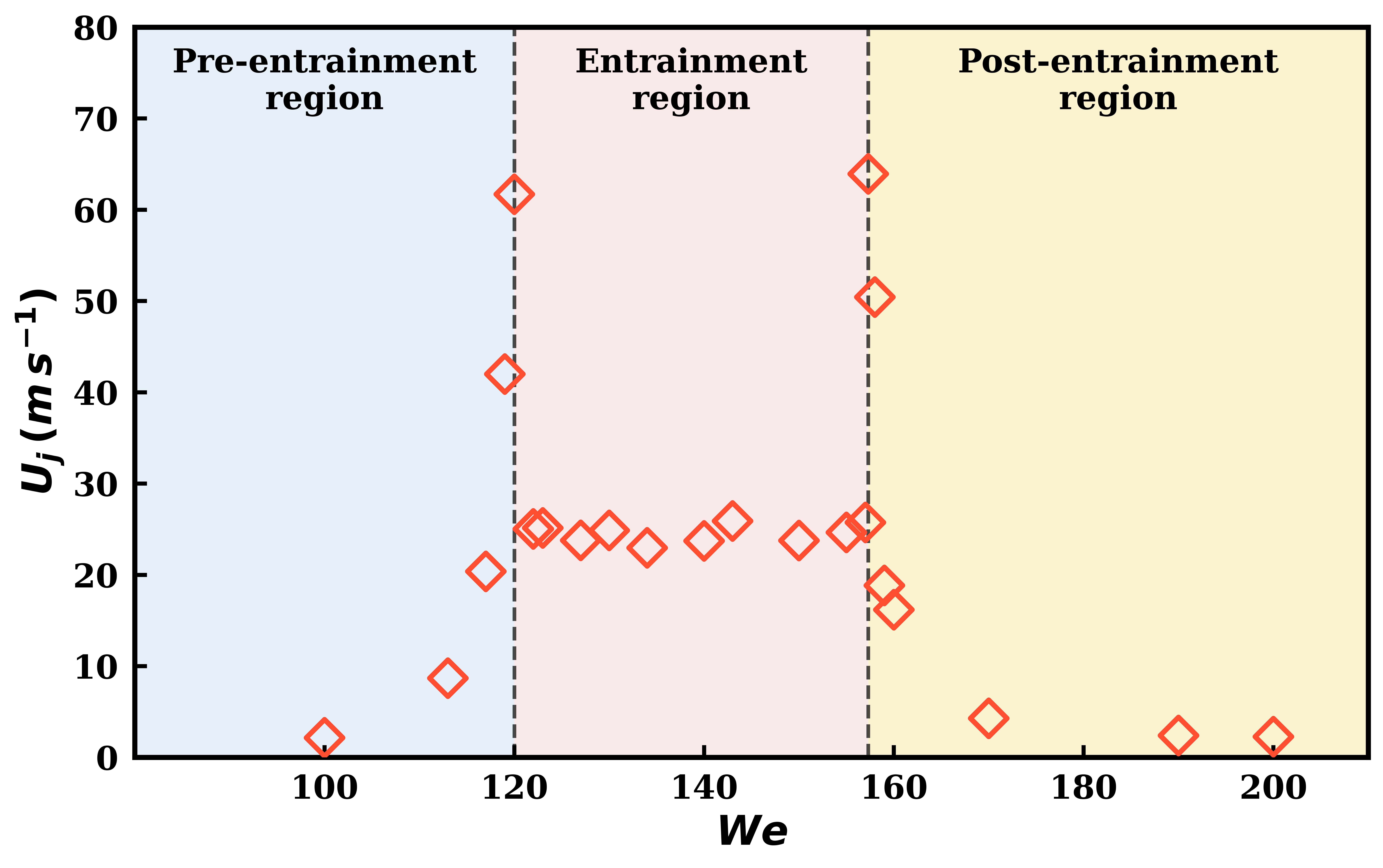}    
  \caption{\footnotesize Jetting velocity $U_j$ versus impact $We$ We, showing the distinct behavior across the pre-entrainment, entrainment, and post-entrainment region.}
  \label{fig:Jetting_velocity}
\end{figure}

The jetting behavior after the drop-impact is classified into distinct regimes, based on bubble entrainment during crater collapse. \cite{morton2000investigation} defined the entrainment regime in which bubble pinch-off occurs, bounded by the pre-entrainment and post-entrainment regions, where pinch-off is absent. \cite{leng2001splash} subsequently showed that the entrainment region is a subset of the high-speed jet region. The high-speed jets occur both before and after the entrainment region. \cite{thoroddsen2018singular} further noted that the fastest jet occurs when the pinch-off singularity is just missed. Following the work of \cite{morton2000investigation} and \cite{leng2001splash}, \cite{thoroddsen2018singular} examined the pre-entrainment and entrainment region and found the fastest singular-jet within the high-speed jet regime, near the lower boundary of the entrainment region, where pinch-off is absent. \cite{michon2017jet} reported jetting velocities over a wide range of $Fr$ for different viscosities. They observed a second peak immediately after the plateau that follows the initial peak. For $\mu = 2.5~\mathrm{mPa\,s}$, the first jetting velocity peak reaches $14U_0$, while the second peak attained about $12U_0$, where $U_0$ is the drop-impact velocity. A similar trend was observed for $\mu = 6.0~\mathrm{mPa\,s}$. We also observe that both conditions for the fastest jet are satisfied near the upper boundary of the entrainment region, where the transition from the entrainment region to the post-entrainment region occurs. However, the upper boundary of the entrainment region for singular-jet formation remains unidentified, motivating us to examine the possibility of singular-jet formation near this boundary. Therefore, we simulate the cases listed in table~\ref {tab:Impact_condition} while keeping the $Re$ of the order of $10^2$ and varying $We$ and $Fr$ to span the upper boundary of the entrainment region and extend into the post-entrainment region. We plot the jetting velocities in figure~\ref{fig:Jetting_velocity} for cases in the pre-entrainment, entrainment, and post-entrainment regions.  We observe a second spike in the jetting velocity near the upper boundary of the entrainment region. This second singular-jet has a jetting velocity of $63.93~\mathrm{m\,s^{-1}}$, which is greater than that of the first singular-jet, which is $61.7~\mathrm{m\,s^{-1}}$, obtained near the lower boundary of the entrainment region. We normalize the jetting velocity by the drop-impact velocity, following the convention used in the literature \citep{michon2017jet}, and find corresponding velocities of $43.99$ and $39.82$ near the lower and upper boundaries of the entrainment region, respectively. The velocity plot shows a trend similar to that reported by \cite{michon2017jet} for $\mu = 2.5~\mathrm{mPa\,s}$ and $\mu = 6.0~\mathrm{mPa\,s}$, but with higher jetting velocities.\\

\begin{figure}
    \centering

    \begin{subfigure}[b]{0.50\textwidth}
        \centering
        \begin{tikzpicture}
            \node[anchor=south west, inner sep=0] (img) 
            {\includegraphics[
                width=\linewidth,
                height=0.22\textheight
            ]{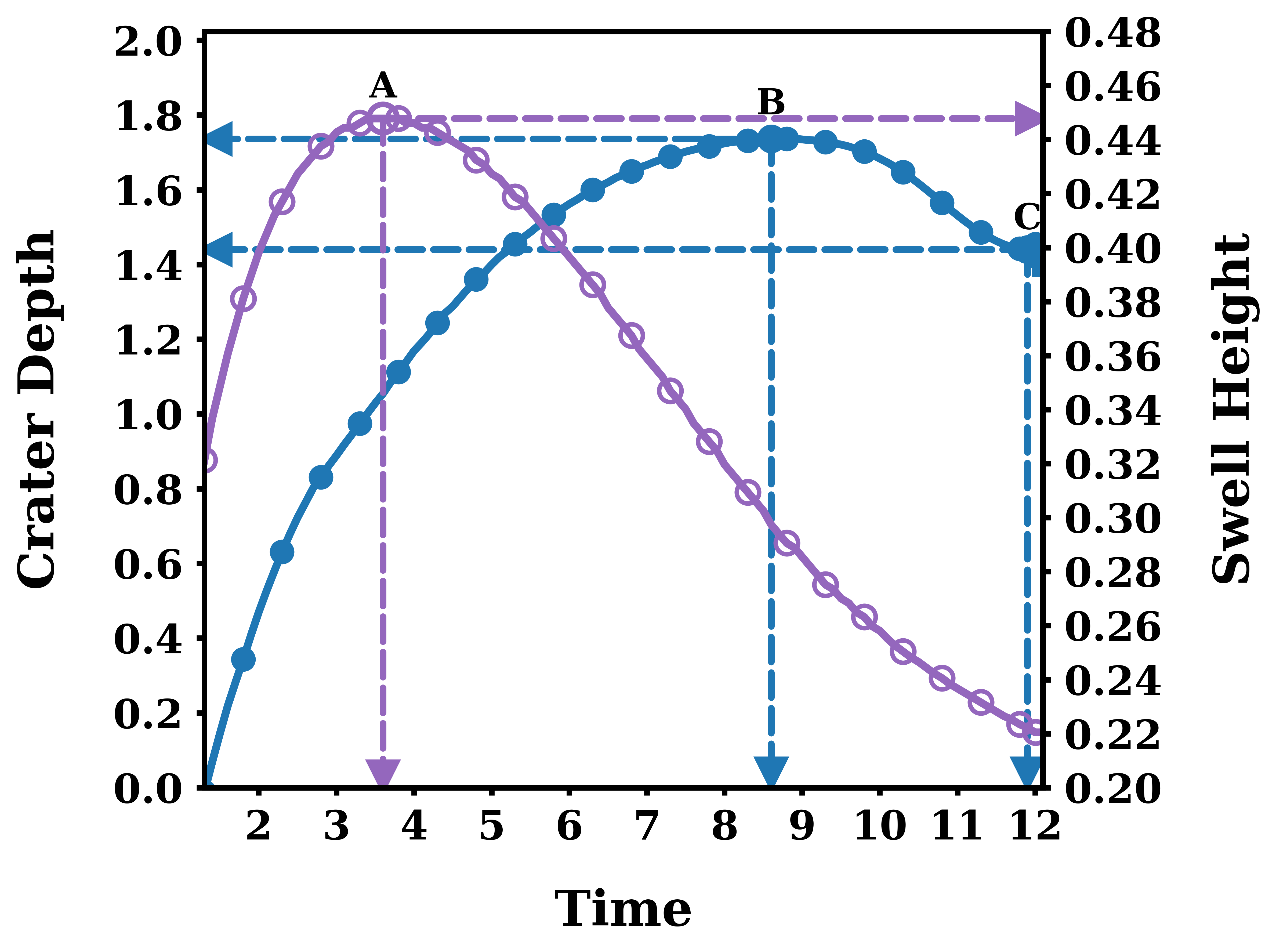}};
            \node[anchor=north west, xshift=-8pt, yshift=0pt] 
            at (img.north west) {\bfseries(a)};
        \end{tikzpicture}
        \label{fig:swell_crater}
    \end{subfigure}

    \vspace{0.2em}

    \begin{subfigure}[b]{\textwidth}
        \centering
        \begin{tikzpicture}
            \node[anchor=south west, inner sep=0] (img) 
            {\includegraphics[
                width=\linewidth,
                height=0.10\textheight
            ]{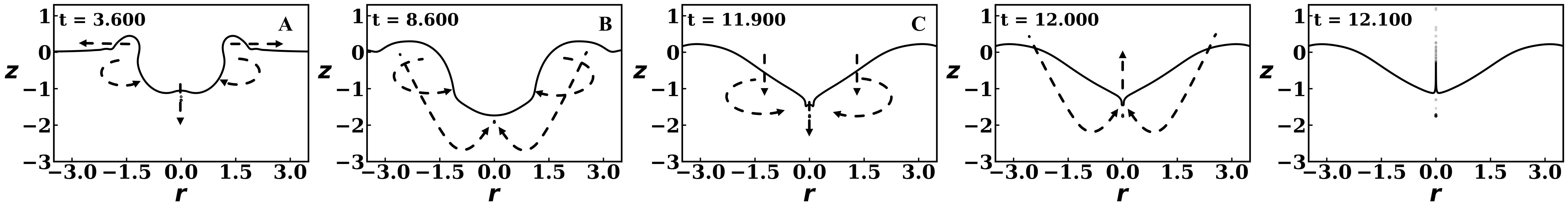}};
            \node[anchor=north west, xshift=-12pt, yshift=0pt] 
            at (img.north west) {\bfseries(b)};
        \end{tikzpicture}
        \label{fig:crater}
    \end{subfigure}

    \vspace{0.3em}

    \begin{subfigure}[b]{\textwidth}
        \centering
        \begin{tikzpicture}
            \node[anchor=south west, inner sep=6] (img) 
            {\includegraphics[
                width=\linewidth,
                height=0.115\textheight,
                keepaspectratio
            ]{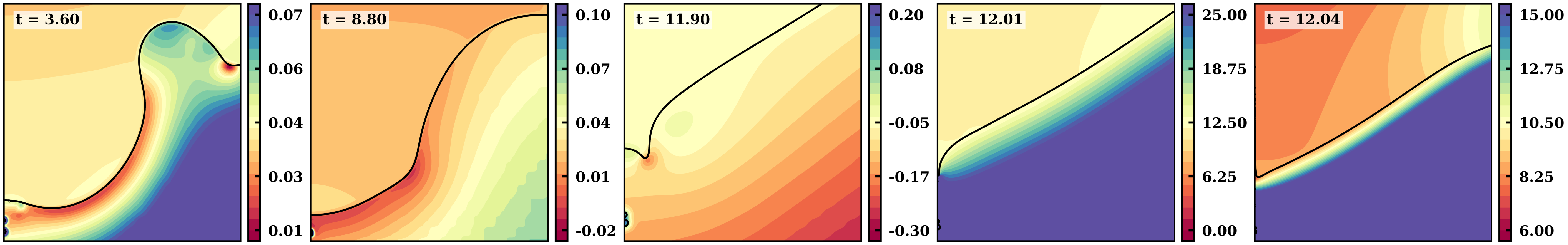}};
            \node[anchor=north west, xshift=-12pt, yshift=-3pt] 
            at (img.north west) {\bfseries (c)};
        \end{tikzpicture}
        \label{fig:pressure}
    \end{subfigure}

    \vspace{0.2em}

    \begin{subfigure}[b]{0.75\textwidth}
        \centering
        \begin{tikzpicture}
            \node[anchor=south west, inner sep=0] (img) 
            {\includegraphics[
                width=\linewidth,
                height=0.30\textheight
            ]{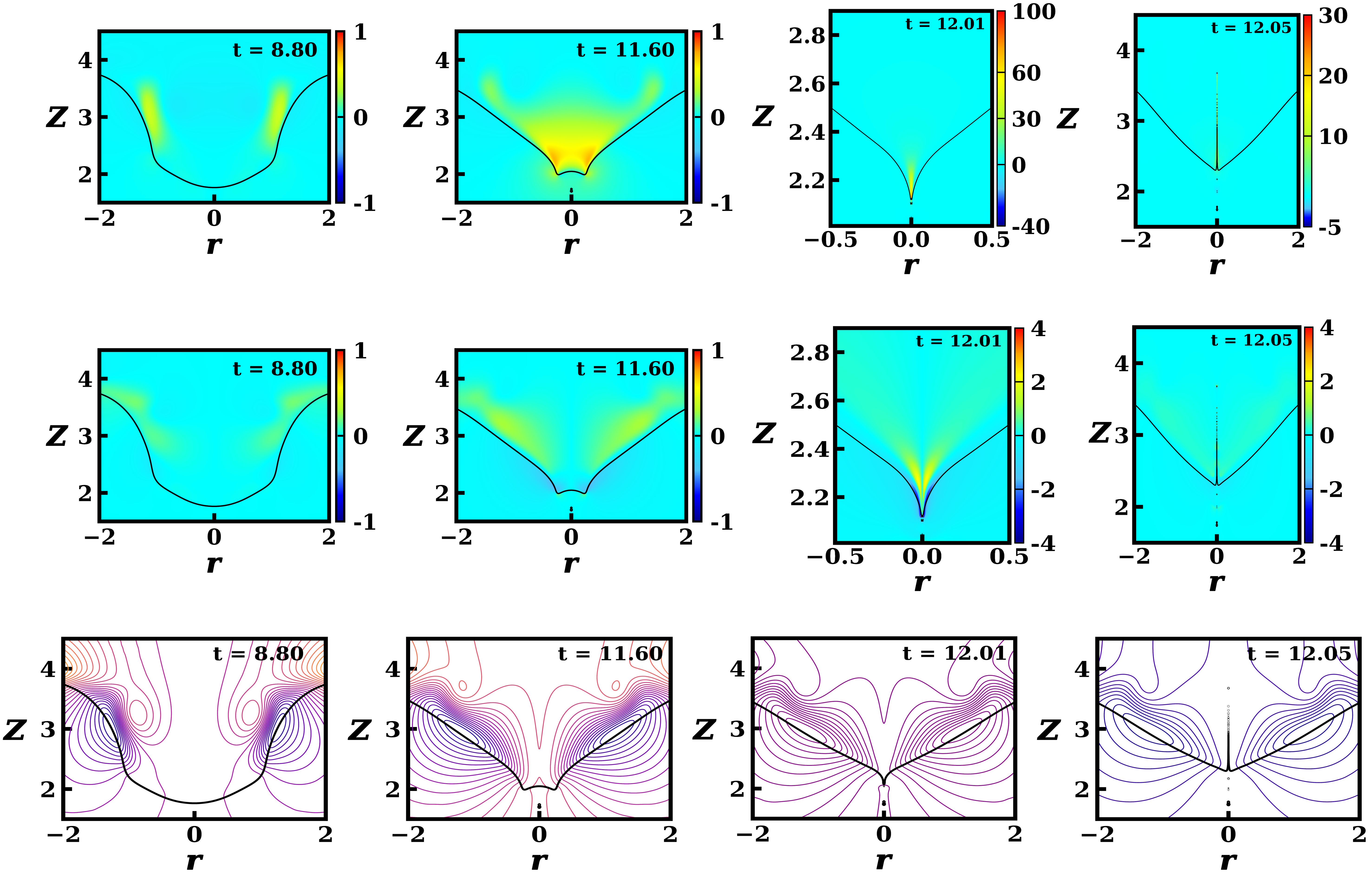}};
            \node[anchor=north west, xshift=-12pt, yshift=-3pt] 
            at (img.north west) {\bfseries (d)};
        \end{tikzpicture}
        \label{fig:velocity}
    \end{subfigure}

    \vspace{0.2em}

    \begin{subfigure}[b]{\textwidth}
        \centering
        \begin{tikzpicture}
            \node[anchor=south west, inner sep=6] (img) 
            {\includegraphics[
                width=\linewidth,
                height=0.115\textheight,
                keepaspectratio
            ]{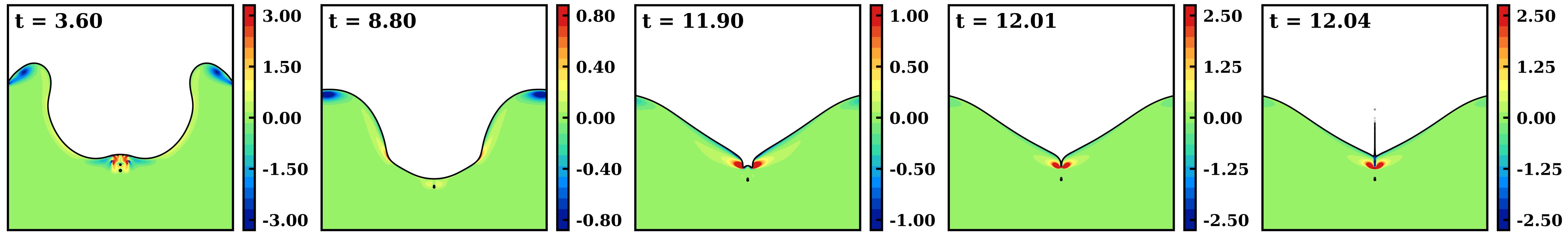}};
            \node[anchor=north west, xshift=-12pt, yshift=-3pt] 
            at (img.north west) {\bfseries (e)};
        \end{tikzpicture}
        \label{fig:vorticity}
    \end{subfigure}

    \caption{
    Evolution of the swell-wave, crater, and associated flow-field quantities during crater collapse.
    (a) Swell-wave and crater depth variation.
    (b) Crater evolution.
    (c) Interface pressure evolution.
    (d) Velocity-field evolution.
    (e) Vorticity-field evolution.
    }
    \label{fig:We_157.3_flowfield}

\end{figure}

To identify the interfacial characteristics associated with higher jetting velocity in the upper boundary of the entrainment regime, we examine the interface morphology and flow fields, as shown in figure~\ref{fig:We_157.3_flowfield}. For $We = 157.3$, the maximum swell-wave (A) amplitude reaches $0.447D$ at a time unit of $3.6$, while the crater reaches a maximum depth (B) of $1.736D$ at $8.8$, as shown in figure~\ref{fig:We_157.3_flowfield}(a). Both the swell amplitude and the crater-depth are greater than those observed in cases $(P1\text{--}P8)$. The crater-base for $We = 157.3$ uplifts by $0.3D$ from its maximum depth before re-expansion, whereas the value observed for $We = 143$ is $0.21D$. The uplift remains confined to the crater-base, while the side walls move radially inward, giving the crater a W-shaped profile as shown in figure~\ref{fig:We_157.3_flowfield}(b). The pronounced central peak limits the inward displacement of the side walls, resulting in a large separation between them at a time unit of $11.9$. This prevents the side walls from collapsing above the crater-base, unlike the $We=143$. During re-expansion, the continuous inward motion of the crater side walls pushes the crater-base downward and flattens the uplifted crater-base. As the crater retracts, it forms a cylindrical dimple with side walls nearly parallel to the central axis, similar to that observed for $We = 120$. The subsequent collapse of the cylindrical dimple misses the pinch-off singularity and produces a thin, fast-moving singular-jet. See the supplementary movie S9 for reference. In flow field analysis, we first examine the pressure fields shown in figure~\ref{fig:We_157.3_flowfield}(c), qualitatively similar to those observed for $We = 143$ at time units of $3.6$ and $8.8$, corresponding to the maximum swell-waves and maximum crater-depth, respectively. The pressure distribution changes at $11.9$, when a high-pressure zone forms at the crater-base. A low-pressure region develops at the corner where the crater side walls meet the crater-base. However, the high-pressure region at the crater-base, as observed in $We = 157.3$, is absent in $We = 143$ at the onset of re-expansion. The pressure gradient at the crater-base suppresses vertical motion, as seen at $11.9$ for $We = 157.3$. This limits the side walls’ inward motion, thereby maintaining a finite separation between them and preventing pinch-off. Thereafter, as the dimple retracts and the jet forms, a larger magnitude high-pressure zone develops throughout the region beneath the interface and persists as the jet grows. This contrasts with $We = 120$, where the high-pressure region is confined to the vicinity of the dimple, and with $We =  127$ and $143$, where it develops at the pinch-off point. \\

In figure~\ref{fig:We_157.3_flowfield}(d), we plot the axial and radial velocities along with the streamlines at different instants. In this case, we observe a prominent outward radial velocity developing on the crater side walls at time unit $8.80$, together with axial velocity. The resulting outward and upward motion redistributes the fluid along the crater side walls. This makes the crater-base flatter earlier than in the previous cases $(P1\text{--}P8)$. The dimple side walls move radially inward, whereas the crater-base corner moves radially outward. However, for $We = 143$,  the inward radial motion of the crater side walls dominates, leading to pinch-off. Again, we observe a synchronous dimple collapse similar to that for $We = 120$, during which radial momentum is completely transferred to axial momentum, forming a high-speed jet. At time $12.01$ units, the dimple side walls remain farther apart than in the $We = 120$ case; however, they are sufficiently close to omit the pinch-off singularity and form a high-speed jet. The streamline contours corroborate that the flow is primarily radial. The streamlines descend along the crater-walls and turn toward the crater-base at time units $8.8$ and $11.6$, which are associated with the crater collapse. Immediately before jet formation, at time unit $12.01$, the streamlines concentrate near the dimple tip, similar to the case of $We = 120$. The rapid change in the streamline patterns during jet growth indicates flow redistribution after singular-jet formation. The vorticity distribution in figure~\ref{fig:We_157.3_flowfield}(e) follows a trend similar to that observed for $We = 120$ and $143$. As the crater grows, vorticity vanishes, whereas the positive vorticity at the crater-base is due to a microbubble entrapped during impact. As the dimple forms, vorticity regenerates due to a rapid change in the interface curvature and persists near the base of the emerging jet. At the onset of the post-entrainment region, the vorticity plots do not resemble those reported in \cite{morton2000investigation,ray2015regimes}. In their vorticity plots, a distinct vortex ring forms at the crater-base as it begins to retract, whereas we do not observe such a vortex ring in this case. Thus, the singular-jet near the upper boundary entrainment region shows some partial characteristics of the post-entrainment regime, such as the development of a high-pressure region at the crater-base, but does not exhibit the vorticity variation.  \\

\FloatBarrier

\begin{figure}
    \centering
    \includegraphics[width=\linewidth,height=0.50\textheight]{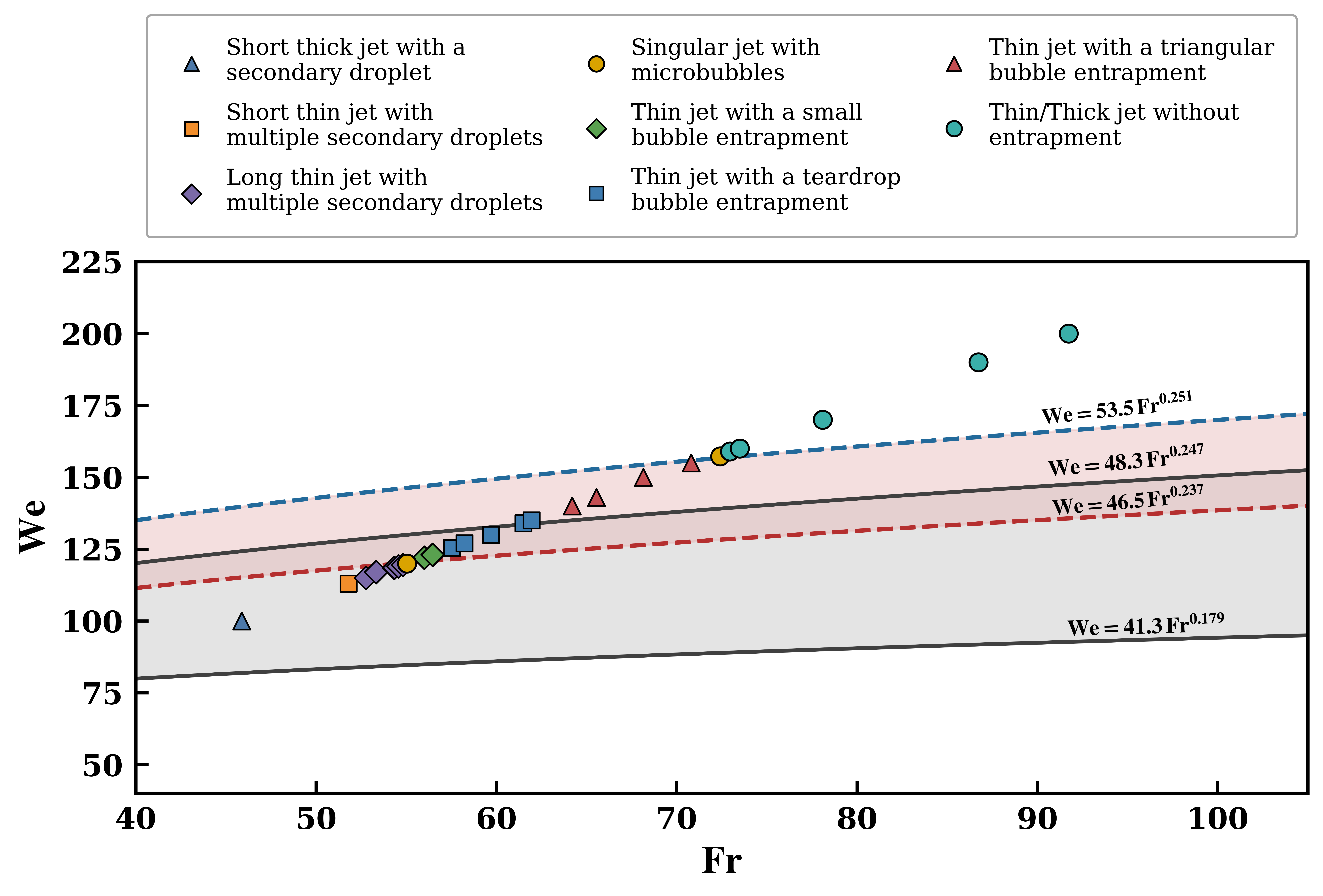}    
    \caption{\footnotesize
   A $We–Fr$ regime plot at $Re = O(10^2)$. The entrainment region reported by \cite{pumphrey1990entrainment} and \cite{oguz1990bubble} 
   is bounded by the black solid curves and shown by the grey-shaded region. The pink-shaded region delineates the best-fit entrainment regime for singular-jet formation obtained from the present numerical simulations, bounded by $We=46Fr^{0.237}$ and $We=53.5Fr^{0.251}$ respectively.}
    \label{fig:Regime_plot}
\end{figure}

We plot a regime map in figure~\ref{fig:Regime_plot} for different values of $We$ and $Fr$. Two solid black curves bound the entrainment region, $We = 48.3\,Fr^{0.247}$ and $We = 41.3\,Fr^{0.179}$, corresponding to the lower and upper boundaries of the entrainment region, proposed by \cite{pumphrey1990entrainment, oguz1990bubble}. Our simulation results do not fully conform to these boundaries, as a few of our cases show no entrainment despite lying within the reported entrainment region. \cite{thoroddsen2018singular} observed a similar characteristic in their experimental dataset, where their data overlap with the upper $We$  range, and a case without bubble entrapment lies within the reported entrainment region. This concludes that the singular-jet phenomenon is highly sensitive to impact conditions and fluid properties. In our simulations, we use highly viscous fluids rather than those used in \cite{pumphrey1990entrainment, oguz1990bubble}. Therefore, we redefine the entrainment-region boundaries based on our simulation data using best-fit curves. For the lower boundary of the entrainment region, we take the first singular-jet case as the reference point, since bubble entrapment begins beyond this case, as indicated by a solid yellow circle in figure~\ref{fig:Regime_plot}. This boundary itself is indicated by a red dashed curve $We = 46\,Fr^{0.237}$. The upper boundary of the entrainment region, represented by the blue dashed curve $We = 53.5\,Fr^{0.251}$, passes exactly through the second singular-jet case. After that, we do not observe pinch-off. The new entrainment region for singular-jet cases is marked in pink on the $We-Fr$ regime plot. \\

\section{Conclusion}\label{sec:conclusion}

We perform DNS to investigate the formation of high-speed singular-jets during the collapse of a drop-impact crater using the open-source volume-of-fluid-based solver Basilisk. \cite{morton2000investigation,ray2015regimes} primarily examined crater dynamics that lead to bubble entrainment, whereas \cite{leng2001splash,michon2017jet} examined jetting behavior under different impact conditions and fluid viscosities. Despite this, the exceptionally high jetting velocity was not observed in their observations. \cite{thoroddsen2018singular,tian2023conical} provide insights into the highest jetting velocities for singular-jets that avoid pinch-off singularity. However, their analyses were limited to jetting velocity. The flow fields associated with jet formation and their impact on jetting velocity have not been examined in detail in previous literature. The present study builds on the numerical investigation of \cite{tian2023conical} and uses the same computational setup to analyze the flow fields. Our numerical results show good qualitative and quantitative agreement with \cite{tian2023conical}. This provides a basis for discussing the flow fields and identifying the underlying mechanism across different regimes over the range $We=100\text{--}200$ and $Fr=45\text{--}92$. We classify the phenomenon into three regimes based on its distinct jetting characteristics. We define these as regime $\mathrm{I}$, the pre-singular-jet regime without bubble entrapment; regime $\mathrm{II}$, the singular-jet regime with microbubbles at the crater-base; and regime $\mathrm{III}$, the post-singular-jet regime with bubble entrapment. \\

The e-ratio ($\textit{e}$) quantifies crater shape across different stages. It begins with the expansion stage, where $e > 1$ indicates an oblate shape until the swell-wave reaches its maximum height. Following this, the e-ratio approaches unity, indicating a nearly hemispherical shape at maximum crater-depth. Thereafter, the e-ratio falls below unity, signifying a prolate crater shape during the final stage of crater collapse. In the singular-jet case, the dimple is cylindrical, with side walls approximately parallel to the central axis during the final collapse. The critical swell angle for the singular-jet case is $109.2^\circ$, consistent with the value reported by \cite{leng2001splash,ray2015regimes}. A swell angle below its critical value produces widely separated dimple side walls, whereas a larger angle promotes pinch-off. Different dimple retractions lead to a transition in jet morphology: a short, thick jet, followed by a high-speed singular-jet, then a long, thin jet with bubble entrapment, as shown in P1--P8. We characterize the different jetting behaviors by their height and thickness, measured relative to the undisturbed pool surface at an instant when the last secondary drop emerges from the jet tip. We observe a trend that both the jet height and thickness increase with increasing $We$. For $We = 100$, the jet remains below the undisturbed pool surface, yielding a negative jet height. Therefore, its thickness is not defined. The last parameter is the diameter of the first secondary drop emanating from the jet tip. We find the largest secondary drop diameters for $We=100$, and the smallest is for the singular-jet case at $We=120$, which produces a maximum count of $20$ secondary droplets as the jet grows.  \\

To explain the singular-jet phenomenon, we analyze the pressure, velocity, and vorticity fields obtained from our simulations. The motion of capillary waves, tracked using the pressure iso-contour, accounts for the geometric changes that arise in crater dynamics across regimes. The rapid change in curvature during dimple collapse generates a high-pressure region, reflecting the pressure-driven deformation of the dimple base. The high-pressure at the base promotes air-sheet entrapment in the singular-jet case. After that, pressure shifts from the base towards the converging side walls, causing pinch-off in regime $\mathrm{III}$. The coupled evolution of interface curvature and pressure accounts for the different pinch-off shapes during dimple collapse. In regime $\mathrm{I}$, axial velocity during dimple collapse dominates, lifting the crater-base before the dimple side walls contract substantially in the radial direction. A synchronous dimple retraction is observed for the singular-jet case in regime $\mathrm{II}$. For the fastest jet, coherent inward motion that drives radial momentum toward the axis results in a strong axial outflow. In regime $\mathrm{III}$, near the junction where the dimple wall and crater side wall meet, both the radial and axial velocities develop, leading to a pinch-off. A substantial amount of momentum transfers to the pinch-off bubble, significantly reducing the jet velocity. Even though the rapid change in curvature generates local vorticity during crater collapse, we find no discernible variation across the regime. \\


We also explored the post-entrainment region and found a singular-jet with a velocity of $63.93~\mathrm{m\,s^{-1}}$ near the upper boundary of the entrainment region. The jet in the post-entrainment region is within a high-speed jetting regime reported by \cite{leng2001splash}. It misses the pinch-off singularity, thereby satisfying both conditions required for singular-jet formation. The initial crater-base uplift resembles post-entrainment characteristics. This prevents crater-side wall collapse and the pinch-off singularity. However, continued inward motion of the dimple side walls builds pressure between them, which suppresses the initial uplift. Similar to the singular-jet in regime $\mathrm{I}$, the dimple in this post-entrainment region case also develops a nearly cylindrical shape, leading to a high-speed jet. Finally, we present a regime map that redefines the conventional entrainment region for the singular-jet cases. The updated lower boundary of the entrainment region is given by $We = 45.5\,Fr^{0.237}$, corresponding to the first singular-jet. The updated upper boundary of the entrainment region is $We = 53.5\,Fr^{0.251}$, at which the second singular-jet is obtained. \\


Our simulation provides a flow-field-based interpretation of singular-jet formation and distinguishes it from the neighboring regimes both qualitatively and quantitatively. The natural extension of the present work is to unravel the unexplored dynamics of singular-jets through three-dimensional simulations. The present simulations assume axisymmetry, whereas the actual phenomenon of crater collapse, jet formation, secondary drop formation, and bubble entrapment are not necessarily axisymmetric. The air sheet entrapped during crater collapse and singular-jet formation excludes azimuthal perturbations. \cite{tian2023conical} raised the question of whether the stability and rupturing of the entrapped air film may influence the subsequent jet dynamics. The axisymmetric formulation suppresses azimuthal deformation of the entrapped air film and precludes rupture either within the film or from its edge. \cite{jian2020air} investigate the contraction of a two-dimensional air sheet in liquid and identify distinct contraction dynamics for different Ohnesorge numbers. They highlight the need for three-dimensional simulations to comprehensively understand air entrapment.  To capture the rim deformation induced by rim instability and vortex shedding, which can eventually lead to bubble shedding, as experimentally observed by \cite{thoroddsen2003air,thoroddsen2012micro}. This highlights a limitation of the present axisymmetric simulations in which azimuthal crater collapse and rupture of the entrapped air film are suppressed. Three-dimensional simulations can address these limitations, but a highly refined mesh is required to resolve the entrapped thin air film and capture the rapidly evolving jets. At such high resolution, the computational cost and simulation time become substantial, requiring high-end computational resources.\\

\appendix

\section{Effect of spatial resolution on jetting velocity}\label{appA}

\begin{figure}
\centering

\begin{minipage}{0.32\textwidth}
\centering
\begin{tikzpicture}
\node[anchor=north west,inner sep=0] (img)
  {\includegraphics[width=\linewidth]{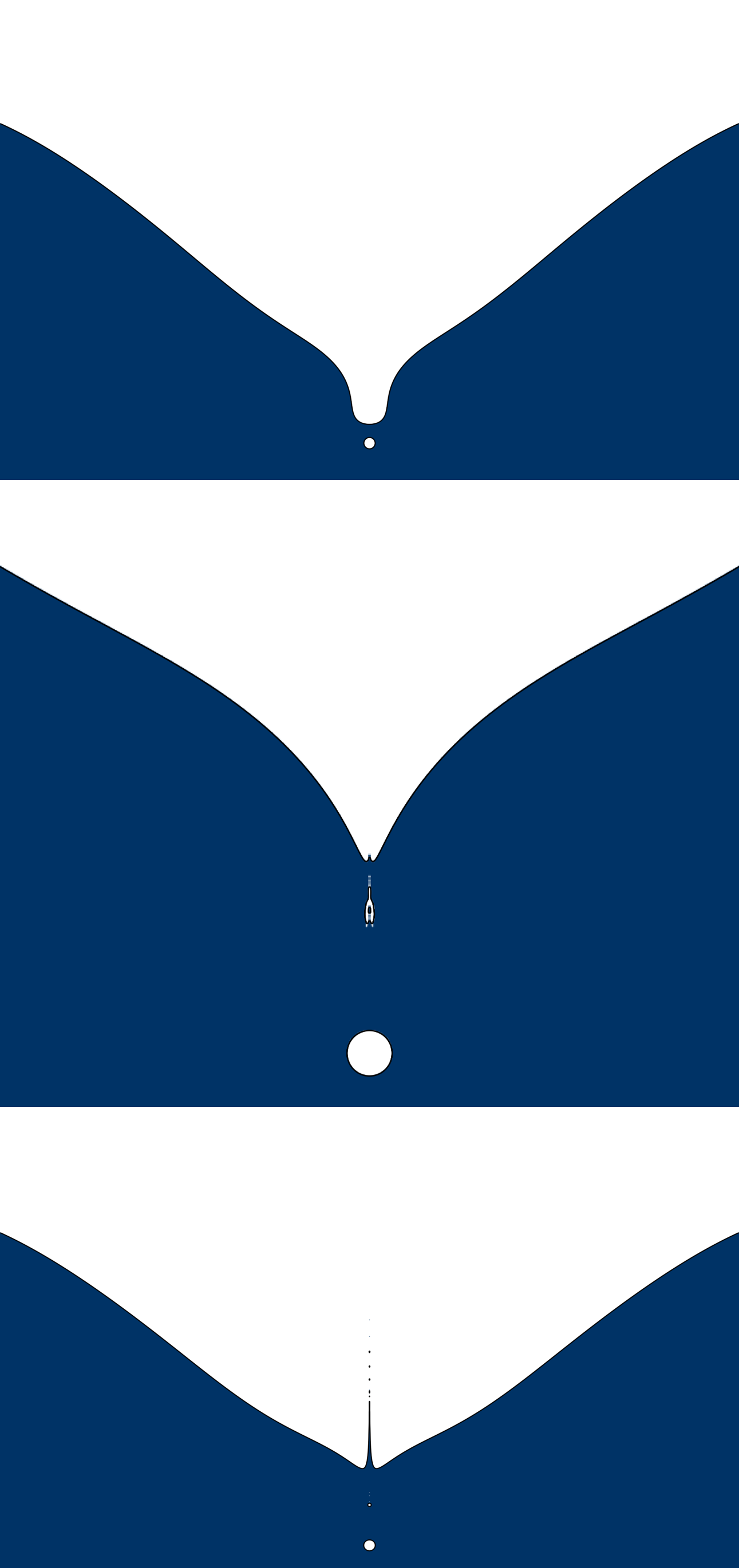}};
\node[anchor=north west,font=\bfseries,fill=white,inner sep=1pt]
  at (img.north west) {(a)};
\end{tikzpicture}
\end{minipage}\hfill
\begin{minipage}{0.32\textwidth}
\centering
\begin{tikzpicture}
\node[anchor=north west,inner sep=0] (img)
  {\includegraphics[width=\linewidth]{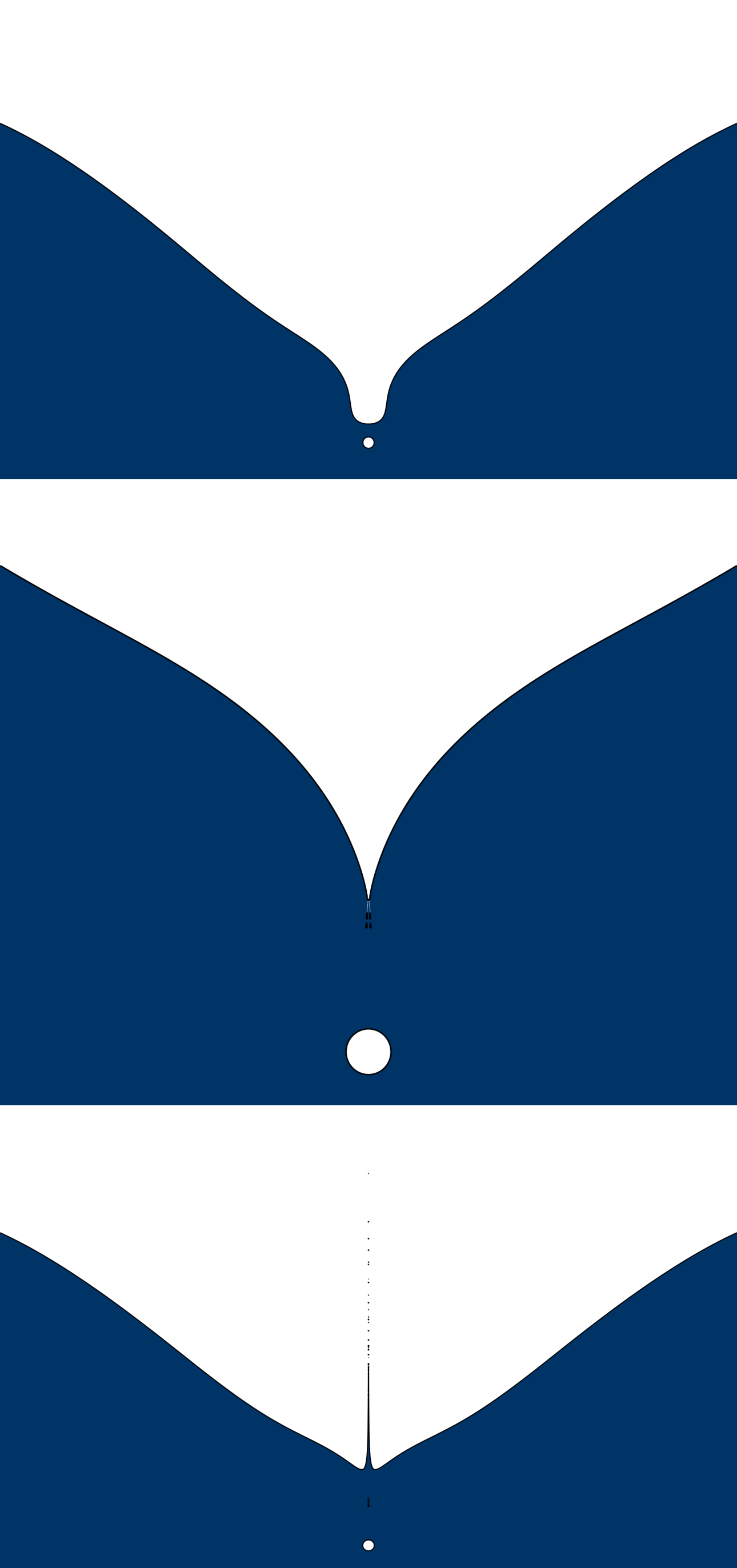}};
\node[anchor=north west,font=\bfseries,fill=white,inner sep=1pt]
  at (img.north west) {(b)};
\end{tikzpicture}
\end{minipage}\hfill
\begin{minipage}{0.32\textwidth}
\centering
\begin{tikzpicture}
\node[anchor=north west,inner sep=0] (img)
  {\includegraphics[width=\linewidth]{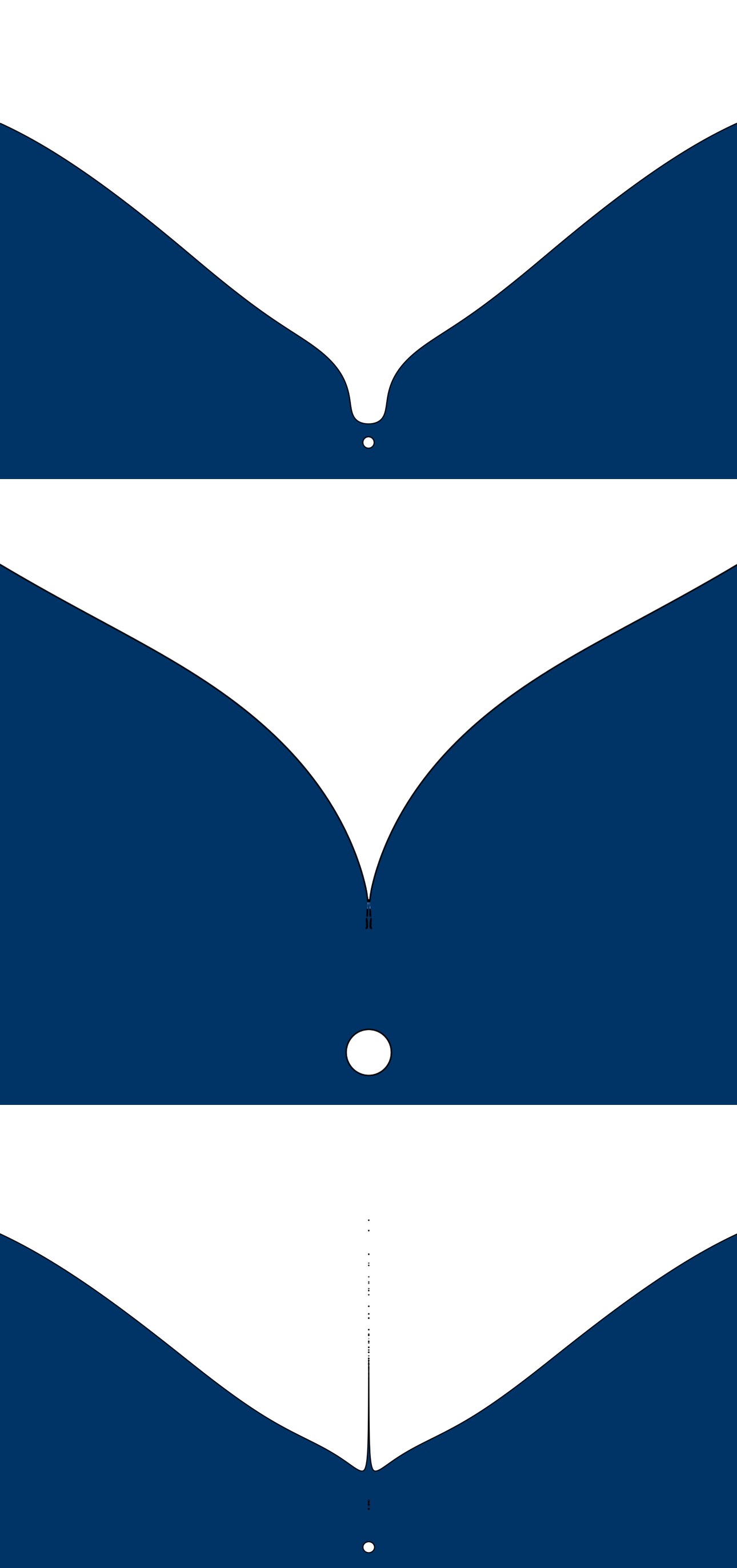}};
\node[anchor=north west,font=\bfseries,fill=white,inner sep=1pt]
  at (img.north west) {(c)};
\end{tikzpicture}
\end{minipage}

\caption{\footnotesize Grid-refinement study showing the effect of the maximum refinement level on dimple collapse and jet formation for (a) \(L_{\max}=12\), (b) \(L_{\max}=14\) and (c) \(L_{\max}=15\). The snapshots show three characteristic stages, selected with reference to the \(L_{\max}=14\) simulation: dimple formation at the crater-base (time unit $= 9.500$), subsequent dimple collapse (time  unit $= 9.603$), and jet formation at the emission of the first secondary droplet from the crater (time unit $= 9.640$). The comparison highlights differences in the impact dynamics and jetting velocity across the refinement levels.}
\label{fig:grid_resolution}
\end{figure}

We examine how spatial resolution influences the dimple collapse and jetting velocity using an adaptive mesh refinement algorithm implemented in Basilisk. The grid independence study is assessed using three different mesh refinements:  $L_{\max}=12$ $(\Delta = 6.22\,\mu\mathrm{m})$, $L_{\max}=14$ $(\Delta = 1.55\,\mu\mathrm{m})$, and $L_{\max}=15$ $(\Delta = 0.78\,\mu\mathrm{m})$. For $L_{\max}=12$, we use the same refinement level throughout the computational domain. However, increasing the refinement beyond this level is computationally expensive; hence, we employed the same strategy discussed in \S~\ref{sec:Refinemet_level}. The maximum refinement level switches to $L_{\max}=14$ or $L_{\max}=15$ from the start of dimple formation until the end of the simulation. \\

Figure ~\ref{fig:grid_resolution} shows the interface evolution at different time units across different refinement levels for $We=120$. The three characteristic stages, selected with reference to $L_{\max}=14$, highlight differences in the impact dynamics and jetting velocity across the refinement levels. These simulation time units correspond to dimple formation at the crater-base, the subsequent dimple collapse, and the jet formation when the first secondary droplet is emitted from the crater. Qualitatively, dimple evolution is consistent across all three refinement levels at a time unit of $9.500$. At time unit $9.603$, for $L_{\max}=12$, we observe a pinch-off, whereas for $L_{\max} = 14$ or $15$ pinch-off is absent and only microbubbles form at the crater-base. Afterward, at time unit $9.640$, the jet and secondary droplet emanate from the jet tip and remain inside the crater for $L_{\max}=12$. For $L_{\max}=14$, the droplet has emerged from the crater, while for $L_{\max}=15$, it remains at the crater opening and emits after two additional time units $(9.642)$. Despite this difference in ejection timing across refinement levels, the jet profile looks similar for $L_{\max}=14$ and $15$. For quantitative assessment, we compute the jetting velocity across different refinement levels. The corresponding jetting velocities are $25.24, 61.7,$ and $57.28\,\mathrm{m/s}$  for $L_{\max}=12,$ $14,$ and $15$, respectively. The jetting velocity for $L_{\max}=12$ is considerably lower than that reported by \cite{tian2023conical} in their experiments, whereas the jetting velocity for $L_{\max}=14$ agrees well with their measurements. The jetting velocity at $L_{\max}=15$ is within acceptable limits. Since the qualitative and quantitative results from $L_{\max} = 14$ and $15$ simulations are similar, we use $L_{\max} = 14$ for the singular-jet cases. For $L_{\max}=12$, the phenomenon is under-resolved. \\

\bibliographystyle{jfm}


\begin{thebibliography}{72}
\expandafter\ifx\csname natexlab\endcsname\relax\def\natexlab#1{#1}\fi
\def\au#1{#1} \def\ed#1{#1} \def\yr#1{#1}\def\at#1{#1}\def\jt#1{\textit{#1}}
  \def\bt#1{#1}\def\bvol#1{\textbf{#1}} \def\vol#1{#1} \def\pg#1{#1}
  \def\publ#1{#1}\def\arxiv#1{#1}\def\org#1{#1}\def\st#1{\textit{#1}}

\bibitem[Bell {\em et~al.\/}(1989)Bell, Colella \& Glaz]{bell1989second}
{\sc \au{Bell, John~B}, \au{Colella, Phillip} \& \au{Glaz, Harland~M}}
  \yr{1989}  \at{A second-order projection method for the incompressible
  navier-stokes equations}.  \jt{Journal of computational physics}
  \bvol{85}~(2),  \pg{257--283}.

\bibitem[Berberovi{\'c} {\em et~al.\/}(2009)Berberovi{\'c}, Van~Hinsberg,
  Jakirli{\'c}, Roisman \& Tropea]{berberovic2009drop}
{\sc \au{Berberovi{\'c}, Edin}, \au{Van~Hinsberg, Nils~P}, \au{Jakirli{\'c},
  Suad}, \au{Roisman, Ilia~V} \& \au{Tropea, Cameron}} \yr{2009}  \at{Drop
  impact onto a liquid layer of finite thickness: Dynamics of the cavity
  evolution}.  \jt{Physical Review E—Statistical, Nonlinear, and Soft Matter
  Physics}  \bvol{79}~(3),  \pg{036306}.

\bibitem[Bierbrauer(2004)]{Bierbrauer2004}
{\sc \au{Bierbrauer, F.}} \yr{2004}  \at{Mathematical modelling of
  water-droplet impact on hot galvanised steel surfaces}. Phd thesis,
  University of Wollongong, Wollongong, Australia, available through University
  of Wollongong Research Online.

\bibitem[Blanchard \& Woodcock(1957)]{blanchard1957bubble}
{\sc \au{Blanchard, DCr} \& \au{Woodcock, A~He}} \yr{1957}  \at{Bubble
  formation and modification in the sea and its meteorological significance}.
  \jt{Tellus}  \bvol{9}~(2),  \pg{145--158}.

\bibitem[Blanchard \& Syzdek(1970)]{Blanchard1970}
{\sc \au{Blanchard, Duncan~C.} \& \au{Syzdek, Lawrence}} \yr{1970}
  \at{Mechanism for the water-to-air transfer and concentration of bacteria}.
  \jt{Science}  \bvol{170}~(3958),  \pg{626--628}.

\bibitem[Brackbill {\em et~al.\/}(1992)Brackbill, Kothe \&
  Zemach]{brackbill1992continuum}
{\sc \au{Brackbill, Jeremiah~U}, \au{Kothe, Douglas~B} \& \au{Zemach, Charles}}
  \yr{1992}  \at{A continuum method for modeling surface tension}.  \jt{Journal
  of computational physics}  \bvol{100}~(2),  \pg{335--354}.

\bibitem[Cai(1989)]{cai1989phenomena}
{\sc \au{Cai, YK}} \yr{1989}  \at{Phenomena of a liquid drop falling to a
  liquid surface}.  \jt{Experiments in fluids}  \bvol{7}~(6),  \pg{388--394}.

\bibitem[Chapman \& Critchlow(1967)]{chapman1967formation}
{\sc \au{Chapman, David~S} \& \au{Critchlow, PR}} \yr{1967}  \at{Formation of
  vortex rings from falling drops}.  \jt{Journal of Fluid Mechanics}
  \bvol{29}~(1),  \pg{177--185}.

\bibitem[Chen {\em et~al.\/}(2006)Chen, Mandre \& Feng]{chen2006partial}
{\sc \au{Chen, Xiaopeng}, \au{Mandre, Shreyas} \& \au{Feng, James~J}} \yr{2006}
   \at{Partial coalescence between a drop and a liquid-liquid interface}.
  \jt{Physics of Fluids}  \bvol{18}~(5).

\bibitem[Cresswell \& Morton(1995)]{cresswell1995drop}
{\sc \au{Cresswell, RW} \& \au{Morton, BR}} \yr{1995}  \at{Drop-formed vortex
  rings—the generation of vorticity}.  \jt{Physics of Fluids}  \bvol{7}~(6),
  \pg{1363--1370}.

\bibitem[Day {\em et~al.\/}(1998)Day, Hinch \& Lister]{day1998self}
{\sc \au{Day, Richard~F}, \au{Hinch, E~John} \& \au{Lister, John~R}} \yr{1998}
  \at{Self-similar capillary pinchoff of an inviscid fluid}.  \jt{Physical
  review letters}  \bvol{80}~(4),  \pg{704}.

\bibitem[Deike {\em et~al.\/}(2018)Deike, Ghabache, Liger-Belair, Das, Zaleski,
  Popinet \& S{\'e}on]{deike2018dynamics}
{\sc \au{Deike, Luc}, \au{Ghabache, Elisabeth}, \au{Liger-Belair, G{\'e}rard},
  \au{Das, Arup~K}, \au{Zaleski, St{\'e}phane}, \au{Popinet, St{\'e}phane} \&
  \au{S{\'e}on, Thomas}} \yr{2018}  \at{Dynamics of jets produced by bursting
  bubbles}.  \jt{Physical Review Fluids}  \bvol{3}~(1),  \pg{013603}.

\bibitem[Eggers(1997)]{eggers1997nonlinear}
{\sc \au{Eggers, Jens}} \yr{1997}  \at{Nonlinear dynamics and breakup of
  free-surface flows}.  \jt{Reviews of modern physics}  \bvol{69}~(3),
  \pg{865}.

\bibitem[Ellison(1950)]{ellison1950soil}
{\sc \au{Ellison, WD}} \yr{1950}  \at{Soil erosion by rainstorms}.
  \jt{Science}  \bvol{111}~(2880),  \pg{245--249}.

\bibitem[Engel(1966)]{engel1966crater}
{\sc \au{Engel, Olive~G}} \yr{1966}  \at{Crater depth in fluid impacts}.
  \jt{Journal of Applied Physics}  \bvol{37}~(4),  \pg{1798--1808}.

\bibitem[Engel(1967)]{engel1967initial}
{\sc \au{Engel, Olive~G}} \yr{1967}  \at{Initial pressure, initial flow
  velocity, and the time dependence of crater depth in fluid impacts}.
  \jt{Journal of applied physics}  \bvol{38}~(10),  \pg{3935--3940}.

\bibitem[Esmailizadeh \& Mesler(1986)]{esmailizadeh1986bubble}
{\sc \au{Esmailizadeh, Ladan} \& \au{Mesler, Russell}} \yr{1986}  \at{Bubble
  entrainment with drops}.  \jt{Journal of colloid and interface science}
  \bvol{110}~(2),  \pg{561--574}.

\bibitem[Fedorchenko \& Wang(2004)]{fedorchenko2004some}
{\sc \au{Fedorchenko, Alexander~I} \& \au{Wang, An-Bang}} \yr{2004}  \at{On
  some common features of drop impact on liquid surfaces}.  \jt{Physics of
  Fluids}  \bvol{16}~(5),  \pg{1349--1365}.

\bibitem[Francois {\em et~al.\/}(2006)Francois, Cummins, Dendy, Kothe, Sicilian
  \& Williams]{francois2006balanced}
{\sc \au{Francois, Marianne~M}, \au{Cummins, Sharen~J}, \au{Dendy, Edward~D},
  \au{Kothe, Douglas~B}, \au{Sicilian, James~M} \& \au{Williams, Matthew~W}}
  \yr{2006}  \at{A balanced-force algorithm for continuous and sharp
  interfacial surface tension models within a volume tracking framework}.
  \jt{Journal of Computational Physics}  \bvol{213}~(1),  \pg{141--173}.

\bibitem[Franz(1959)]{franz1959splashes}
{\sc \au{Franz, GJ}} \yr{1959}  \at{Splashes as sources of sound in liquids}.
  \jt{The Journal of the Acoustical Society of America}  \bvol{31}~(8),
  \pg{1080--1096}.

\bibitem[Fuster \& Popinet(2018)]{fuster2018all}
{\sc \au{Fuster, Daniel} \& \au{Popinet, St{\'e}phane}} \yr{2018}  \at{An
  all-mach method for the simulation of bubble dynamics problems in the
  presence of surface tension}.  \jt{Journal of Computational Physics}
  \bvol{374},  \pg{752--768}.

\bibitem[Ga{\~n}{\'a}n-Calvo(2017)]{ganan2017revision}
{\sc \au{Ga{\~n}{\'a}n-Calvo, Alfonso~M}} \yr{2017}  \at{Revision of bubble
  bursting: universal scaling laws of top jet drop size and speed}.
  \jt{Physical review letters}  \bvol{119}~(20),  \pg{204502}.

\bibitem[Gonnermann \& Manga(2007)]{gonnermann2007fluid}
{\sc \au{Gonnermann, Helge~M} \& \au{Manga, Michael}} \yr{2007}  \at{The fluid
  mechanics inside a volcano}.  \jt{Annu. Rev. Fluid Mech.}  \bvol{39}~(1),
  \pg{321--356}.

\bibitem[Hirt \& Nichols(1981)]{hirt1981volume}
{\sc \au{Hirt, Cyril~W} \& \au{Nichols, Billy~D}} \yr{1981}  \at{Volume of
  fluid (vof) method for the dynamics of free boundaries}.  \jt{Journal of
  computational physics}  \bvol{39}~(1),  \pg{201--225}.

\bibitem[Jian {\em et~al.\/}(2020)Jian, Deng \& Thoraval]{jian2020air}
{\sc \au{Jian, Zhen}, \au{Deng, Peng} \& \au{Thoraval, Marie-Jean}} \yr{2020}
  \at{Air sheet contraction}.  \jt{Journal of Fluid Mechanics}  \bvol{899},
  \pg{A7}.

\bibitem[Kientzler {\em et~al.\/}(1954)Kientzler, Arons, Blanchard \&
  Woodcock]{kientzler1954photographic}
{\sc \au{Kientzler, CF}, \au{Arons, Arnold~B}, \au{Blanchard, Duncan~C} \&
  \au{Woodcock, AHhttps}} \yr{1954}  \at{Photographic investigation of the
  projection of droplets by bubbles bursting at a water surface}.  \jt{Tellus}
  \bvol{6}~(1),  \pg{1--7}.

\bibitem[Leng(2001)]{leng2001splash}
{\sc \au{Leng, Liow~Jong}} \yr{2001}  \at{Splash formation by spherical drops}.
   \jt{Journal of Fluid Mechanics}  \bvol{427},  \pg{73--105}.

\bibitem[Lewis \& Schwartz(2004)]{lewis2004sea}
{\sc \au{Lewis, Ernie~R} \& \au{Schwartz, Stephen~E}} \yr{2004} {\em Sea salt
  aerosol production: mechanisms, methods, measurements, and models\/}, ,
  \vol{vol. 152}.  \publ{American geophysical union}.

\bibitem[Lohse(2022)]{lohse2022fundamental}
{\sc \au{Lohse, Detlef}} \yr{2022}  \at{Fundamental fluid dynamics challenges
  in inkjet printing}.  \jt{Annual review of fluid mechanics}  \bvol{54}~(1),
  \pg{349--382}.

\bibitem[Marrucci(1969)]{marrucci1969theory}
{\sc \au{Marrucci, G}} \yr{1969}  \at{A theory of coalescence}.  \jt{Chemical
  engineering science}  \bvol{24}~(6),  \pg{975--985}.

\bibitem[Medwin {\em et~al.\/}(1992)Medwin, Nystuen, Jacobus, Ostwald \&
  Snyder]{medwin1992anatomy}
{\sc \au{Medwin, Herman}, \au{Nystuen, Jeffrey~A}, \au{Jacobus, Peter~W},
  \au{Ostwald, Leo~H} \& \au{Snyder, David~E}} \yr{1992}  \at{The anatomy of
  underwater rain noise}.  \jt{The Journal of the Acoustical Society of
  America}  \bvol{92}~(3),  \pg{1613--1623}.

\bibitem[Michon {\em et~al.\/}(2017)Michon, Josserand \&
  S{\'e}on]{michon2017jet}
{\sc \au{Michon, Guy-Jean}, \au{Josserand, Christophe} \& \au{S{\'e}on,
  Thomas}} \yr{2017}  \at{Jet dynamics post drop impact on a deep pool}.
  \jt{Physical review fluids}  \bvol{2}~(2),  \pg{023601}.

\bibitem[Morton {\em et~al.\/}(2000)Morton, Rudman \&
  Jong-Leng]{morton2000investigation}
{\sc \au{Morton, David}, \au{Rudman, Murray} \& \au{Jong-Leng, Liow}} \yr{2000}
   \at{An investigation of the flow regimes resulting from splashing drops}.
  \jt{Physics of Fluids}  \bvol{12}~(4),  \pg{747--763}.

\bibitem[Oguz \& Prosperetti(1990)]{oguz1990bubble}
{\sc \au{Oguz, Hasan~N} \& \au{Prosperetti, Andrea}} \yr{1990}  \at{Bubble
  entrainment by the impact of drops on liquid surfaces}.  \jt{Journal of Fluid
  Mechanics}  \bvol{219},  \pg{143--179}.

\bibitem[Popinet(2003)]{popinet2003gerris}
{\sc \au{Popinet, St{\'e}phane}} \yr{2003}  \at{Gerris: a tree-based adaptive
  solver for the incompressible euler equations in complex geometries}.
  \jt{Journal of computational physics}  \bvol{190}~(2),  \pg{572--600}.

\bibitem[Popinet(2009)]{popinet2009accurate}
{\sc \au{Popinet, St{\'e}phane}} \yr{2009}  \at{An accurate adaptive solver for
  surface-tension-driven interfacial flows}.  \jt{Journal of Computational
  Physics}  \bvol{228}~(16),  \pg{5838--5866}.

\bibitem[Popinet(2015)]{popinet2015quadtree}
{\sc \au{Popinet, St{\'e}phane}} \yr{2015}  \at{A quadtree-adaptive multigrid
  solver for the serre--green--naghdi equations}.  \jt{Journal of Computational
  Physics}  \bvol{302},  \pg{336--358}.

\bibitem[Popinet(2018)]{popinet2018numerical}
{\sc \au{Popinet, St{\'e}phane}} \yr{2018}  \at{Numerical models of surface
  tension}.  \jt{Annual Review of Fluid Mechanics}  \bvol{50},  \pg{49--75}.

\bibitem[Popinet \& Collaborators(2013--2025)]{Popinet2013}
{\sc \au{Popinet, S.} \& \au{Collaborators}} \yr{2013--2025}  \at{Basilisk} .

\bibitem[Prosperetti {\em et~al.\/}(1989)Prosperetti, Crum \&
  Pumphrey]{prosperetti1989underwater}
{\sc \au{Prosperetti, Andrea}, \au{Crum, Lawrence~A} \& \au{Pumphrey, HC}}
  \yr{1989}  \at{The underwater noise of rain}.  \jt{Journal of Geophysical
  Research: Oceans}  \bvol{94}~(C3),  \pg{3255--3259}.

\bibitem[Pumphrey {\em et~al.\/}(1989)Pumphrey, Crum \&
  Bjo/Rno/]{pumphrey1989underwater}
{\sc \au{Pumphrey, Hugh~C}, \au{Crum, LA} \& \au{Bjo/Rno/, L}} \yr{1989}
  \at{Underwater sound produced by individual drop impacts and rainfall}.
  \jt{The Journal of the Acoustical Society of America}  \bvol{85}~(4),
  \pg{1518--1526}.

\bibitem[Pumphrey \& Elmore(1990)]{pumphrey1990entrainment}
{\sc \au{Pumphrey, Hugh~C} \& \au{Elmore, Paul~A}} \yr{1990}  \at{The
  entrainment of bubbles by drop impacts}.  \jt{Journal of Fluid Mechanics}
  \bvol{220},  \pg{539--567}.

\bibitem[Ray {\em et~al.\/}(2012)Ray, Biswas \& Sharma]{ray2012bubble}
{\sc \au{Ray, Bahni}, \au{Biswas, Gautam} \& \au{Sharma, Ashutosh}} \yr{2012}
  \at{Bubble pinch-off and scaling during liquid drop impact on liquid pool}.
  \jt{Physics of Fluids}  \bvol{24}~(8).

\bibitem[Ray {\em et~al.\/}(2015)Ray, Biswas \& Sharma]{ray2015regimes}
{\sc \au{Ray, Bahni}, \au{Biswas, Gautam} \& \au{Sharma, Ashutosh}} \yr{2015}
  \at{Regimes during liquid drop impact on a liquid pool}.  \jt{Journal of
  Fluid Mechanics}  \bvol{768},  \pg{492--523}.

\bibitem[Rein(1993)]{rein1993phenomena}
{\sc \au{Rein, Martin}} \yr{1993}  \at{Phenomena of liquid drop impact on solid
  and liquid surfaces}.  \jt{Fluid dynamics research}  \bvol{12}~(2),
  \pg{61--93}.

\bibitem[Rein(1996)]{rein1996transitional}
{\sc \au{Rein, Martin}} \yr{1996}  \at{The transitional regime between
  coalescing and splashing drops}.  \jt{Journal of Fluid Mechanics}
  \bvol{306},  \pg{145--165}.

\bibitem[Renardy \& Renardy(2002)]{renardy2002prost}
{\sc \au{Renardy, Yuriko} \& \au{Renardy, Michael}} \yr{2002}  \at{Prost: a
  parabolic reconstruction of surface tension for the volume-of-fluid method}.
  \jt{Journal of computational physics}  \bvol{183}~(2),  \pg{400--421}.

\bibitem[Rodriguez \& Mesler(1988)]{rodriguez1988penetration}
{\sc \au{Rodriguez, Francisco} \& \au{Mesler, Russell}} \yr{1988}  \at{The
  penetration of drop-formed vortex rings into pools of liquid}.  \jt{Journal
  of colloid and interface science}  \bvol{121}~(1),  \pg{121--129}.

\bibitem[Rudman(1998)]{rudman1998volume}
{\sc \au{Rudman, Murray}} \yr{1998}  \at{A volume-tracking method for
  incompressible multifluid flows with large density variations}.
  \jt{International Journal for numerical methods in fluids}  \bvol{28}~(2),
  \pg{357--378}.

\bibitem[S{\'e}on \& Liger-Belair(2017)]{seon2017effervescence}
{\sc \au{S{\'e}on, Thomas} \& \au{Liger-Belair, G{\'e}rard}} \yr{2017}
  \at{Effervescence in champagne and sparkling wines: From bubble bursting to
  droplet evaporation}.  \jt{The European Physical Journal Special Topics}
  \bvol{226}~(1),  \pg{117--156}.

\bibitem[Sun {\em et~al.\/}(2023)Sun, Zeng, Deng, Zhang \&
  Zhang]{sun2023droplet}
{\sc \au{Sun, Zhengnan}, \au{Zeng, Xu}, \au{Deng, Xu}, \au{Zhang, Xiaosheng} \&
  \au{Zhang, Yi}} \yr{2023}  \at{Droplet interface in additive manufacturing:
  from process to application}.  \jt{Droplet}  \bvol{2}~(2),  \pg{e57}.

\bibitem[Thomson \& Newall(1886)]{thomson1886v}
{\sc \au{Thomson, Joseph~John} \& \au{Newall, Hugh~Frank}} \yr{1886}  \at{On
  the formation of vortex rings by drops falling into liquids, and some allied
  phenomena}.  \jt{Proceedings of the royal society of London}
  \bvol{39}~(239-241),  \pg{417--436}.

\bibitem[Thoraval {\em et~al.\/}(2012)Thoraval, Takehara, Etoh, Popinet, Ray,
  Josserand, Zaleski \& Thoroddsen]{thoraval2012karman}
{\sc \au{Thoraval, Marie-Jean}, \au{Takehara, Kohsei}, \au{Etoh,
  Takeharu~Goji}, \au{Popinet, St{\'e}phane}, \au{Ray, Pascal}, \au{Josserand,
  Christophe}, \au{Zaleski, St{\'e}phane} \& \au{Thoroddsen, Sigurdur~T}}
  \yr{2012}  \at{von k{\'a}rm{\'a}n vortex street within an impacting drop}.
  \jt{Physical review letters}  \bvol{108}~(26),  \pg{264506}.

\bibitem[Thoroddsen {\em et~al.\/}(2003)Thoroddsen, Etoh \&
  Takehara]{thoroddsen2003air}
{\sc \au{Thoroddsen, ST}, \au{Etoh, TG} \& \au{Takehara, K}} \yr{2003}  \at{Air
  entrapment under an impacting drop}.  \jt{Journal of Fluid Mechanics}
  \bvol{478},  \pg{125--134}.

\bibitem[Thoroddsen {\em et~al.\/}(2007{\natexlab{{\em a\/}}})Thoroddsen, Etoh
  \& Takehara]{thoroddsen2007experiments}
{\sc \au{Thoroddsen, ST}, \au{Etoh, TG} \& \au{Takehara, K}}
  \yr{2007{\natexlab{{\em a\/}}}}  \at{Experiments on bubble pinch-off}.
  \jt{Physics of Fluids}  \bvol{19}~(4).

\bibitem[Thoroddsen {\em et~al.\/}(2007{\natexlab{{\em b\/}}})Thoroddsen, Etoh
  \& Takehara]{thoroddsen2007microjetting}
{\sc \au{Thoroddsen, ST}, \au{Etoh, TG} \& \au{Takehara, K}}
  \yr{2007{\natexlab{{\em b\/}}}}  \at{Microjetting from wave focusing on
  oscillating drops}.  \jt{Physics of Fluids}  \bvol{19}~(5).

\bibitem[Thoroddsen {\em et~al.\/}(2018)Thoroddsen, Takehara, Nguyen \&
  Etoh]{thoroddsen2018singular}
{\sc \au{Thoroddsen, Sigurdur~T}, \au{Takehara, K}, \au{Nguyen, HD} \&
  \au{Etoh, TG}} \yr{2018}  \at{Singular jets during the collapse of
  drop-impact craters}.  \jt{Journal of Fluid Mechanics}  \bvol{848},  \pg{R3}.

\bibitem[Thoroddsen {\em et~al.\/}(2012)Thoroddsen, Thoraval, Takehara \&
  Etoh]{thoroddsen2012micro}
{\sc \au{Thoroddsen, Sigurdur~T}, \au{Thoraval, M-J}, \au{Takehara, K} \&
  \au{Etoh, TG}} \yr{2012}  \at{Micro-bubble morphologies following drop
  impacts onto a pool surface}.  \jt{Journal of fluid mechanics}  \bvol{708},
  \pg{469--479}.

\bibitem[Tian {\em et~al.\/}(2023)Tian, Yang \& Thoroddsen]{tian2023conical}
{\sc \au{Tian, Yuan~Si}, \au{Yang, Zi~Qiang} \& \au{Thoroddsen, Sigurdur~T.}}
  \yr{2023}  \at{Conical focusing: mechanism for singular jetting from
  collapsing drop-impact craters}.  \jt{Journal of Fluid Mechanics}
  \bvol{958},  \pg{R1}.

\bibitem[Tran {\em et~al.\/}(2016)Tran, Lee, Lee, Woo, Han, Kim \&
  Hwang]{tran2016hydrodynamic}
{\sc \au{Tran, Thanh~Tinh}, \au{Lee, Eun~Gyo}, \au{Lee, In~Su}, \au{Woo,
  Nam~Sub}, \au{Han, Sang~Mok}, \au{Kim, Young~Ju} \& \au{Hwang, Wook~Ryol}}
  \yr{2016}  \at{Hydrodynamic extensional stress during the bubble bursting
  process for bioreactor system design}.  \jt{Korea-Australia Rheology Journal}
   \bvol{28}~(4),  \pg{315--326}.

\bibitem[Tsimplis \& Thorpe(1989)]{tsimplis1989wave}
{\sc \au{Tsimplis, M} \& \au{Thorpe, SA}} \yr{1989}  \at{Wave damping by rain}.
   \jt{Nature}  \bvol{342}~(6252),  \pg{893--895}.

\bibitem[Van~Hooft {\em et~al.\/}(2018)Van~Hooft, Popinet, Van~Heerwaarden,
  Van~der Linden, De~Roode \& Van~de Wiel]{van2018towards}
{\sc \au{Van~Hooft, J~Antoon}, \au{Popinet, St{\'e}phane}, \au{Van~Heerwaarden,
  Chiel~C}, \au{Van~der Linden, Steven~JA}, \au{De~Roode, Stephan~R} \&
  \au{Van~de Wiel, Bas~JH}} \yr{2018}  \at{Towards adaptive grids for
  atmospheric boundary-layer simulations}.  \jt{Boundary-layer meteorology}
  \bvol{167}~(3),  \pg{421--443}.

\bibitem[Veron(2015)]{veron2015ocean}
{\sc \au{Veron, Fabrice}} \yr{2015}  \at{Ocean spray}.  \jt{Annual Review of
  Fluid Mechanics}  \bvol{47}~(1),  \pg{507--538}.

\bibitem[Walls {\em et~al.\/}(2014)Walls, Bird \& Bourouiba]{walls2014moving}
{\sc \au{Walls, Peter~LL}, \au{Bird, James~C} \& \au{Bourouiba, Lydia}}
  \yr{2014}  \at{Moving with bubbles: a review of the interactions between
  bubbles and the microorganisms that surround them}.  \jt{American Zoologist}
  \bvol{54}~(6),  \pg{1014--1025}.

\bibitem[Walls {\em et~al.\/}(2015)Walls, Henaux \& Bird]{walls2015jet}
{\sc \au{Walls, Peter~LL}, \au{Henaux, Louis} \& \au{Bird, James~C}} \yr{2015}
  \at{Jet drops from bursting bubbles: How gravity and viscosity couple to
  inhibit droplet production}.  \jt{Physical Review E}  \bvol{92}~(2),
  \pg{021002}.

\bibitem[Walls {\em et~al.\/}(2017)Walls, McRae, Natarajan, Johnson, Antoniou
  \& Bird]{walls2017quantifying}
{\sc \au{Walls, Peter~LL}, \au{McRae, Oliver}, \au{Natarajan, Venkatesh},
  \au{Johnson, Chris}, \au{Antoniou, Chris} \& \au{Bird, James~C}} \yr{2017}
  \at{Quantifying the potential for bursting bubbles to damage suspended
  cells}.  \jt{Scientific reports}  \bvol{7}~(1),  \pg{15102}.

\bibitem[Wang {\em et~al.\/}(2023)Wang, Liu, Bayeul-Lain{\'e}, Murphy, Katz \&
  Coutier-Delgosha]{wang2023analysis}
{\sc \au{Wang, Hui}, \au{Liu, Shuo}, \au{Bayeul-Lain{\'e}, Annie-Claude},
  \au{Murphy, David}, \au{Katz, Joseph} \& \au{Coutier-Delgosha, Olivier}}
  \yr{2023}  \at{Analysis of high-speed drop impact onto deep liquid pool}.
  \jt{Journal of Fluid Mechanics}  \bvol{972},  \pg{A31}.

\bibitem[Weymouth \& Yue(2010)]{weymouth2010conservative}
{\sc \au{Weymouth, Gabriel~D} \& \au{Yue, Dick K-P}} \yr{2010}
  \at{Conservative volume-of-fluid method for free-surface simulations on
  cartesian-grids}.  \jt{Journal of Computational Physics}  \bvol{229}~(8),
  \pg{2853--2865}.

\bibitem[Williams(1973)]{williams1973combustion}
{\sc \au{Williams, Alan}} \yr{1973}  \at{Combustion of droplets of liquid
  fuels: a review}.  \jt{Combustion and flame}  \bvol{21}~(1),  \pg{1--31}.

\bibitem[Worthington(1908)]{worthington1908study}
{\sc \au{Worthington, Arthur~Mason}} \yr{1908} {\em A study of splashes\/}.
  \publ{Longmans, Green, and Company}.

\bibitem[Yarin(2006)]{yarin2006drop}
{\sc \au{Yarin, Alexander~L}} \yr{2006}  \at{Drop impact dynamics: Splashing,
  spreading, receding, bouncing…}.  \jt{Annu. Rev. Fluid Mech.}
  \bvol{38}~(1),  \pg{159--192}.

\bibitem[Zeff {\em et~al.\/}(2000)Zeff, Kleber, Fineberg \&
  Lathrop]{zeff2000singularity}
{\sc \au{Zeff, Benjamin~W}, \au{Kleber, Benjamin}, \au{Fineberg, Jay} \&
  \au{Lathrop, Daniel~P}} \yr{2000}  \at{Singularity dynamics in curvature
  collapse and jet eruption on a fluid surface}.  \jt{Nature}
  \bvol{403}~(6768),  \pg{401--404}.

\end{thebibliography}

\end{document}